\documentclass[twocolumn, showpacs,amsmath,amssymb,aps,pra,floatfix,longbibliography, 10pt]{revtex4-1}

\usepackage[latin1]{inputenc}
\usepackage[american]{babel}
\usepackage[T1]{fontenc}
\usepackage{graphicx}
\usepackage{amsfonts}
\usepackage{amsmath}
\usepackage{bbold}
\usepackage{mathrsfs}
\usepackage{hyperref}
\usepackage{epstopdf}
\usepackage{bm, epsfig}

\usepackage[tbtags]{mathtools}
\usepackage{afterpage}

\DeclarePairedDelimiterX\braket[2]{\langle}{\rangle}{#1 \delimsize\vert #2}
\hypersetup{colorlinks=true, urlcolor=blue, citecolor=blue, filecolor=blue, linkcolor=blue}
\newcommand*{\diff}{\mathop{}\!\mathrm{d}}

\newcommand*{\Imm}{\mathop{}\!\mathbf{Im}}

\newcommand*{\Mtau}{M_{\tau}}
\newcommand{\uimm}{\mathrm{i}}
\newcommand{\eu}{\mathrm{e}}
\newcommand{\diag}{\mathrm{diag}}

\newcommand{\daga}{^{\dagger}}

\newcommand*{\sinc}{\mathop{}\!\mathrm{sinc}}

\begin{document}

\title{Light-induced states in the transient-absorption spectrum \\ of a periodically pumped strong-field-excited system}

\author{Juliane~Haug}
\altaffiliation[On leave from: ]{Eberhard Karls Universit\"at T\"ubingen, Geschwister-Scholl-Platz, 72074 T\"ubingen, Germany}
\affiliation{Max-Planck-Institut f\"ur Kernphysik, Saupfercheckweg 1, 69117 Heidelberg, Germany}
\author{Stefano~M.~Cavaletto}
\email[Email: ]{smcavaletto@gmail.com}
\affiliation{Max-Planck-Institut f\"ur Kernphysik, Saupfercheckweg 1, 69117 Heidelberg, Germany}
\date{\today}
\begin{abstract}
The transient-absorption spectrum of a $V$-type three-level system is investigated, when this is periodically excited by a train of equally spaced, $\delta$-like pump pulses as, e.g., from an optical-frequency-comb laser. We show that, even though the probe pulse is not assumed to be much shorter than the pump pulses, light-induced states appear in the absorption spectrum. The frequency- and time-delay-dependent features of the absorption spectra are investigated as a function of several laser control parameters, such as the number of pump pulses used, their pulse area, and the pulse-to-pulse phase shift. We show that the frequencies of the light-induced states and the time-delay-dependent features of the spectra contain information on the action of the intense pulses exciting the system, which can thus complement the information on light-imposed amplitude and phase changes encoded in the absorption line shapes.
%\pacs{32.80.Qk, 32.80.Wr, 42.65.Re}
\end{abstract}

%letteraaa

\maketitle

\section{Introduction}

With the advent of femto- and attosecond pulses, transient-absorption spectroscopy (TAS) \cite{doi:10.1146/annurev.pc.43.100192.002433, 0953-4075-49-6-062003} has established itself as a powerful method to study strong-field quantum dynamics in atoms \cite{PhysRevLett.98.143601, GoulielmakisNature466, PhysRevLett.105.143002, Wirth195, PhysRevLett.106.123601, Sabbar-NaturePhys}, molecules \cite{WARRICK2017408, ReduzziJPB, PhysRevA.94.023403}, and solids \cite{Schultze-Nature, Schultze1348, Lucchini916, Moulet1134}. First experiments employed a traditional pump--probe setup, where a short pump pulse is used to excite strong-field dynamics in the system, and its time response is observed by measuring the absorption spectrum of a probe pulse at different time delays \cite{Mathies06051988, GoulielmakisNature466}. However, increasing attention has been received in recent years by theoretical studies and experiments in which the probe pulse either precedes or overlaps with the strong pump pulse \cite{PhysRevA.86.063408, Chini-SciRep, PhysRevA.96.013430}. In several attosecond transient-absorption-spectroscopy (ATAS) experiments, for example, the spectrum of an attosecond extreme-ultraviolet (XUV) pulse is observed in the presence of a subsequent strong femtosecond infrared (IR) pulse dressing the states of the atomic system. In the absence of the IR pulse, the XUV spectrum consists of lines centered on the transition energies between the ground state and the so-called bright states directly excited by the XUV pulse. However, in the presence of a strong IR pulse coupling these bright states to other dark levels, light-induced states (LISs) appear in the spectrum during overlap, associated with the dressed states of the system. 

The modification of the absorption line shapes and the appearance of LISs in the spectrum have emerged as key ingredients to understand and control strong-field quantum dynamics. As recently demonstrated, the line shapes contain the full information about the temporal response of a strongly driven quantum system, enabling its reconstruction without scanning over time delays \cite{stooss_reconstructing}, as long as the dynamics are initiated and probed by a sufficiently short pulse. However, for LISs to appear during overlap, and in order to enable real-time reconstruction of dipole responses directly from absorption spectra, it is necessary that the probe pulse be much shorter than the timescale of the observed strong-field-induced dynamics.

If the duration of the probe pulse is comparable with that of the pump pulse, the absorption spectra do not provide the resolution necessary to access the real-time dynamics taking place \textit{during} the pump pulse. This is for example the case when optical transitions have to be studied or controlled with optical femtosecond pump and probe pulses of equal duration \cite{PhysRevLett.115.033003, Liu-SciRep, PhysRevA.95.043413, 0953-4075-51-3-035501}. The Rabi oscillations responsible for the appearance of LISs take place only within the pump pulse, and a probe pulse of equal duration does not have the required resolution to distinguish them. Therefore, no LISs appear in the spectrum as a signature of strong-field dynamics in such case. However, the absorption lines of the bright states still carry information about the total (integrated and non-time-resolved) action of the pulse on the atomic system, which is imprinted in the associated line shapes and can be understood in terms of light-imposed amplitude and phase changes \cite{PhysRevLett.109.073601, PhysRevA.88.033409, Ott10052013, PhysRevLett.112.103001, Meyer22122015}. 

\begin{figure}
\centering
\includegraphics[width=\linewidth]{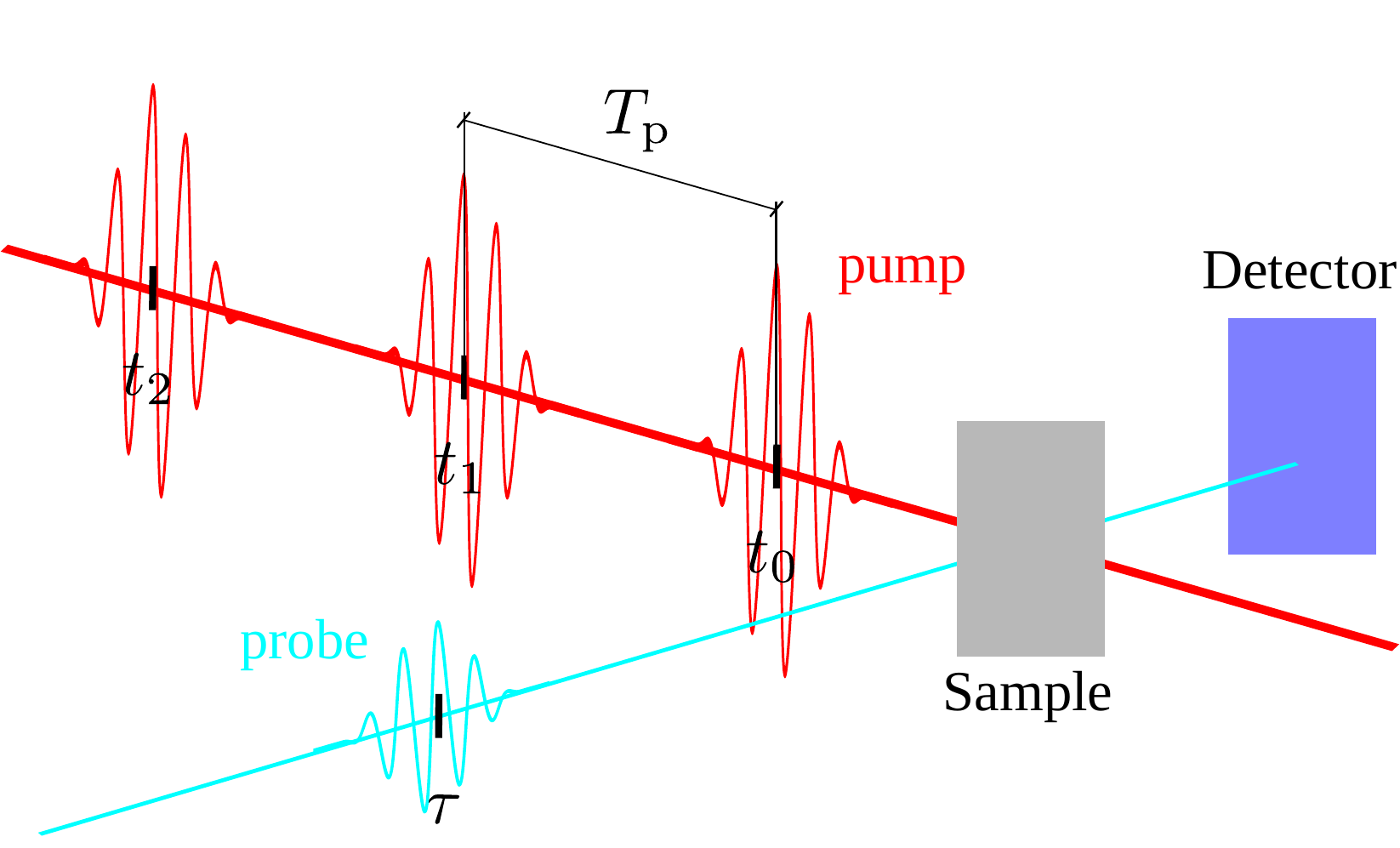}
\caption{Experimental setup for the detection of the optical-density transient-absorption spectrum of a transmitted probe pulse (light blue) in the presence of an additional pump field (red) consisting of a train of pulses in a noncollinear geometry.}
\label{fig:Experiment}
\end{figure}

Here, we show that, even when probe and pump pulses have the same duration, LISs appear in the absorption spectrum of the probe pulse if a train of pump pulses is used, employing the TAS setup shown in Fig.~\ref{fig:Experiment}. Periodic trains of intense optical pulses, as provided by optical-frequency-comb lasers \cite{Nature.416.233, 0022-3727-35-8-201, RevModPhys.75.325}, have found numerous applications in precision spectroscopy \cite{PhysRevLett.82.3568}, the development of all-optical atomic clocks \cite{Diddams03082001}, and attosecond science \cite{Nature.421.611}. Furthermore, by exploiting coherent pulse accumulation and quantum interference
effects, they have also been employed for the control of atomic coherences \cite{PhysRevLett.96.153001, PhysRevLett.98.113004, PhysRevLett.100.203001} and united time--frequency spectroscopy \cite{Marian17122004}. Control of the x-ray transient-absorption spectrum with an optical frequency comb has also been put forward for the generation of an x-ray frequency comb \cite{nphoton.2014.113, 1367-2630-16-9-093005}. 

We investigate the dynamics of a $V$-type three-level system, modeling optical transitions in atomic Rb, when excited by ultrashort optical pump and probe pulses. We model the pump field as a periodic train of $N$ identically spaced pulses, showing that, for sufficiently large values of $N$, LISs appear in the absorption spectrum of the probe pulse. The spectra are investigated as a function of the delay between the probe and pump fields, in the case of a probe pulse preceding, in between, or following the pump pulses. We show that the strong-field action of the pump pulses is encoded in the central frequencies of the LISs appearing in the spectrum, and in their time-delay-dependent periodic properties. This enables the extraction of information about the intensity-dependent action of the pump pulses on the system directly from the frequency of the LISs. It can thus be used to complement the information obtained in the case of a single pump pulse, where no LISs appear and the action of the intense pump pulse is exclusively encoded in the line shapes of the bright states.

The paper is organized as follows. Section~\ref{Theoretical Model} introduces the theoretical model used to describe the $V$-type three-level system and its interaction with the pump and probe fields (\ref{Three-level model and equations of motion}), and the transient-absorption spectrum (\ref{Transient-absorption spectrum}). In Sec.~\ref{Dynamics of the system and associated spectrum}, the dynamics of the system and the associated spectra are calculated for a train of $N$ equally distant $\delta$-like pump pulses, for a probe--pump (\ref{Probe--pump setup}), pump--probe (\ref{Pump--probe setup}), and pump--probe--pump setup (\ref{Pump--probe--pump setup}), depending on the position of the probe pulse with respect to the train of pump pulses. The resulting spectra are presented and discussed in Sec.~\ref{Results and discussion} for different values of the laser control parameters. In particular, we investigate the appearance of LISs for an increasing number of pump pulses (\ref{Dependence on the number of pump pulses}), focusing on $N\rightarrow \infty$ (\ref{Dependence on the pulse-to-pulse phase shift and pulse area for infinitely many pump pulses}), for which we highlight the  frequency- and time-delay-dependent features exhibited by the spectra, in Secs.~\ref{Frequency-dependent features and position of the additional spectral lines} and \ref{Time-delay-dependent features for given values of the pulse-to-pulse phase shift}, respectively. Additional mathematical details are included in the Appendixes. Atomic units are used throughout unless otherwise stated.

\section{Theoretical Model}
\label{Theoretical Model}

\subsection{Three-level model and equations of motion}
\label{Three-level model and equations of motion}

The TAS geometry is displayed in Fig.~\ref{fig:Experiment}. It features a probe pulse, whose absorption spectrum is detected upon transmission through the atomic sample, and an additional pump field, consisting of a train of pulses, which modifies the dipole response of the atomic system. The pulses considered in the following have the form \cite{diels2006ultrashort}
\begin{equation}
\begin{aligned}
\boldsymbol{\mathcal{E}}(t, t_{\mathrm{c}}, \phi) &= \mathcal{E}(t, t_{\mathrm{c}}, \phi)\,\hat{\boldsymbol{e}}_z\\ 
&=\mathcal{E}_{0}(t - t_{\mathrm{c}})\, \cos[\omega_{\mathrm{c}}(t - t_{\mathrm{c}}) + \phi]\,\hat{\boldsymbol{e}}_z,
\end{aligned}
\label{eq:singlepulse}
\end{equation}
where $\mathcal{E}(t, t_{\mathrm{c}}, \phi)$ is the amplitude of the field and $\hat{\boldsymbol{e}}_z$ is the direction of linear polarization. Here, we have introduced the central time of the pulse $t_{\mathrm{c}}$, its carrier frequency $\omega_{\mathrm{c}}$, envelope function $\mathcal{E}_{0}(t)$, and carrier-envelope phase (CEP) $\phi$.

\begin{figure}
\centering
\includegraphics[width=\linewidth]{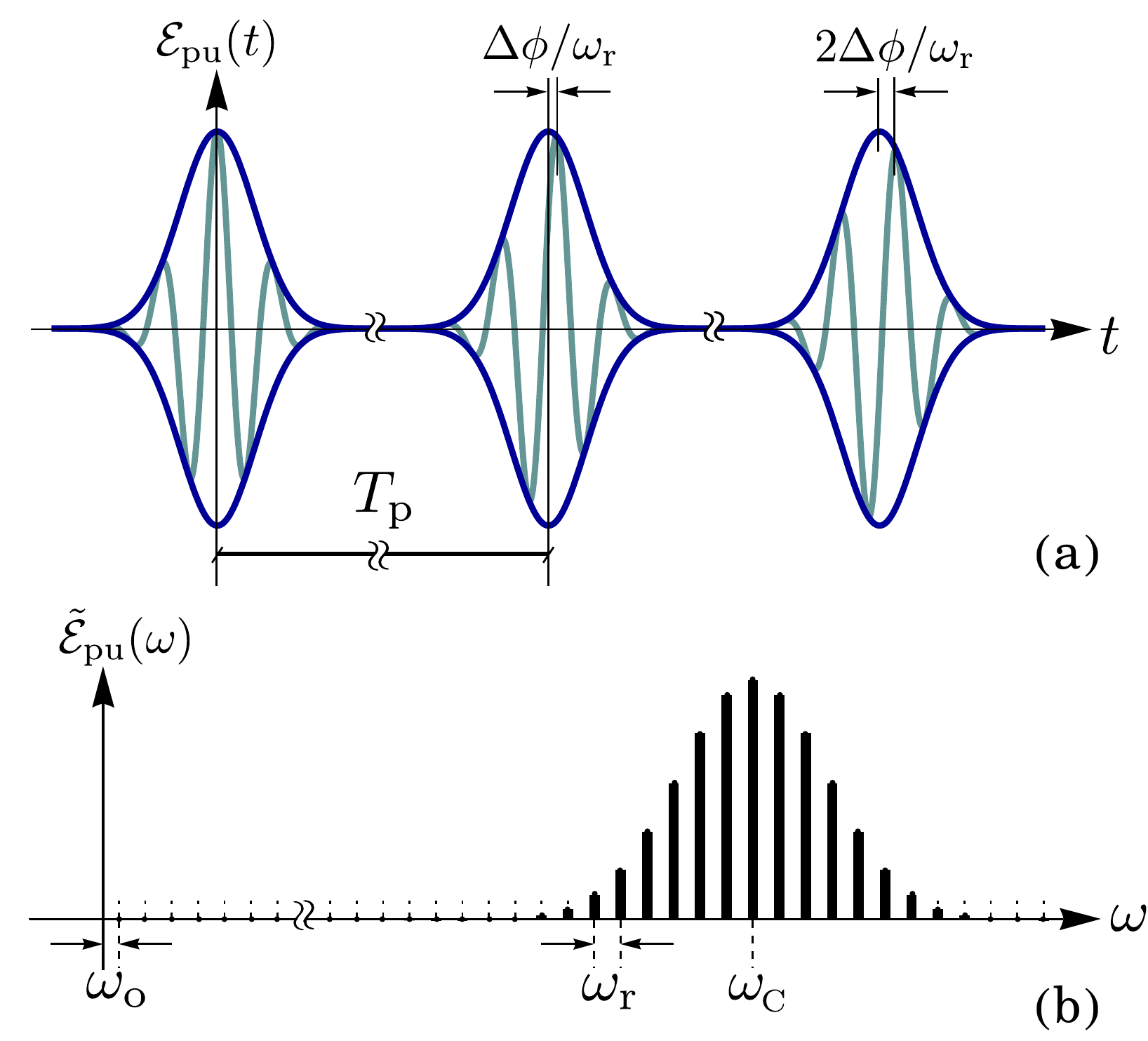}
\caption{(a) Time-dependent pump field (light blue) with envelope function (dark blue). (b) Spectrum of a pump-pulse train for $N\rightarrow \infty$.}
\label{fig:CombTimeAndFreq}
\end{figure}

The time-dependent pump field \cite{Nature.416.233, 0022-3727-35-8-201, RevModPhys.75.325}
\begin{equation}
\begin{aligned}
&\boldsymbol{\mathcal{E}}_{\mathrm{pu}}(t) = \mathcal{E}_{\mathrm{pu}}(t)\,\hat{\boldsymbol{e}}_z =\\% \sum_{n=0}^{N-1} \boldsymbol{\mathcal{E}}(t, n T_{\mathrm{p}}, \phi_{0,\mathrm{pu}} + n \Delta\phi)= \\ 
&\sum_{n=0}^{N-1} \mathcal{E}_{0,\mathrm{pu}}(t - n T_{\mathrm{p}})\, \cos[\omega_{\mathrm{c}}(t - n T_{\mathrm{p}}) + \phi_{0,\mathrm{pu}} + n \Delta\phi]\,\hat{\boldsymbol{e}}_z,
\end{aligned}
\label{eq:pumpfield}
\end{equation}
consists of a train of $N$ equally spaced pulses, centered at times $t_n = n T_{\mathrm{p}}$, $n\in\{0,\,1,\,\ldots,N-1\}$, separated by a repetition period $T_{\mathrm{p}}$, and with envelope function $\mathcal{E}_{0,\mathrm{pu}}(t)$ as shown in Fig.~\ref{fig:CombTimeAndFreq}(a). In the following, we will refer to a pulse centered on $t_n$ as the $n$th pulse---for instance, the $0$th pulse will always denote the first-arriving pump pulse centered on $t_0$. The CEP of the $n$th pulse is given by $\phi_{0,\mathrm{pu}} + n \Delta\phi$, where the CEP $\phi_{0,\mathrm{pu}}$ of the initial $0$th pulse and the constant pulse-to-pulse phase shift $\Delta\phi$ are both $\phi_0,\ \Delta\phi \in [0,\, 2\pi]$.

We define the Fourier transform of a generic time-dependent function $g(t)$ as
\begin{equation}
\tilde{g}(\omega) = \int_{-\infty}^{\infty}g(t)\,\eu^{-\uimm \omega t}\,\diff t.
\end{equation}
For a single pulse, $N=1$, the Fourier transform $\tilde{\mathcal{E}}_{\mathrm{pu}}(\omega)$ of the pump field $\mathcal{E}_{\mathrm{pu}}(t)$ is obtained by the Fourier transform $\tilde{\mathcal{E}}_{0,\mathrm{pu}}(\omega - \omega_{\mathrm{c}})$ of the envelope function shifted by the carrier frequency $\omega_{\mathrm{c}}$. However, for an infinite train of pulses, $N\rightarrow \infty$, $\tilde{\mathcal{E}}_{\mathrm{pu}}(\omega)$ consists of a set of equally spaced lines centered on the frequencies
\begin{equation}
\omega_m = \omega_{\mathrm{o}} + m\omega_{r},\ m\in\mathbb{Z}
\label{eq:omegam}
\end{equation}
with the repetition frequency and offset frequency
\begin{equation}
\omega_{\mathrm{r}} = \frac{2\pi}{T_{\mathrm{p}}},\ \ \ \ \ \omega_{\mathrm{o}} = \frac{\Delta \phi}{T_{\mathrm{p}}},
\label{eq:omegarandomegao}
\end{equation}
respectively \cite{Nature.416.233, 0022-3727-35-8-201, RevModPhys.75.325}. The strength of the lines is modulated by $\tilde{\mathcal{E}}_{0,\mathrm{pu}}(\omega - \omega_{\mathrm{c}})$. This is shown in Fig.~\ref{fig:CombTimeAndFreq}(b) and further discussed in Appendix~\ref{Appendix:Time--frequency description of the train of pump pulses}.

In addition to a train of pump pulses, a weak probe pulse is used,
\begin{equation}
\begin{aligned}
\boldsymbol{\mathcal{E}}_{\mathrm{pr}}(t) %&= \boldsymbol{\mathcal{E}}(t, \tau,\phi_{0,\mathrm{pr}} )\\ &
=  \mathcal{E}_{0,\mathrm{pr}}(t - \tau)\, \cos[\omega_{\mathrm{c}}(t - \tau)+\phi_{0,\mathrm{pr}}]\,\hat{\boldsymbol{e}}_z,
\end{aligned}
\end{equation}
whose absorption spectrum is measured upon interaction with the atomic sample, as shown in Fig.~\ref{fig:Experiment}. The probe pulse is assumed to be linearly polarized, with envelope $\mathcal{E}_{0,\mathrm{pr}}(t)$, CEP $\phi_{0,\mathrm{pr}}$, and is centered on $\tau$. This represents the time delay between $\boldsymbol{\mathcal{E}}_{\mathrm{pr}}(t)$ and the initial pulse in the train of pulses $\boldsymbol{\mathcal{E}}_{\mathrm{pu}}(t)$. A negative time delay $\tau<0$ models a probe--pump experimental setup in which the probe pulse precedes the train of pump pulses. In contrast, positive time delays can either model a pump--probe--pump setup, in which the probe pulse is preceded and followed by pump pulses; or a pump--probe setup, where the probe pulse excites the system after the total (and finite) number $N$ of pump pulses.

\begin{figure}
\centering
\includegraphics[width=\linewidth]{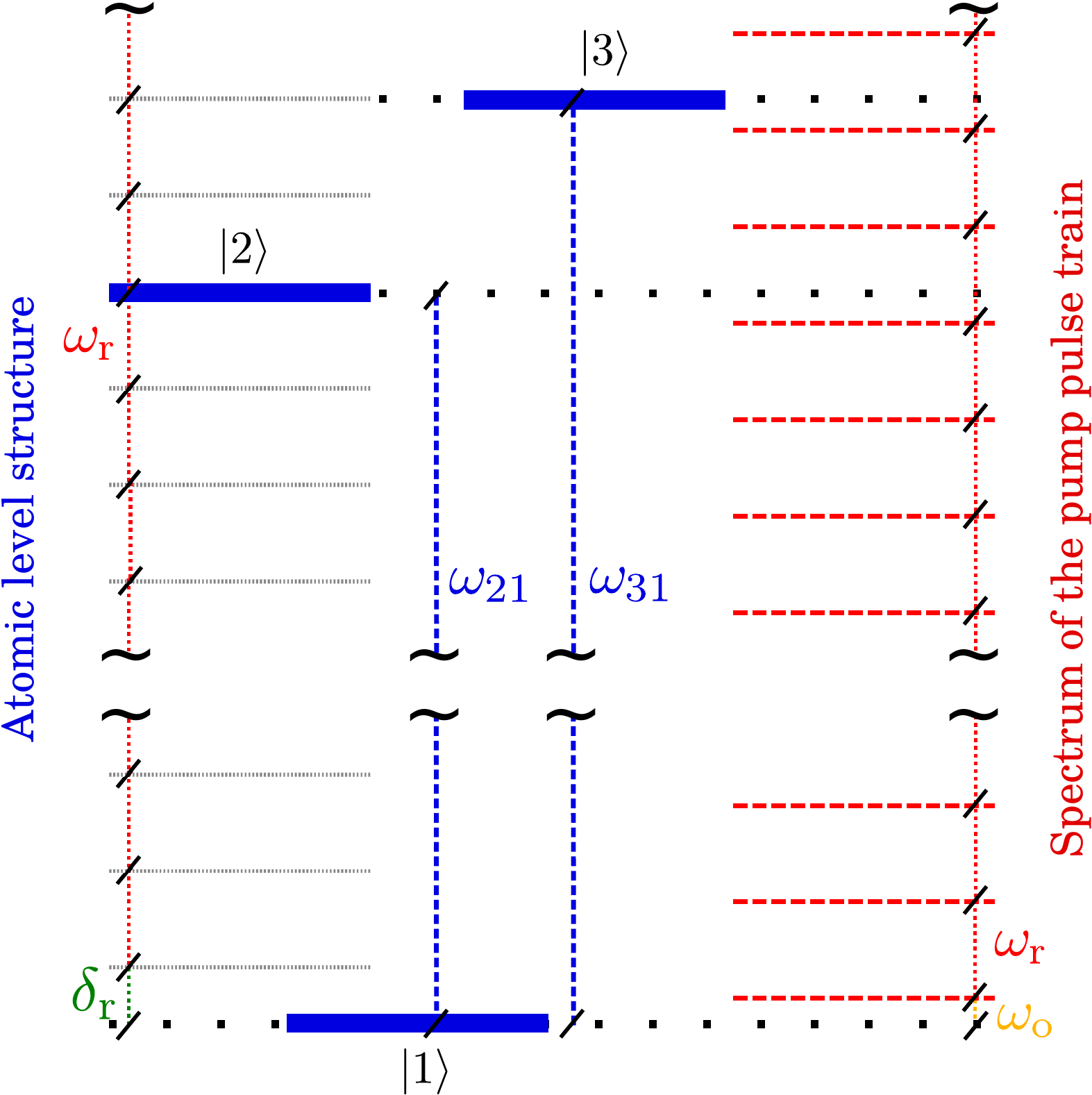}
\caption{$V$-type three-level scheme (blue), with transition energies
$\omega_{21}$ and $\omega_{31}$, used to model Rb atoms interacting with
broadband laser pulses. The red lines on the right display the spectrum of the pump field in the case of a train of $\delta$~pulses equally separated by the repetition period $T_{\mathrm{p}} = 2\pi/\omega_{\mathrm{r}}$. The black lines on the left introduce the $\omega_{\mathrm{r}}$-dependent effective detuning $\delta_{\mathrm{r}}$.}
\label{fig:MoreLevelSystem}
\end{figure}

The pulses excite the $V$-type three level system shown in Fig.~\ref{fig:MoreLevelSystem}, with electric-dipole-($E1$-)allowed transitions $|1\rangle\rightarrow |k\rangle$, $k\in\{2,\,3\}$. This is here used to model $5s\,^2S_{1/2}\rightarrow 5p\,^2P_{1/2}$ and $5s\,^2S_{1/2}\rightarrow 5p\,^2P_{3/2}$ transitions in Rb atoms between the ground state and the fine-structure-split excited states \cite{PhysRevA.30.2881, PhysRevA.69.022509}, with transition energies $\omega_{k1} = \omega_k - \omega_1$ and dipole-moment matrix elements $\boldsymbol{D}_{1k}= D_{1k}\hat{\boldsymbol{e}}_z$. For the Rb atomic implementation, $\omega_{21} = 1.56\,\mathrm{eV}$ and $\omega_{31}= 1.59\,\mathrm{eV}$,  whereas $D_{1k}$ are well approximated by their nonrelativistic values \cite{johnson2007atomic}, $D_{13} = D_{12}\sqrt{2}$, i.e., $D_{12} = 1.75\,\mathrm{a.u.}$ and $D_{13} = 2.47\,\mathrm{a.u.}$ \cite{PhysRevA.30.2881, PhysRevA.69.022509}. The left side of Fig.~\ref{fig:MoreLevelSystem} introduces the effective detuning
\begin{equation}
\delta_{\mathrm{r}} = \omega_{21} - \left\lfloor \frac{\omega_{21}}{\omega_{\mathrm{r}}}\right\rfloor\,\omega_{\mathrm{r}},
\label{eq:effective-detuning}
\end{equation}
where $\lfloor x \rfloor$ denotes the floor function. Notice that $\lfloor \omega_{21}/\omega_{\mathrm{r}}\rfloor\,\omega_{\mathrm{r}}$ is the greatest frequency consisting of a multiple of $\omega_{\mathrm{r}}$ which, at the same time, is smaller than or equal to $\omega_{21}$. For comparison, the central frequencies of the lines in the spectrum of the pump-pulse train are shown on the right side of Fig.~\ref{fig:MoreLevelSystem}.

The time evolution of the state of the system
$|\psi(t)\rangle = \sum_{i=1}^3 c_i(t) |i\rangle$
is determined by the Schr\"odinger equation
\begin{equation}
\uimm \frac{\diff |\psi(t)\rangle}{\diff t} = \hat{H}(t)|\psi(t)\rangle,
\end{equation}
with the total Hamiltonian
$\hat{H}(t) = \hat{H}_0 + \hat{H}^{\mathrm{tot}}_{\mathrm{int}}(t)$
consisting of the unperturbed atomic Hamiltonian $\hat{H}_0 = \sum_{i=1}^3 \omega_i |i\rangle\langle i|$ and the total $E1$ light-matter interaction Hamiltonian $\hat{H}^{\mathrm{tot}}_{\mathrm{int}}(t)$ in the rotating-wave approximation \cite{Scully:QuantumOptics, Foot:AtomicPhysics, Kiffner_review}. For a single pulse as described by Eq.~(\ref{eq:singlepulse}), the interaction Hamiltonian reads
\begin{equation}
\hat{H}_{\mathrm{int}}(t,t_{\mathrm{c}}, \phi) = -\frac{1}{2}\sum_{k=2}^3 \varOmega_{k}(t - t_{\mathrm{c}})\,\eu^{\uimm \omega_{\mathrm{c}}(t - t_{\mathrm{c}})} \,\eu^{\uimm\phi}\,|1\rangle \langle k| + \mathrm{H.c.},
\label{eq:singlepulseintHam}
\end{equation}
where we have introduced the time-dependent Rabi frequencies 
\begin{equation}
\varOmega_{k}(t) = D_{1k}\,\mathcal{E}_{0}(t).
\end{equation} 
The total interaction Hamiltonian
\begin{equation}
\hat{H}^{\mathrm{tot}}_{\mathrm{int}}(t) = \hat{H}_{\mathrm{pr}}(t)  +  \sum_{n = 0}^{N-1} \hat{H}_{\mathrm{pu,}n}(t),
\end{equation}
including the action of the probe and pump pulses,
\begin{equation}
\hat{H}_{\mathrm{pr}}(t) = -\frac{1}{2}\sum_{k=2}^3 \varOmega_{\mathrm{pr},k}(t -\tau)\,\eu^{\uimm \omega_{\mathrm{c}}(t-\tau)}\,\eu^{\uimm \phi_{0,\mathrm{pr}}}\, |1\rangle \langle k| + \mathrm{H.c.},
\end{equation}
\begin{equation}
\begin{aligned}
&\hat{H}_{\mathrm{pu},n}(t) =\\
& -\frac{1}{2}\sum_{k=2}^3 \varOmega_{\mathrm{pu},k}(t -n T_{\mathrm{p}})\,\eu^{\uimm \omega_{\mathrm{c}}(t-nT_{\mathrm{p}})}\,\eu^{\uimm \phi_{0,\mathrm{pu}}}\,\eu^{\uimm n\Delta\phi}\, |1\rangle \langle k| + \mathrm{H.c.},
\end{aligned}
\end{equation}
respectively, can then be defined in terms of the probe and pump Rabi frequencies $\varOmega_{\mathrm{pr/pu},k}(t) = D_{1k} \mathcal{E}_{0,\mathrm{pr/pu}}(t)$. 

In the following, we will analytically study the case of ultrashort pulses approximated by Dirac $\delta$~peaks,
\begin{equation}
\varOmega_{k}(t) = \vartheta_k\,\delta(t),
\end{equation}
with pulse areas 
\begin{equation}
\vartheta_k = \int \varOmega_{k}(t)\,\diff t.
\end{equation}
For the three-level system of interest, we introduce the effective pulse area
\begin{equation}
\vartheta = \sqrt{\vartheta_2^2 + \vartheta_3^3}
\end{equation}
and the angle
\begin{equation}
\alpha = \arctan(\vartheta_2 / \vartheta_3),
\end{equation}
such that
\begin{equation}
\vartheta_2 = \vartheta \sin(\alpha),\ \ \ \vartheta_3 = \vartheta\cos(\alpha).
\end{equation}

The time evolution of $|\psi(t)\rangle = \hat{U}(t,t_0)|\psi(t_0)\rangle$ from an initial state at time $t_0$, $t>t_0$, can be expressed as the action of an evolution operator $\hat{U}(t,t_0) = \sum_{i,j = 1}^3 U_{ij}(t,t_0)|i\rangle\langle j|$, of elements $U_{ij}(t,t_0)$, which is a solution of
\begin{equation}
\uimm \frac{\diff \hat{U}(t,t_0)}{\diff t} = \hat{H}(t)\hat{U}(t,t_0),\ \ \ \hat{U}(t_0,t_0)=\hat{I},
\label{eq:operatorSchequation}
\end{equation}
where $\hat{I}$ is the identity matrix. In the case of interest, consisting of pump and probe $\delta$~pulses, the evolution of the system can be split into intervals of free evolution, characterized by the operator
\begin{equation}
\hat{V}(t) = \eu^{-\uimm\hat{H}_0 t} = \diag(1,\,\eu^{-\uimm\omega_{21} t},\,\eu^{- \uimm\omega_{31}t}),
\end{equation}
separated by the instantaneous action of a $\delta$~pulse \cite{PhysRevA.95.043413}. The pump- and probe-pulse interaction operators describing this instantaneous action can be obtained after rewriting the interaction Hamiltonian in Eq.~(\ref{eq:singlepulseintHam}) in the case of a single $\delta$~pulse as
\begin{equation}
\hat{H}_{\mathrm{int}}(t,t_{\mathrm{c}},\phi) = \hat{F}\daga(\phi)\,\hat{B}(\vartheta,\alpha)\,\hat{F}(\phi)\,\delta(t-t_{\mathrm{c}}),
\end{equation}
in terms of the unitary matrix
\begin{equation}
\hat{F}(\phi) = \diag(1,\,\eu^{\uimm\phi},\,\eu^{\uimm\phi})
\end{equation}
accounting for the phase of the pulse, and the operator
\begin{equation}
\hat{B}(\vartheta,\alpha) = - \frac{\vartheta}{2}
\begin{pmatrix}
0 &\sin(\alpha) &\cos(\alpha)\\
\sin(\alpha) & 0 & 0\\
\cos(\alpha) & 0 & 0
\end{pmatrix}
\end{equation}
including the dependence upon the pulse strength. An explicit solution of Eq.~(\ref{eq:operatorSchequation}) in this single-pulse case %is given by
% \begin{equation}
% \begin{aligned}
% \hat{U}(t,t_0) = &\,\hat{V}(t-t_{\mathrm{c}})\,\Bigl[\hat{F}\daga(\phi)\,\eu^{-\uimm\hat{B}(\vartheta,\alpha)\,[\theta(t-t_{\mathrm{c}}) - \theta(t_0-t_{\mathrm{c}})]}\,\hat{F}(\phi)\Bigr]\\
% &\times\,\hat{V}(t_{\mathrm{c}}-t_0),
% \end{aligned}
% \label{eq:spbrack}
% \end{equation}
% with the Heaviside step function $\theta(x)$. Equation~(\ref{eq:spbrack}) shows that, whenever $t>t_{\mathrm{c}}>t_0$, the instantaneous action of a  pulse centered at $t = t_{\mathrm{c}}$ is given by the central term in between square brackets. One can therefore 
allows one to introduce the probe- and pump-pulse interaction operators
\begin{widetext}
\begin{align}
\hat{U}_{\mathrm{pr}}(\vartheta_{\mathrm{pr}},\alpha, \phi_{0,\mathrm{pr}}) &= \hat{F}\daga(\phi_{0,\mathrm{pr}})\,\hat{A}(\vartheta_{\mathrm{pr}},\alpha) \,\hat{F}(\phi_{0,\mathrm{pr}}),\\
\hat{U}_{\mathrm{pu},n}(\vartheta_{\mathrm{pu}},\alpha, \phi_{0,\mathrm{pu}}, \Delta\phi)& =\hat{F}\daga(\phi_{0,\mathrm{pu}})\,\bigl[\hat{F}\daga(\Delta\phi)\bigr]^n\,\hat{A}(\vartheta_{\mathrm{pu}},\alpha)\,\bigl[\hat{F}(\Delta\phi)\bigr]^n\,\hat{F}(\phi_{0,\mathrm{pu}}),
\end{align}
respectively, both modeling the instantaneous action of the associated $\delta$~pulse and defined in terms of \cite{PhysRevA.86.013423}
\begin{equation}
\hat{A}(\vartheta,\alpha) = \eu^{-\uimm\hat{B}(\vartheta,\alpha)} = 
\begin{pmatrix}
\cos(\vartheta/2) &\uimm\,\sin(\alpha)\,\sin(\vartheta/2) &\uimm\,\cos(\alpha)\,\sin(\vartheta/2)\\
\uimm\,\sin(\alpha)\,\sin(\vartheta/2) & \sin^2(\alpha)\,\cos(\vartheta/2) + \cos^2(\alpha)& \sin(\alpha)\,\cos(\alpha)\,[\cos(\vartheta/2)-1]\\
\uimm\,\cos(\alpha)\,\sin(\vartheta/2) & \sin(\alpha)\,\cos(\alpha)\,[\cos(\vartheta/2)-1] & \sin^2(\alpha)+\cos^2(\alpha)\cos(\vartheta/2)
\end{pmatrix}.
\label{eq:pulseaction}
\end{equation}
\end{widetext}

As described in the following, for the calculation of the absorption spectrum it is convenient to introduce the associated density matrix $\hat{\rho}(t) = \sum_{i,j=1}^3 \rho_{ij} |i\rangle \langle j|$, of elements $\rho_{ij}$ and given by $\hat{\rho}(t) = |\psi(t)\rangle \langle \psi(t)| = \hat{U}(t,t_0)\rho(t_0) \hat{U}\daga(t,t_0)$ in the case of a pure state. By defining the nine-dimensional column vector $\vec{R} = (\rho_{11},\,\rho_{12},\,\rho_{13},\,\rho_{21},\,\rho_{22},\,\rho_{23},\,\rho_{31},\,\rho_{32},\,\rho_{33})^{\mathrm{T}}$, i.e., the row-ordered vectorization of the density matrix, with elements $R_i(t)$, $i\in\{1,\,\ldots,9\}$, its time evolution $\vec{R}(t) = \hat{\mathcal{U}}(t,t_0) \vec{R}(t_0)$ can be written in terms of the $9\times 9$ matrix
\begin{equation}
\hat{\mathcal{U}}(t,t_0) = \hat{U}(t,t_0) \otimes \hat{U}^*(t,t_0),
\end{equation}
where $\hat{U}^*$ is the complex conjugate of $\hat{U}$ and where $\otimes$ denotes the Kronecker product \cite{horn_johnson_1991}
$$\hat{\mathcal{U}} = 
\begin{pmatrix}
U_{11}\hat{U}^* &U_{12}\hat{U}^* &U_{13}\hat{U}^*\\
U_{21}\hat{U}^* &U_{22}\hat{U}^* &U_{23}\hat{U}^*\\
U_{31}\hat{U}^* &U_{32}\hat{U}^* &U_{33}\hat{U}^*
\end{pmatrix}.
$$
Due to the mixed-product property, $(\hat{U}_1\hat{U}_2)\otimes (\hat{U}_1^*\hat{U}_2^*)= (\hat{U}_1\otimes\hat{U}_1^*)(\hat{U}_2\otimes\hat{U}_2^*)$, whenever the evolution operator $\hat{U} = \hat{U}_1\hat{U}_2$ is equal to the product of two terms $\hat{U}_1$ and $\hat{U}_2$, then $\hat{\mathcal{U}}=\hat{U}\otimes \hat{U}^* = \hat{\mathcal{U}}_1 \hat{\mathcal{U}}_2$ is also equal to the product of the associated matrices $\hat{\mathcal{U}}_1 = \hat{U}_1\otimes\hat{U}_1^*$ and $\hat{\mathcal{U}}_2 = \hat{U}_2\otimes\hat{U}_2^*$.

\subsection{Transient-absorption spectrum}
\label{Transient-absorption spectrum}

Experimental optical-density absorption spectra can be simulated via calculation of the single-particle dipole response of the system \cite{0953-4075-49-6-062003}
\begin{equation}
S(\bar{\omega},\tau) \propto -\omega\Imm\biggl[\sum_{k=2}^3 D_{1k}^* \int_{-\infty}^{\infty}\rho_{1k}(t,\tau)\,\eu^{-\uimm\bar{\omega} (t-\tau)}\,\diff t\biggr],
\label{eq:spdefinition}
\end{equation}
with the Fourier transform centered on the arrival time of the measured probe pulse. The above expression is valid for low densities and small medium lengths, where the effect of the propagation of the pulses through the medium can be neglected. The transient-absorption spectrum provides access to the dipole response of the system via the coherences $\rho_{1k}(t,\tau)$, i.e., off-diagonal terms of the density matrix. 

In order to effectively account for broadening effects in the experiment, which determine the finite linewidth of the absorption lines, the Fourier transform in Eq.~(\ref{eq:spdefinition}) will be evaluated at the complex frequency $\bar{\omega} = \omega - \uimm \gamma/2$. Here, $\omega$ is the real frequency of the photons detected by the spectrometer, while $\gamma$ accounts for the experimental linewidth. Evaluating Eq.~(\ref{eq:spdefinition}) at this complex frequency is equivalent to calculating the Fourier transform of $ \rho_{1k}(t,\tau)\,\eu^{-\gamma(t-\tau)/2}$, i.e., of an effectively decaying dipole. This is also equivalent to convolving $S_1(\omega,\tau)$ with a Lorentzian function of width $\gamma/2$. It is also important to stress that the poles of $S(\bar{\omega},\tau)$ lie on the real axis, as we will show in Sec.~\ref{Dynamics of the system and associated spectrum} and Appendix~\ref{Appendix:Central frequency of the spectral lines}. If we evaluated $S(\bar{\omega},\tau)$ for $\bar{\omega}=\omega\in\mathbb{R}$, the spectrum would diverge at the frequencies corresponding to these poles. By evaluating the spectra at the complex frequency $\bar{\omega} = \omega - \uimm \gamma/2$, however, these divergences reduce to peaks of width $\gamma/2$. The poles of $S(\bar{\omega},\tau)$ are then associated with the central frequencies of the peaks appearing in the spectrum. 

In the following, we will set $\gamma\ll\varGamma_k$, i.e., much smaller than the spontaneous decay rates $\varGamma_k$ of the excited states to the ground state. As a result, during the time scales of interest as defined by the exponential function  $\eu^{-\gamma(t-\tau)/2}$, spontaneous decay can be safely neglected in the equations of motion, thus justifying the pure-state approach used to derive the equations of motion of $\hat{\rho}(t,\tau)$. At the same time, we will set $\gamma\tau\ll 1$, $\gamma T_{\mathrm{p}}\ll 1$, such that the dipole response of the system can be controlled by the sequence of pump pulses within its decay.

Alternatively, one could have effectively included broadening effects via an atomic Hamiltonian $\hat{H}_0$ with complex eigenenergies $\omega_k - \uimm \gamma_k/2$, i.e., by including the effective decay of the coherences $\rho_{1k}(t,\tau)$ directly in the equations of motion. However, for the parameters chosen, and in particular when $\gamma\tau\ll 1$, we tested that there is no appreciable difference between results obtained with these two alternative approaches. Using Eq.~(\ref{eq:spdefinition}) with a complex frequency $\bar{\omega}$ will allow us to significantly simplify the presentation of the analytical calculations in Sec.~\ref{Dynamics of the system and associated spectrum} and Appendix~\ref{Appendix:Central frequency of the spectral lines}.

We finally notice that in the following we will calculate and show spectra $\mathcal{S}(\bar{\omega},\tau)$ assuming the noncollinear geometry depicted in Fig.~\ref{fig:Experiment}. In transient-absorption spectroscopy experiments, this geometry is employed to measure the spectrum of the probe pulse independent of the pump pulse, and thus separate the contributions from pulses with the same laser frequency. In this geometry, however, fast oscillations of the absorption spectrum as a function of time delay are effectively averaged out in an experiment \cite{PhysRevLett.115.033003, 0953-4075-51-3-035501}. We will account for this by identifying and selectively removing fast time-delay-dependent oscillating terms in the resulting single-particle absorption spectra $S(\bar{\omega},\tau)$:
\begin{equation}
\mathcal{S}(\bar{\omega},\tau) = \langle S(\bar{\omega},\tau)\rangle_{\tau},
\end{equation}
where $\langle \cdots\rangle_{\tau}$ denotes averaging over $\tau$.

\subsection{Dynamics of the system and associated spectrum}
\label{Dynamics of the system and associated spectrum}

In the following, we will obtain analytical expressions for the time evolution of the state $\vec{R}(t)$, when it is excited by a probe pulse centered on $\tau$ and by a sequence of $N$ pump pulses centered on $t_n = n T_{\mathrm{p}}$, $n\in\{0,\,\ldots,\,N-1\}$, up to the limit of $N\rightarrow \infty$. The initial state of the system is 
\begin{equation}
\begin{aligned}
\vec{R}_0 &=(1,\,0,\,0)^{\mathrm{T}}\,\otimes\,(1,\,0,\,0)^{\mathrm{T}} \\
&=(1,\,0,\,0,\,0,\,0,\,0,\,0,\,0,\,0)^{\mathrm{T}},
\end{aligned}
\label{eq:groundstate}
\end{equation}
i.e., the system is initially in its ground state. These analytical expressions will then be used to calculate the associated transient-absorption spectrum via Eq.~(\ref{eq:spdefinition}), after eliminating fast oscillations in $\tau$. For this purpose, we introduce the $9\times 9$ operators
\begin{equation}
\hat{\mathcal{A}}(\vartheta,\alpha) = \hat{A}(\vartheta,\alpha)\otimes \hat{A}^*(\vartheta,\alpha)
\end{equation}
and
\begin{equation}
\begin{aligned}
\hat{\mathcal{F}}(\phi) &= \hat{F}(\phi)\otimes \hat{F}^*(\phi) \\
& =\diag(1,\,\eu^{-\uimm\phi},\,\eu^{-\uimm\phi},\,\eu^{\uimm\phi},\,1,\,1,\,\eu^{\uimm\phi},\,1,\,1).
\end{aligned}
\end{equation}
The instantaneous interaction with the pump and probe $\delta$~pulses can then be modeled by the interaction operators
\begin{equation}
\hat{\mathcal{U}}_{\mathrm{pr}} = \hat{U}_{\mathrm{pr}}\otimes \hat{U}_{\mathrm{pr}}^*= \hat{\mathcal{F}}_{0,\mathrm{pr}}\daga\,\hat{\mathcal{A}}_{\mathrm{pr}}\,\hat{\mathcal{F}}_{0,\mathrm{pr}}
\end{equation}
and
\begin{equation}
\begin{aligned}
\hat{\mathcal{U}}_{\mathrm{pu},n} = \hat{U}_{\mathrm{pu},n}\otimes \hat{U}_{\mathrm{pu},n}^*= &\,\hat{\mathcal{F}}_{0,\mathrm{pu}}\daga\,(\hat{\mathcal{F}}_{\Delta}\daga)^n\,\hat{\mathcal{A}}_{\mathrm{pu}}\,(\hat{\mathcal{F}}_{\Delta})^n\,\hat{\mathcal{F}}_{0,\mathrm{pu}},
\end{aligned}
\end{equation}
where we have simplified the notation by introducing
\begin{align}
\hat{\mathcal{F}}_{0,\mathrm{pr/pu}} &= \hat{\mathcal{F}}(\phi_{0,\mathrm{pr/pu}}),\\
\hat{\mathcal{F}}_{\Delta} &= \hat{\mathcal{F}}(\Delta\phi),\\
\hat{\mathcal{A}}_{\mathrm{pr/pu}} & = \hat{\mathcal{A}}(\vartheta_{\mathrm{pr/pu}},\alpha).
\end{align}
The free evolution of the system between two consecutive pulses is modeled by the $9\times 9$ free-evolution operator 
\begin{equation}
\begin{aligned}
&\hat{\mathcal{V}}(t) = \hat{V}(t)\otimes \hat{V}^*(t)\\
=&\,\diag(1,\,\eu^{\uimm\omega_{21}t},\,\eu^{\uimm\omega_{31}t},\,\eu^{-\uimm\omega_{21}t},\,1,\,\eu^{\uimm\omega_{32}t},\,\eu^{-\uimm\omega_{31}t},\,\eu^{-\uimm\omega_{32}t},\,1);
\end{aligned}
\end{equation}
in order to describe the free-evolution in the period $T_{\mathrm{p}}$ between two pump pulses we define
\begin{equation}
\hat{\mathcal{V}}_{\mathrm{p}} = \hat{\mathcal{V}}(T_{\mathrm{p}}).
\end{equation}

Depending on the position of the probe pulse, three experimental setups can be distinguished. When $\tau<0$, the probe pulse completely precedes the sequence of pump pulses, while it fully follows the train of pump pulses when $\tau>(N-1)T_{\mathrm{p}}$. The general structure of the absorption spectrum for these two experimental setups was previously investigated for a single pump pulse \cite{PhysRevLett.115.033003, PhysRevA.95.043413}, also in the presence of an intense probe pulse \cite{0953-4075-51-3-035501}. In the following, we will show how the formulas presented therein can be modified in order to account for a sequence of pump pulses, and how this is imprinted in the shape of the absorption spectra for increasing values of $N$. For the case of a train of pump pulses, a new pump--probe--pump setup also exists for $0<\tau<(N-1)T_{\mathrm{p}}$, i.e., whenever the probe pulse lies in between two pump pulses. We will show that the structure of the spectrum in this general case shares several elements with the pump--probe and probe--pump setups mentioned previously. For all the above cases, we will show that the pulse-to-pulse phase shift $\Delta\phi$ provides an important additional degree of freedom to shape the absorption spectrum and gain understanding of the evolution of the system in the presence of a periodic external excitation from absorption line shapes.

\subsubsection{Probe--pump setup $(\tau<0)$}
\label{Probe--pump setup}

For negative time delays, when the probe pulse precedes the train of pump pulses, the evolution of the system $\vec{R}(t)$ from the initial state $\vec{R}_0$ [Eq.~(\ref{eq:groundstate})] in the presence of a finite number $N$ of pump pulses reads:
\begin{widetext}
\begin{equation}
% \begin{aligned}
\vec{R}(t) =
\left\{
\begin{aligned}
&\vec{R}_0, & t<\tau, \\
&\hat{\mathcal{V}}(t-\tau)\,\hat{\mathcal{U}}_{\mathrm{pr}}\,\vec{R}_0, & \tau<t<0,\\ 
&\hat{\mathcal{V}}(t - lT_{\mathrm{p}})\,\hat{\mathcal{F}}_{0,\mathrm{pu}}\daga\,(\hat{\mathcal{F}}_{\Delta}\daga)^l\,\hat{\mathcal{A}}_{\mathrm{pu}}\,(\hat{\mathcal{F}}_{\Delta}\,\hat{\mathcal{V}}_{\mathrm{p}}\,\hat{\mathcal{A}}_{\mathrm{pu}})^l\,\hat{\mathcal{F}}_{0,\mathrm{pu}}\,\hat{\mathcal{V}}(-\tau)\,\hat{\mathcal{U}}_{\mathrm{pr}}\,\vec{R}_0, & lT_{\mathrm{p}}<t<(l+1)T_{\mathrm{p}},\\
&\hat{\mathcal{V}}(t - (N-1)T_{\mathrm{p}})\,\hat{\mathcal{F}}_{0,\mathrm{pu}}\daga\,(\hat{\mathcal{F}}_{\Delta}\daga)^{N-1}\,\hat{\mathcal{A}}_{\mathrm{pu}}\,(\hat{\mathcal{F}}_{\Delta}\,\hat{\mathcal{V}}_{\mathrm{p}}\,\hat{\mathcal{A}}_{\mathrm{pu}})^{N-1}\,\hat{\mathcal{F}}_{0,\mathrm{pu}}\,\hat{\mathcal{V}}(-\tau)\,\hat{\mathcal{U}}_{\mathrm{pr}}\,\vec{R}_0, & t>(N-1)T_{\mathrm{p}},
\end{aligned}
\right.
% \end{aligned}
\label{eq:evolutionprpu}
\end{equation}
\end{widetext}
where the third line describes the dynamics of the system in the interval $[t_{l},\,t_{l+1}]$, $l\in\{0,\,\ldots,\,N-2\}$, in between the $l$th and the $(l+1)$th pulse. In Appendix~\ref{Appendix:Position-dependent action of the pump pulse}, we present the evolution of a system between a general $a$th and a general $b$th pulse, with $0\leq a \leq b \leq N-1$. The third (fourth) line in Eq.~(\ref{eq:evolutionprpu}) are thus obtained from Eq.~(\ref{eq:finitesequence}) with $a = 0$ and $b = l$ ($b = N-1$). The last two lines are affected by the number $N$ of pump pulses. The third line is only present for $N>1$, since it describes the dynamics of the system in between two pump pulses. The fourth line is only present for a finite number of pulses, since it describes the free evolution of the system following interaction with the last pulse centered on $t_{N-1}$.

The two density-matrix elements $\rho_{1k}(t) = R_k(t)= \vec{v}_k \vec{R}(t)$, $k\in\{2,\,3\}$ of interest for the calculation of the absorption spectrum are then obtained by multiplying the row-vector 
\begin{equation}
\begin{aligned}
\vec{v}_k =& (1,\,0,\,0)\otimes(0,\,\delta_{k2},\,\delta_{k3})\\
=& (0,\,\delta_{k2},\,\delta_{k3},\,0,\,0,\,0,\,0,\,0,\,0) 
\end{aligned}
\end{equation}
with $\vec{R}(t)$, as shown in Eq.~(\ref{eq:evolutionprpu-k}). The integral in Eq.~(\ref{eq:spdefinition}) can then be performed in each one of the intervals identified in Eq.~(\ref{eq:evolutionprpu}), leading to
\begin{widetext}
% \begin{equation}
% S_N(\bar{\omega},\tau) \propto -\omega\Imm\biggl\{\sum_{k=2}^3 D_{1k}^*\biggl[\int_{\tau}^{0}R_{k}(t)\,\eu^{-\uimm\bar{\omega} (t-\tau)}\,\diff t+ \sum_{l=0}^{N-2}\int_{lT_{\mathrm{p}}}^{(l+1)T_{\mathrm{p}}}R_{k}(t)\,\eu^{-\uimm\bar{\omega} (t-\tau)}\,\diff t + \int_{(N-1)T_{\mathrm{p}}}^{\infty}R_{k}(t)\,\eu^{-\uimm\bar{\omega} (t-\tau)}\,\diff t\biggr]\biggr\},
% \end{equation}
% leading to
\begin{equation}
\begin{aligned}
&\,\mathcal{S}_N(\bar{\omega},\tau)\propto -\omega\Imm\biggl\{\sum_{k=2}^3 \frac{D_{1k}^*}{\uimm(\bar{\omega} - \omega_{k1})}\,\vec{v}_k\,\biggl[\bigl(1-\eu^{\uimm(\bar{\omega} - \omega_{k1})\tau}\bigr)\\
&+ \bigl(1-\eu^{-\uimm(\bar{\omega} - \omega_{k1})T_{\mathrm{p}}}\bigr)\,\hat{\mathcal{A}}_{\mathrm{pu}}\,\Bigl[\hat{\mathcal{I}} - \bigl(\eu^{-\uimm (\bar{\omega}T_{\mathrm{p}}- \Delta\phi)}\,\hat{\mathcal{F}}_{\Delta}\,\hat{\mathcal{V}}_{\mathrm{p}}\,\hat{\mathcal{A}}_{\mathrm{pu}}\bigr)^{N-1}\Bigr]\,\bigl(\hat{\mathcal{I}} -\eu^{-\uimm (\bar{\omega}T_{\mathrm{p}}- \Delta\phi)}\,\hat{\mathcal{F}}_{\Delta}\,\hat{\mathcal{V}}_{\mathrm{p}}\,\hat{\mathcal{A}}_{\mathrm{pu}}\bigr)^{-1}\,\eu^{\uimm\bar{\omega}\tau}\,\hat{\mathcal{G}}(\tau)\\
&+ \hat{\mathcal{A}}_{\mathrm{pu}}\,\bigl(\eu^{-\uimm (\bar{\omega}T_{\mathrm{p}}- \Delta\phi)}\,\hat{\mathcal{F}}_{\Delta}\,\hat{\mathcal{V}}_{\mathrm{p}}\,\hat{\mathcal{A}}_{\mathrm{pu}}\bigr)^{N-1}\,\eu^{\uimm\bar{\omega}\tau}\,\hat{\mathcal{G}}(\tau)\biggr]\,\hat{\mathcal{U}}_{\mathrm{pr}}\,\vec{R}_0 \biggr\},
\end{aligned}
\label{eq:spectrumprpu}
\end{equation}
where we have introduced $\hat{\mathcal{I}} = \hat{I}\otimes \hat{I}$ and used the fact that
\begin{equation}
% \begin{aligned}
\sum_{l=0}^{N-2}\bigl(\eu^{-\uimm (\bar{\omega}T_{\mathrm{p}}- \Delta\phi)}\,\hat{\mathcal{F}}_{\Delta}\,\hat{\mathcal{V}}_{\mathrm{p}}\,\hat{\mathcal{A}}_{\mathrm{pu}}\bigr)^l=\Bigl[\hat{\mathcal{I}}-\bigl(\eu^{-\uimm (\bar{\omega}T_{\mathrm{p}}- \Delta\phi)}\,\hat{\mathcal{F}}_{\Delta}\,\hat{\mathcal{V}}_{\mathrm{p}}\,\hat{\mathcal{A}}_{\mathrm{pu}}\bigr)^{N-1}\Bigr]\,\bigl(\hat{\mathcal{I}}-\eu^{-\uimm (\bar{\omega}T_{\mathrm{p}}- \Delta\phi)}\,\hat{\mathcal{F}}_{\Delta}\,\hat{\mathcal{V}}_{\mathrm{p}}\,\hat{\mathcal{A}}_{\mathrm{pu}}\bigr)^{-1},
% \end{aligned}
\label{eq:inverseoperator}
\end{equation}
\end{widetext}
where $\bar{\omega} = \omega - \uimm\gamma/2$. The subscript $N$ in $S_N(\bar{\omega}, \tau)$ and $\mathcal{S}_N(\bar{\omega},\tau)$ indicates their dependence upon the number of pulses. In $\mathcal{S}_N(\bar{\omega},\tau)$ we have also averaged over fast oscillations as a function of $\tau$, i.e., removed fast time-delay oscillating terms $\eu^{\pm\uimm\omega_{k1}\tau}$ appearing in $S_N(\bar{\omega}, \tau)$ for the frequencies of interest $\bar{\omega}\approx\omega_{k1} - \uimm\gamma/2$. This is accounted for by the operator 
\begin{equation}
\begin{aligned}
\eu^{\uimm\bar{\omega}\tau}\,\hat{\mathcal{G}}(\tau) &\doteq \langle\eu^{\uimm\bar{\omega}\tau}\,\hat{\mathcal{V}}(-\tau)\rangle_{\tau}\\
&= \eu^{\uimm\bar{\omega}\tau}\,\diag(0,\,\eu^{-\uimm\omega_{21}\tau},\,\eu^{-\uimm\omega_{31}\tau},\,0,\,0,\,0,\,0,\,0,\,0).
\end{aligned}
\end{equation}
We notice that the resulting spectrum is independent of the initial-pump-pulse CEP $\phi_{0,\mathrm{pu}}$ due to
$
\eu^{\uimm\phi_{0,\mathrm{pu}}}\,\hat{\mathcal{F}}_{0,\mathrm{pu}}\,\hat{\mathcal{G}}(\tau) = \hat{\mathcal{G}}(\tau)
$. By introducing the operator
\begin{equation}
\begin{aligned}
&\hat{\mathcal{D}}_N(\bar{\omega}) = \hat{\mathcal{A}}_{\mathrm{pu}}\,\Bigl\{\hat{\mathcal{I}} -\bigl(\eu^{-\uimm (\bar{\omega}T_{\mathrm{p}}- \Delta\phi)}\,\hat{\mathcal{F}}_{\Delta}\,\hat{\mathcal{V}}_{\mathrm{p}}\,\hat{\mathcal{A}}_{\mathrm{pu}}\bigr)^N\\
&\ \ \ \ -\eu^{-\uimm(\bar{\omega} - \omega_{k1})T_{\mathrm{p}}}\,\Bigl[\hat{\mathcal{I}} - \bigl(\eu^{-\uimm (\bar{\omega}T_{\mathrm{p}}- \Delta\phi)}\,\hat{\mathcal{F}}_{\Delta}\,\hat{\mathcal{V}}_{\mathrm{p}}\,\hat{\mathcal{A}}_{\mathrm{pu}}\bigr)^{N-1}\Bigr] \Bigr\}\\
&\ \ \ \ \times\bigl(\hat{\mathcal{I}} -\eu^{-\uimm (\bar{\omega}T_{\mathrm{p}}- \Delta\phi)}\,\hat{\mathcal{F}}_{\Delta}\,\hat{\mathcal{V}}_{\mathrm{p}}\,\hat{\mathcal{A}}_{\mathrm{pu}}\bigr)^{-1},
\end{aligned}
\label{eq:actionpumpafterprobe}
\end{equation}
the spectrum can be written as
\begin{equation}
\begin{aligned}
&\mathcal{S}_N(\bar{\omega},\tau)\propto -\omega\Imm\biggl\{\sum_{k=2}^3 \frac{D_{1k}^*}{\uimm(\bar{\omega} - \omega_{k1})}\,\vec{v}_k\\
&\ \ \times \biggl[\bigl(1-\eu^{\uimm(\bar{\omega} - \omega_{k1})\tau}\bigr)+ \hat{\mathcal{D}}_{N}(\bar{\omega})\,\eu^{\uimm\bar{\omega}\tau}\,\hat{\mathcal{G}}(\tau)\biggr]\,\hat{\mathcal{U}}_{\mathrm{pr}}\,\vec{R}_0 \biggr\},
\end{aligned}
\label{eq:spectrumprpushort}
\end{equation}
with the second term in the sum highlighting how the sequence of $N$ pump pulses acts on the system and shapes the resulting absorption spectrum. For a single pump pulse, $\hat{\mathcal{D}}_N(\bar{\omega})$ reduces to $\hat{\mathcal{D}}_1(\bar{\omega}) = \hat{\mathcal{A}}_{\mathrm{pu}}$, whereas for an infinite train of pump pulses it reads
\begin{equation}
\begin{aligned}
\hat{\mathcal{D}}_{\infty}(\bar{\omega}) &= \bigl(1-\eu^{-\uimm(\bar{\omega} - \omega_{k1})T_{\mathrm{p}}}\bigr)\,\hat{\mathcal{A}}_{\mathrm{pu}}\\
&\ \times\bigl(\hat{\mathcal{I}} -\eu^{-\uimm (\bar{\omega}T_{\mathrm{p}}- \Delta\phi)}\,\hat{\mathcal{F}}_{\Delta}\,\hat{\mathcal{V}}_{\mathrm{p}}\,\hat{\mathcal{A}}_{\mathrm{pu}}\bigr)^{-1}.
\end{aligned}
\label{eq:actioninfinitelymanypumpafterprobe}
\end{equation}
For large numbers of pump pulses, and particularly in the limit $N\rightarrow \infty$, the frequency-dependent operator $\hat{\mathcal{D}}_{N}(\bar{\omega})$ causes the appearance of LISs in the spectrum. These additional peaks are due to the presence of the inverse operator $\bigl(\hat{\mathcal{I}} -\eu^{-\uimm (\bar{\omega}T_{\mathrm{p}}- \Delta\phi)}\,\hat{\mathcal{F}}_{\Delta}\,\hat{\mathcal{V}}_{\mathrm{p}}\,\hat{\mathcal{A}}_{\mathrm{pu}}\bigr)^{-1}$ and are therefore centered on frequencies which are determined by the eigenvalues of $\hat{\mathcal{F}}_{\Delta}\,\hat{\mathcal{V}}_{\mathrm{p}}\,\hat{\mathcal{A}}_{\mathrm{pu}}$. The appearance of these additional lines is the main signature of the pump-pulse-induced periodic excitation of the system in the probe--pump setup: in this case, the initial dipole generated by the probe pulse is subsequently modified by the periodic sequence of pump pulses, and these strong-field periodic dynamics are imprinted into the spectrum via the appearance of LISs.

\subsubsection{Pump--probe setup $[\tau>(N-1)T_{\mathrm{p}}]$}
\label{Pump--probe setup}

For a finite number $N$ of pump pulses, a pump--probe setup is possible, in which the probe pulse encounters the atomic system at $\tau>(N-1)T_{\mathrm{p}}$, following the complete pump-pulse sequence. In this case, the evolution of the system is given by
\begin{widetext}
\begin{equation}
% \begin{aligned}
\vec{R}(t) =
\left\{
\begin{aligned}
&\vec{R}_0, & t<0, \\
&\hat{\mathcal{V}}(t - lT_{\mathrm{p}})\,\hat{\mathcal{F}}_{0,\mathrm{pu}}\daga\,(\hat{\mathcal{F}}_{\Delta}\daga)^l\,\hat{\mathcal{A}}_{\mathrm{pu}}\,(\hat{\mathcal{F}}_{\Delta}\,\hat{\mathcal{V}}_{\mathrm{p}}\,\hat{\mathcal{A}}_{\mathrm{pu}})^l\,\hat{\mathcal{F}}_{0,\mathrm{pu}}\,\vec{R}_0, & lT_{\mathrm{p}}<t<(l+1)T_{\mathrm{p}},\\
&\hat{\mathcal{V}}(t - (N-1)T_{\mathrm{p}})\,\hat{\mathcal{F}}_{0,\mathrm{pu}}\daga\,(\hat{\mathcal{F}}_{\Delta}\daga)^{N-1}\,\hat{\mathcal{A}}_{\mathrm{pu}}\,(\hat{\mathcal{F}}_{\Delta}\,\hat{\mathcal{V}}_{\mathrm{p}}\,\hat{\mathcal{A}}_{\mathrm{pu}})^{N-1}\,\hat{\mathcal{F}}_{0,\mathrm{pu}}\,\vec{R}_0, & (N-1)T_{\mathrm{p}}<t<\tau,\\
&\hat{\mathcal{V}}(t-\tau)\,\hat{\mathcal{U}}_{\mathrm{pr}}\,\hat{\mathcal{V}}(\tau - (N-1)T_{\mathrm{p}})\,\hat{\mathcal{F}}_{0,\mathrm{pu}}\daga\,(\hat{\mathcal{F}}_{\Delta}\daga)^{N-1}\,\hat{\mathcal{A}}_{\mathrm{pu}}\,(\hat{\mathcal{F}}_{\Delta}\,\hat{\mathcal{V}}_{\mathrm{p}}\,\hat{\mathcal{A}}_{\mathrm{pu}})^{N-1}\,\hat{\mathcal{F}}_{0,\mathrm{pu}}\,\vec{R}_0, & t>\tau,\\ 
\end{aligned}
\right.
% \end{aligned}
\label{eq:evolutionpupr}
\end{equation}
\end{widetext}
where the second line, describing the dynamics of the system in the interval $[t_{l},\,t_{l+1}]$, $l\in\{0,\,\ldots,\,N-2\}$, is present only if $N>1$. We first observe that $\hat{\mathcal{F}}_{0,\mathrm{pu}}\,\vec{R}_0 = \vec{R}_0$ for the initial state in Eq.~(\ref{eq:groundstate}). By further multiplying the row-vector $\vec{v}_k$ with $\vec{R}(t)$, as shown in Eq.~(\ref{eq:evolutionpupr-k}), the integral in Eq.~(\ref{eq:spdefinition}) can then be performed in each one of the intervals identified in Eq.~(\ref{eq:evolutionpupr}).
% \begin{equation}
% S_N(\bar{\omega},\tau) \propto -\omega\Imm\biggl\{\sum_{k=2}^3 D_{1k}^*\biggl[ \sum_{l=0}^{N-2}\int_{lT_{\mathrm{p}}}^{(l+1)T_{\mathrm{p}}}R_{k}(t)\,\eu^{-\uimm\bar{\omega} (t-\tau)}\,\diff t + \int_{(N-1)T_{\mathrm{p}}}^{\tau}R_{k}(t)\,\eu^{-\uimm\bar{\omega} (t-\tau)}\,\diff t + \int_{\tau}^{\infty}R_{k}(t)\,\eu^{-\uimm\bar{\omega} (t-\tau)}\,\diff t\biggr]\biggr\}.
% \label{eq:spectrumasasumpupr}
% \end{equation}
Integrals in $[t_{l},\,t_{l+1}]$ and in $[t_{N-1},\, \tau]$ feature fast time-delay-dependent oscillations for $\bar{\omega}\approx \omega_{k1} - \uimm\gamma/2$,
% \begin{equation}
% \begin{aligned}
% &\int_{lT_{\mathrm{p}}}^{(l+1)T_{\mathrm{p}}} \eu^{\uimm\omega_{k1}(t - lT_{\mathrm{p}})}\,\eu^{-\uimm\bar{\omega} (t-\tau)}\,\diff t \\
% = &\,\eu^{\uimm\bar{\omega}\tau}\,\frac{1- \eu^{-\uimm(\bar{\omega} - \omega_{k1})T_{\mathrm{p}}}}{\uimm (\bar{\omega} - \omega_{k1})}\,\eu^{-\uimm\bar{\omega}l T_{\mathrm{p}}},
% \end{aligned}
% \end{equation}
% \begin{equation}
% \begin{aligned}
% &\int_{(N-1)T_{\mathrm{p}}}^{\tau} \eu^{\uimm\omega_{k1}[t - (N-1)T_{\mathrm{p}}]}\,\eu^{-\uimm\bar{\omega} (t-\tau)}\,\diff t \\
% = &\,\eu^{\uimm\bar{\omega}\tau}\,\frac{1- \eu^{-\uimm(\bar{\omega} - \omega_{k1})[\tau - (N-1)T_{\mathrm{p}}]}}{\uimm (\bar{\omega} - \omega_{k1})}\,\eu^{-\uimm\bar{\omega}(N-1) T_{\mathrm{p}}}
% \end{aligned}
% \end{equation}
due to the fast oscillating factor $\eu^{\uimm\bar{\omega}\tau}$, and therefore do not contribute to $\mathcal{S}_N(\bar{\omega}, \tau)$. The spectrum thus results from the dynamics of the system only for $t>\tau$, with the periodic sequence of pump pulses determining the state of the system encountered by the probe pulse at $\tau^-$. This is a typical feature of the spectra in a pump--probe setup for a noncollinear geometry, which was already recognized for the single-pump-pulse case \cite{PhysRevA.95.043413, 0953-4075-51-3-035501}. In contrast to the probe--pump case, where the periodic excitation of the system following the probe pulse causes the appearance of LISs, here the train of pulses preceding the probe pulse only determines the state in which the system is prepared. The spectrum thus reads
\begin{equation}
\begin{aligned}
\mathcal{S}_N(\bar{\omega},\tau)&\propto -\omega\Imm\biggl[\sum_{k=2}^3 \frac{D_{1k}^*}{\uimm(\bar{\omega} - \omega_{k1})}\,\vec{v}_k\,\hat{\mathcal{U}}_{\mathrm{pr}}\\
&\times\,\hat{\mathcal{W}}(\tau - (N-1)T_{\mathrm{p}})\,\hat{\mathcal{A}}_{\mathrm{pu}}\,(\hat{\mathcal{F}}_{\Delta}\,\hat{\mathcal{V}}_{\mathrm{p}}\,\hat{\mathcal{A}}_{\mathrm{pu}})^{N-1}\,\vec{R}_0 \biggr],
\end{aligned}
\label{eq:spectrumpupr}
\end{equation}
where we have removed the fast time-delay-dependent oscillations by introducing 
\begin{equation}
\begin{aligned}
\hat{\mathcal{W}}(\tau) &\doteq \langle \hat{\mathcal{V}}(\tau) \rangle_{\tau} \\
&= \diag(1,\,0,\,0,\,0,\,1,\,\eu^{\uimm\omega_{32}t},\,0,\,\eu^{-\uimm\omega_{32}t},\,1)
\end{aligned}
\end{equation}
and where we have taken advantage of
$$
\hat{\mathcal{W}}(\tau - (N-1)T_{\mathrm{p}})\,\hat{\mathcal{F}}_{0,\mathrm{pu}}\daga\,(\hat{\mathcal{F}}_{\Delta}\daga)^{N-1} = \hat{\mathcal{W}}(\tau - (N-1)T_{\mathrm{p}}),
$$
i.e., the resulting spectrum is also in this case independent of the initial-pump-pulse CEP $\phi_{0,\mathrm{pu}}$.

The pump--probe setup described above is present only if the pump field consists of a finite number of pulses. In this case, the operator $\hat{\mathcal{A}}_{\mathrm{pu}}\,(\hat{\mathcal{F}}_{\Delta}\,\hat{\mathcal{V}}_{\mathrm{p}}\,\hat{\mathcal{A}}_{\mathrm{pu}})^{N-1}$ in Eq.~(\ref{eq:spectrumpupr}) contains all the information on the action of the train of pump pulses which is encoded in the absorption spectrum. This operator clearly reduces to the single-pump-pulse operator $\hat{\mathcal{A}}_{\mathrm{pu}}$ for $N = 1$. We stress again that the above formulas can be used only if $\gamma\tau\ll 1$, i.e., for time delays that allow one to neglect the amplitude change of the dipole response due to the decay rate $\gamma$.

\subsubsection{Pump--probe--pump setup $[0<\tau<(N-1)T_{\mathrm{p}}]$}
\label{Pump--probe--pump setup}

The final setup we are going to consider, present only for $N>1$, consists of a probe pulse exciting the system in between two pump pulses in $\mathcal{E}_{\mathrm{pu}}(t)$. This pump--probe--pump setup shares features with both cases discussed above: as in the pump--probe case, also here the action of the pump pulses preceding the probe pulse is encoded in the state of the system encountered by the probe pulse; in analogy with the probe--pump term, the pump pulses following the probe pulse actively modify the dipole response of the system and shape the absorption spectrum into additional LISs. 

In order to highlight these properties, we first consider the dynamics in a pump--probe--pump system, which can be divided into different intervals as follows:
\begin{widetext}
\begin{equation}
\begin{aligned}
&\ \ \ \, \vec{R}(t) =\\
&\left\{
\begin{aligned}
&\vec{R}_0, & t<0, \\
&\hat{\mathcal{V}}(t - pT_{\mathrm{p}})\,\hat{\mathcal{F}}_{0,\mathrm{pu}}\daga\,(\hat{\mathcal{F}}_{\Delta}\daga)^p\,\hat{\mathcal{A}}_{\mathrm{pu}}\,(\hat{\mathcal{F}}_{\Delta}\,\hat{\mathcal{V}}_{\mathrm{p}}\,\hat{\mathcal{A}}_{\mathrm{pu}})^p\,\hat{\mathcal{F}}_{0,\mathrm{pu}}\,\vec{R}_0, & pT_{\mathrm{p}}<t<(p+1)T_{\mathrm{p}},\\
&\hat{\mathcal{V}}(t - (\Mtau-1)T_{\mathrm{p}})\,\hat{\mathcal{F}}_{0,\mathrm{pu}}\daga\,(\hat{\mathcal{F}}_{\Delta}\daga)^{\Mtau-1}\,\hat{\mathcal{A}}_{\mathrm{pu}}\,(\hat{\mathcal{F}}_{\Delta}\,\hat{\mathcal{V}}_{\mathrm{p}}\,\hat{\mathcal{A}}_{\mathrm{pu}})^{\Mtau-1}\,\hat{\mathcal{F}}_{0,\mathrm{pu}}\,\vec{R}_0, & (\Mtau-1)T_{\mathrm{p}}<t<\tau,\\
&\hat{\mathcal{V}}(t-\tau)\,\hat{\mathcal{U}}_{\mathrm{pr}}\,\hat{\mathcal{V}}(\tau - (\Mtau-1)T_{\mathrm{p}})\,\hat{\mathcal{F}}_{0,\mathrm{pu}}\daga\,(\hat{\mathcal{F}}_{\Delta}\daga)^{\Mtau-1}\,\hat{\mathcal{A}}_{\mathrm{pu}}\,(\hat{\mathcal{F}}_{\Delta}\,\hat{\mathcal{V}}_{\mathrm{p}}\,\hat{\mathcal{A}}_{\mathrm{pu}})^{\Mtau-1}\,\hat{\mathcal{F}}_{0,\mathrm{pu}}\,\vec{R}_0, & \tau<t<\Mtau T_{\mathrm{p}},\\ 
&\left.
\hspace{-1.5 mm}\begin{aligned}
&\hat{\mathcal{V}}(t-qT_{\mathrm{p}})\,\hat{\mathcal{F}}_{0,\mathrm{pu}}\daga\,(\hat{\mathcal{F}}_{\Delta}\daga)^q\,\hat{\mathcal{A}}_{\mathrm{pu}}\,(\hat{\mathcal{F}}_{\Delta}\,\hat{\mathcal{V}}_{\mathrm{p}}\,\hat{\mathcal{A}}_{\mathrm{pu}})^{q-\Mtau}\,(\hat{\mathcal{F}}_{\Delta})^{\Mtau}\,\hat{\mathcal{F}}_{0,\mathrm{pu}}\\
&\times \hat{\mathcal{V}}(\Mtau T_{\mathrm{p}}-\tau)\,\hat{\mathcal{U}}_{\mathrm{pr}}\,\hat{\mathcal{V}}(\tau - (\Mtau-1)T_{\mathrm{p}})\,\hat{\mathcal{F}}_{0,\mathrm{pu}}\daga\,(\hat{\mathcal{F}}_{\Delta}\daga)^{\Mtau-1}\,\hat{\mathcal{A}}_{\mathrm{pu}}\,(\hat{\mathcal{F}}_{\Delta}\,\hat{\mathcal{V}}_{\mathrm{p}}\,\hat{\mathcal{A}}_{\mathrm{pu}})^{\Mtau-1}\,\hat{\mathcal{F}}_{0,\mathrm{pu}}\,\vec{R}_0, 
\end{aligned}
\right\}
& qT_{\mathrm{p}}<t<(q+1)T_{\mathrm{p}},\\
&\left.
\hspace{-1.5 mm}\begin{aligned}
&\hat{\mathcal{V}}(t-(N-1)T_{\mathrm{p}})\,\hat{\mathcal{F}}_{0,\mathrm{pu}}\daga\,(\hat{\mathcal{F}}_{\Delta}\daga)^{N-1}\,\hat{\mathcal{A}}_{\mathrm{pu}}\,(\hat{\mathcal{F}}_{\Delta}\,\hat{\mathcal{V}}_{\mathrm{p}}\,\hat{\mathcal{A}}_{\mathrm{pu}})^{N-\Mtau-1}\,(\hat{\mathcal{F}}_{\Delta})^{\Mtau}\,\hat{\mathcal{F}}_{0,\mathrm{pu}}\\
&\times \hat{\mathcal{V}}(\Mtau T_{\mathrm{p}}-\tau)\,\hat{\mathcal{U}}_{\mathrm{pr}}\,\hat{\mathcal{V}}(\tau - (\Mtau-1)T_{\mathrm{p}})\,\hat{\mathcal{F}}_{0,\mathrm{pu}}\daga\,(\hat{\mathcal{F}}_{\Delta}\daga)^{\Mtau-1}\,\hat{\mathcal{A}}_{\mathrm{pu}}\,(\hat{\mathcal{F}}_{\Delta}\,\hat{\mathcal{V}}_{\mathrm{p}}\,\hat{\mathcal{A}}_{\mathrm{pu}})^{\Mtau-1}\,\hat{\mathcal{F}}_{0,\mathrm{pu}}\,\vec{R}_0, 
\end{aligned}
\right\}
& t>(N-1) T_{\mathrm{p}}.
\end{aligned}
\right.
\end{aligned}
\label{eq:evolutionpuprpu}
\end{equation}
\end{widetext}
In Eq.~(\ref{eq:evolutionpuprpu}), we have introduced
\begin{equation}
\Mtau - 1= \lfloor \tau/T_{\mathrm{p}}\rfloor
\end{equation}
with the floor function $\lfloor x \rfloor$. The first four lines in Eq.~(\ref{eq:evolutionpuprpu}) are identical to the pump--probe case analyzed previously. Here, however, the second line is present only if $N>2$ and $\tau>T_{\mathrm{p}}$, since it describes the dynamics of the system in the interval $[t_{p},\,t_{p+1}]$, with $p\in\{0,\,\ldots,\,\Mtau-2\}$. The fifth line accounts for the dynamics of the system in the interval $[t_{q},\,t_{q+1}]$, where now the index $q\in\{\Mtau,\,\ldots,\,N-2\}$ is associated with one of the pump pulses following the probe pulse, provided that $N>2$ and $\tau<(N-2)T_{\mathrm{p}}$. The sixth line describes the free evolution of the system after interaction with the whole train of pump pulses, present only if $N$ is finite. The fifth (sixth) line has been obtained from Eq.~(\ref{eq:finitesequence}) with $a = \Mtau$ and $b = q$ ($b = N-1$). The dipole response of the system is provided in Eq.~(\ref{eq:evolutionpuprpu-k}).

For the same reasons described for the pump--probe setup, the integral of Eq.~(\ref{eq:spdefinition}) in $t<\tau$ does not contribute to the absorption spectrum after averaging over fast time-delay-dependent oscillations, while the contribution for $t\in[\tau,\,t_{\Mtau}]$ can be obtained by following the same steps leading to Eq.~(\ref{eq:spectrumpupr}). To account for the terms in the spectrum resulting from the integrals in $[t_q,\,t_{q+1}]$ and for $t>t_{N-1}$,
% \begin{widetext}
% \begin{equation}
% \int_{0}^{T_{\mathrm{p}}}\eu^{-\uimm(\bar{\omega}-\omega_{k1})t}\,\diff t\,\eu^{\uimm\phi_{0,\mathrm{pu}}}\,\vec{v}_k\,\hat{\mathcal{A}}_{\mathrm{pu}}\,(\eu^{-\uimm (\bar{\omega}T_{\mathrm{p}} - \Delta\phi)}\,\hat{\mathcal{F}}_{\Delta}\,\hat{\mathcal{V}}_{\mathrm{p}}\,\hat{\mathcal{A}}_{\mathrm{pu}})^{q-M}\,\eu^{\uimm  M\Delta\phi}\,(\hat{\mathcal{F}}_{\Delta})^M\,\hat{\mathcal{F}}_{0,\mathrm{pu}}\, \eu^{-\uimm  \bar{\omega}(MT_{\mathrm{p}}-\tau)}\,\hat{\mathcal{V}}(MT_{\mathrm{p}}-\tau)\,\hat{\mathcal{U}}_{\mathrm{pr}}\,\hat{\mathcal{V}}(\tau - (M-1)T_{\mathrm{p}})\,\hat{\mathcal{F}}_{0,\mathrm{pu}}\daga\,(\hat{\mathcal{F}}_{\Delta}\daga)^{M-1}\,\hat{\mathcal{A}}_{\mathrm{pu}}\,(\hat{\mathcal{F}}_{\Delta}\,\hat{\mathcal{V}}_{\mathrm{p}}\,\hat{\mathcal{A}}_{\mathrm{pu}})^{M-1}\,\vec{R}_0
% \end{equation}
% \end{widetext}
we first introduce the operator
\begin{equation}
\eu^{-\uimm\bar{\omega}(T_{\mathrm{p}}-\tau')}\,\hat{\mathcal{Z}}(\tau')\doteq \langle\eu^{-\uimm\bar{\omega}(T_{\mathrm{p}}-\tau')}\,\hat{\mathcal{V}}(T_{\mathrm{p}}-\tau')\,\hat{\mathcal{U}}_{\mathrm{pr}}\,\hat{\mathcal{V}}(\tau')\rangle_{\tau'},
\label{eq:Z}
\end{equation}
where $\tau' = \tau -  (\Mtau-1)T_{\mathrm{p}} = \tau-\lfloor \tau/T_{\mathrm{p}}\rfloor T_{\mathrm{p}}$, $\tau'\in[0,\,T_{\mathrm{p}}]$. Due to averaging over fast time-delay-dependent oscillations, several matrix elements of $\hat{\mathcal{Z}}(\tau')$ vanish, as shown in Appendix~\ref{Appendix:The operator Z} explicitly. 
% \vspace{5 mm}
%From Eq.~(\ref{eq:spdefinition}), the contribution of the fifth and sixth lines in Eq.~(\ref{eq:evolutionpuprpu-k}) to the spectrum, after removing the fast oscillations in $\tau$, is provided by the following integrals,
% \begin{equation}
% \begin{aligned}
% &\,\int_{0}^{t_{\mathrm{fin},q'}} \eu^{-\uimm(\bar{\omega}-\omega_{k1})t}\,\diff t\,\vec{v}_k\,\hat{\mathcal{A}}_{\mathrm{pu}}\,(\eu^{-\uimm (\bar{\omega}T_{\mathrm{p}} - \Delta\phi)}\,\hat{\mathcal{F}}_{\Delta}\,\hat{\mathcal{V}}_{\mathrm{p}}\,\hat{\mathcal{A}}_{\mathrm{pu}})^{q'-M}\\
% &\times\,\eu^{-\uimm  \bar{\omega}(T_{\mathrm{p}}-\tau')}\,\hat{\mathcal{Z}}(\tau')\,\hat{\mathcal{F}}_{\Delta}\,\hat{\mathcal{A}}_{\mathrm{pu}}\,(\hat{\mathcal{F}}_{\Delta}\,\hat{\mathcal{V}}_{\mathrm{p}}\,\hat{\mathcal{A}}_{\mathrm{pu}})^{M-1}\,\vec{R}_0\\
% =&\,\frac{1-\eu^{-\uimm(\bar{\omega}-\omega_{k1})t_{\mathrm{fin},q'}}}{\uimm(\bar{\omega} - \omega_{k1})}\,\vec{v}_k\,\hat{\mathcal{A}}_{\mathrm{pu}}\,(\eu^{-\uimm (\bar{\omega}T_{\mathrm{p}} - \Delta\phi)}\,\hat{\mathcal{F}}_{\Delta}\,\hat{\mathcal{V}}_{\mathrm{p}}\,\hat{\mathcal{A}}_{\mathrm{pu}})^{q'-M}\\
% &\times\,\eu^{-\uimm  \bar{\omega}(T_{\mathrm{p}}-\tau')}\,\hat{\mathcal{Z}}(\tau')\,\hat{\mathcal{F}}_{\Delta}\,\hat{\mathcal{A}}_{\mathrm{pu}}\,(\hat{\mathcal{F}}_{\Delta}\,\hat{\mathcal{V}}_{\mathrm{p}}\,\hat{\mathcal{A}}_{\mathrm{pu}})^{M-1}\,\vec{R}_0
% \end{aligned}
% \end{equation}
% where $q'\in\{M,\,\ldots,\,N-1\}$ and $t_{\mathrm{fin},q'}\rightarrow \infty$ if $q'=N-1$, $t_{\mathrm{fin},q'} = T_{\mathrm{p}}$ otherwise. 
Using Eq.~(\ref{eq:inverseoperator}), the spectrum finally reads
% \vspace{45 mm}
\begin{widetext}
\begin{equation}
\begin{aligned}
&\,\mathcal{S}_N(\bar{\omega},\tau)\propto -\omega\Imm\biggl\{\sum_{k=2}^3 \frac{D_{1k}^*}{\uimm(\bar{\omega} - \omega_{k1})}\,\vec{v}_k\,\biggl[\bigl(1-\eu^{-\uimm(\bar{\omega} - \omega_{k1})(T_{\mathrm{p}}-\tau')}\bigr)\,\hat{\mathcal{U}}_{\mathrm{pr}}\,\hat{\mathcal{W}}(\tau')\\
&+ \bigl(1-\eu^{-\uimm(\bar{\omega} - \omega_{k1})T_{\mathrm{p}}}\bigr)\,\hat{\mathcal{A}}_{\mathrm{pu}}\,\Bigl[\hat{\mathcal{I}} - \bigl(\eu^{-\uimm (\bar{\omega}T_{\mathrm{p}}- \Delta\phi)}\,\hat{\mathcal{F}}_{\Delta}\,\hat{\mathcal{V}}_{\mathrm{p}}\,\hat{\mathcal{A}}_{\mathrm{pu}}\bigr)^{N-\Mtau-1}\Bigr]\,\bigl(\hat{\mathcal{I}} -\eu^{-\uimm (\bar{\omega}T_{\mathrm{p}}- \Delta\phi)}\,\hat{\mathcal{F}}_{\Delta}\,\hat{\mathcal{V}}_{\mathrm{p}}\,\hat{\mathcal{A}}_{\mathrm{pu}}\bigr)^{-1}\,\eu^{-\uimm  \bar{\omega}(T_{\mathrm{p}}-\tau')}\,\hat{\mathcal{Z}}(\tau')\,\hat{\mathcal{F}}_{\Delta}\\
&+ \hat{\mathcal{A}}_{\mathrm{pu}}\,\bigl(\eu^{-\uimm (\bar{\omega}T_{\mathrm{p}}- \Delta\phi)}\,\hat{\mathcal{F}}_{\Delta}\,\hat{\mathcal{V}}_{\mathrm{p}}\,\hat{\mathcal{A}}_{\mathrm{pu}}\bigr)^{N-\Mtau-1}\,\eu^{-\uimm  \bar{\omega}(T_{\mathrm{p}}-\tau')}\,\hat{\mathcal{Z}}(\tau')\,\hat{\mathcal{F}}_{\Delta}\biggr]\,\hat{\mathcal{A}}_{\mathrm{pu}}\,(\hat{\mathcal{F}}_{\Delta}\,\hat{\mathcal{V}}_{\mathrm{p}}\,\hat{\mathcal{A}}_{\mathrm{pu}})^{\Mtau-1}\,\vec{R}_0 \biggr\},
\end{aligned}
\label{eq:spectrumpuprpu}
\end{equation}
\end{widetext}
where we have used the fact that
\begin{equation*}
\begin{aligned}
&\,\eu^{\uimm\phi_{0,\mathrm{pu}}}\,\eu^{\uimm \Mtau\Delta\phi}\,(\hat{\mathcal{F}}_{\Delta})^{\Mtau}\,\hat{\mathcal{F}}_{0,\mathrm{pu}}\,\hat{\mathcal{Z}}(\tau')\,\hat{\mathcal{F}}_{0,\mathrm{pu}}\daga\,(\hat{\mathcal{F}}_{\Delta}\daga)^{\Mtau-1} \\
= &\,\hat{\mathcal{Z}}(\tau')\,(\hat{\mathcal{F}}_{\Delta}\daga)^{-1} = \hat{\mathcal{Z}}(\tau')\,\hat{\mathcal{F}}_{\Delta},
\end{aligned}
\end{equation*}
such that also in this case the CEP $\phi_{0,\mathrm{pu}}$ does not influence the absorption spectrum in a noncollinear geometry. 

In analogy to the pump--probe case, the operator $\hat{\mathcal{A}}_{\mathrm{pu}}\,(\hat{\mathcal{F}}_{\Delta}\,\hat{\mathcal{V}}_{\mathrm{p}}\,\hat{\mathcal{A}}_{\mathrm{pu}})^{\Mtau-1}$ describes the state of the system prepared by the initial sequence of $\Mtau$ pump pulses preceding the probe pulse. The term in the first line of Eq.~(\ref{eq:spectrumpuprpu}) has then the same structure as the pump--probe spectrum of Eq.~(\ref{eq:spectrumpupr}), with the factor $\bigl(1-\eu^{-\uimm(\bar{\omega} - \omega_{k1})(T_{\mathrm{p}}-\tau')}\bigr)$ due to the finite duration of the interval $[\tau,\,t_{\Mtau} ]$. The second and third lines in Eq.~(\ref{eq:spectrumpuprpu}) clearly show a structure similar to the probe--pump spectrum of Eq.~(\ref{eq:spectrumprpu}), which becomes even more apparent by using the operator $\hat{\mathcal{D}}_N(\bar{\omega})$ defined in Eq.~(\ref{eq:actionpumpafterprobe}) to write the pump--probe--pump spectrum as
\begin{equation}
\begin{aligned}
\mathcal{S}_N(\bar{\omega},\tau)&\propto -\omega\Imm\biggl\{\sum_{k=2}^3 \frac{D_{1k}^*}{\uimm(\bar{\omega} - \omega_{k1})}\,\vec{v}_k\\
&\ \ \ \times\,\bigl[\bigl(1-\eu^{-\uimm(\bar{\omega} - \omega_{k1})(T_{\mathrm{p}}-\tau')}\bigr)\,\hat{\mathcal{U}}_{\mathrm{pr}}\,\hat{\mathcal{W}}(\tau') \\
&\ \ \ \ \ \ \ + \hat{\mathcal{D}}_{N-\Mtau}(\bar{\omega})\,\eu^{-\uimm  \bar{\omega}(T_{\mathrm{p}}-\tau')}\,\hat{\mathcal{Z}}(\tau')\,\hat{\mathcal{F}}_{\Delta}\bigr]\\
&\ \ \ \times\,\hat{\mathcal{A}}_{\mathrm{pu}}\,(\hat{\mathcal{F}}_{\Delta}\,\hat{\mathcal{V}}_{\mathrm{p}}\,\hat{\mathcal{A}}_{\mathrm{pu}})^{\Mtau-1}\,\vec{R}_0 \biggr\}.
\end{aligned}
\label{eq:spectrumpuprpushort}
\end{equation}
Similarly to the probe--pump case, also here the periodic excitation of the system by $N-\Mtau$ pulses following the probe pulse shapes the absorption spectrum, causing the appearance of LISs. We stress again that the above formulas can be used only if $\gamma\tau\ll 1$ and $\gamma T_{\mathrm{p}}\ll 1$, i.e., for time delays and repetition periods that allow one to neglect the amplitude change of the dipole response due to the decay rate $\gamma$.

\section{Results and discussion}
\label{Results and discussion}

\begin{figure*}
\centering
\includegraphics[width=\textwidth]{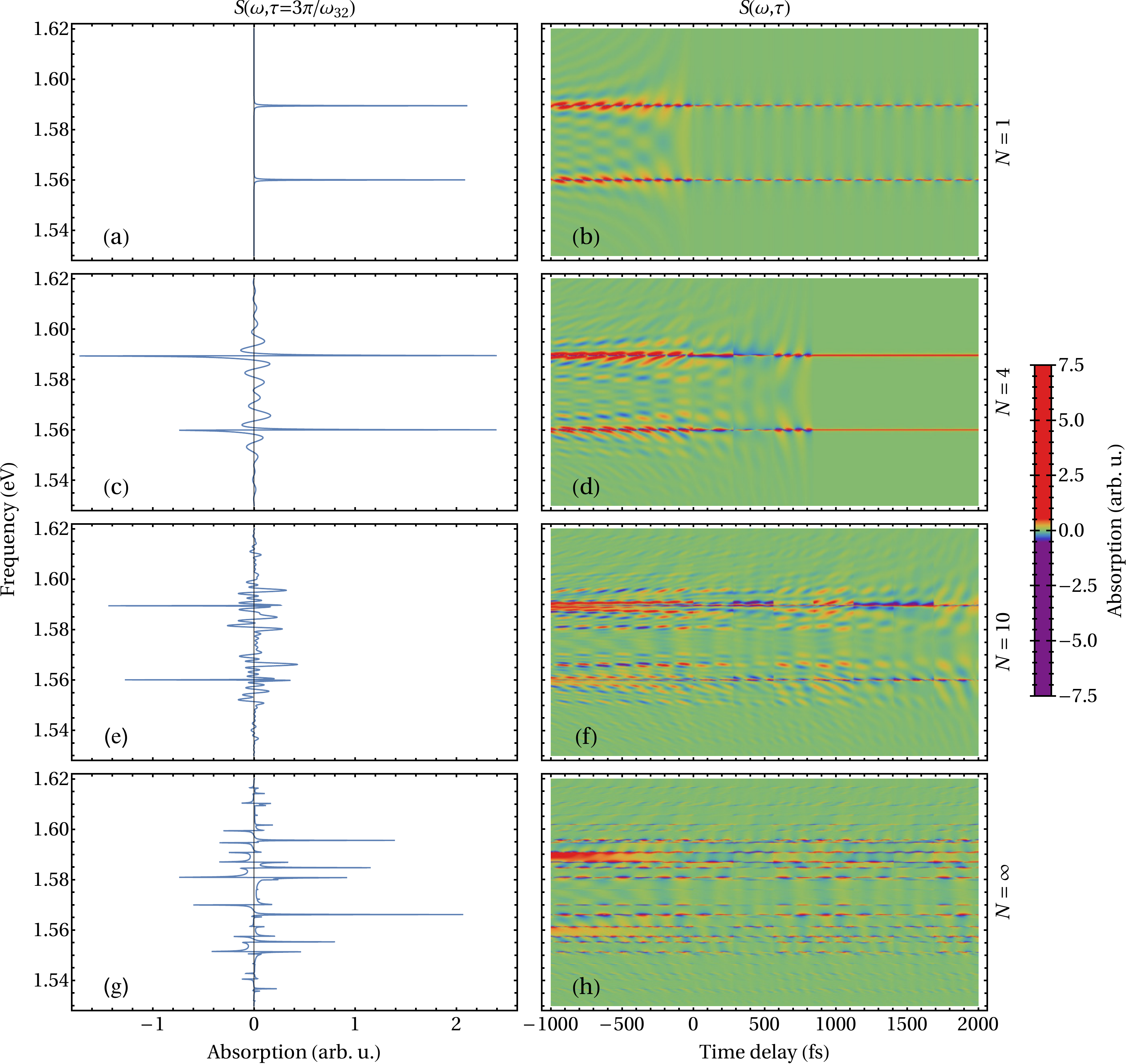}
\caption{Transient-absorption spectra for different numbers $N$ of pump pulses, for fixed pulse area $\vartheta = 3\pi/2$ and pulse-to-pulse phase shift $\Delta\phi = 0$. The number of pump pulses is [(a),(b)] $N=1$, [(c),(d)] $N=4$, [(e),(f)] $N=10$, and [(g),(h)] $N \rightarrow \infty$. The left column [(a),(c),(e),(g)] presents spectral line shapes evaluated at a time delay of $\tau = 3\pi/\omega_{32} = 210\,\mathrm{fs}$, while the two-dimensional spectra on the right column [(b),(d),(f),(h)] are exhibited as a function of frequency $\omega$ and time delay $\tau$.}
\label{fig:TransitionGridPlot1}
\end{figure*}

The formulas obtained in the previous section will be used in the following to characterize the main features of the transient-absorption spectra in the presence of a periodic pump excitation for different setups, i.e., different values of the time delay $\tau$. We assume a repetition frequency of the train of pulses $\omega_{\mathrm{r}} = \omega_{32}/2$, corresponding to a period $T_{\mathrm{p}} \doteq 2\pi/\omega_{\mathrm{r}} = 4\pi/\omega_{32} = 280\,\mathrm{fs}$, and pulse areas $\vartheta\in[0,\,2\pi]$. We also notice that, for a $\delta$~pulse, the spectrum $\tilde{\mathcal{E}}_{0,\mathrm{pu}}(\omega)$ is a constant function, so that the spectrum of a train of $\delta$~pulses is given by a set of equally spaced, equally intense lines as shown in Fig.~\ref{fig:MoreLevelSystem}. The modulation of the spectrum around $\omega_{\mathrm{c}}$ displayed in Fig.~\ref{fig:CombTimeAndFreq}(b) is absent in our case, which explains why the formulas obtained in Sec.~\ref{Theoretical Model} are independent of the carrier frequency. The results, however, do depend explicitly on $\delta_{\mathrm{r}}$ and $\omega_{\mathrm{o}}$. If Gaussian pulses with a duration of $30\,\mathrm{fs}$ were considered instead of $\delta$~pulses, then a pulse area of $2\pi$ would correspond to a peak intensity of $\approx 4\times 10^{10}\,\mathrm{W/cm^2}$.

The spectra are studied in the interval $\tau\in[-1\,\mathrm{ps},\,2\,\mathrm{ps}]$, assuming an experimental width of $\gamma = 0.2\,\mathrm{ps}^{-1}$, such that both $\gamma\tau\ll 1$ and $\gamma T_{\mathrm{p}}\ll 1$ hold. For the atomic implementation in Rb, where $D_{1k}$ are well approximated by their nonrelativistic values \cite{johnson2007atomic}, $D_{13} = D_{12}\sqrt{2}$, it follows that $\alpha = \arctan(\sqrt{2}/2)$. We assume a weak probe pulse with vanishing CEP, i.e., $\vartheta_{\mathrm{pr}}\ll 1$ and $\phi_{0,\mathrm{pr}} = 0$, described by the interaction operator
\begin{equation}
\hat{U}_{\mathrm{pr}} = 
\begin{pmatrix}
1 & \uimm\, \frac{\vartheta_{\mathrm{pr}}}{2}\,\sin(\alpha) &\uimm\, \frac{\vartheta_{\mathrm{pr}}}{2}\,\cos(\alpha)\\
\uimm\,\frac{\vartheta_{\mathrm{pr}}}{2}\,\sin(\alpha) & 1 &0\\
\uimm\, \frac{\vartheta_{\mathrm{pr}}}{2}\,\cos(\alpha) &0 &1
\end{pmatrix},
\label{eq:probepulseapprox}
\end{equation}
where we have neglected terms of second or higher order in $\vartheta_{\mathrm{pr}}$. Since there is no ambiguity, in the following and in the Appendixes we drop the subscript in $\vartheta_{\mathrm{pu}}$, so that $\vartheta$ always refers to the pump-pulse area.

\subsection{Appearance of light-induced states for an increasing number of pump pulses}
\label{Dependence on the number of pump pulses}
% \label{Transient absorption spectra for one, several and infinitely many pump pulses}

In Fig.~\ref{fig:TransitionGridPlot1}, the time-delay-dependent absorption spectra are displayed for fixed values of the pulse area $\vartheta = 3\pi/2$ and pulse-to-pulse phase shift $\Delta\phi = 0$, for an increasing number $N$ of pump pulses. For a single $\delta$~like pump pulse centered on $t=0$, Fig.~\ref{fig:TransitionGridPlot1}(b) shows the modification of the absorption line shapes of a probe pulse centered on $t=\tau$. The main features in this single-pulse case were already thoroughly described in Refs.~\cite{PhysRevLett.115.033003, 0953-4075-51-3-035501}. In particular, the two absorption lines, centered on the atomic transition energies $\omega_{21} = 1.56\,\mathrm{eV}$ and $\omega_{31} = 1.59\,\mathrm{eV}$, respectively, exhibit oscillations as a function of time delay, with a periodicity of $2\pi/\omega_{32} = 140\,\mathrm{fs}$ determined by the beating frequency $\omega_{32}$. At negative time delays, when the evolution of the atomic dipole between the first-arriving probe pulse and the subsequent pump pulse influences the spectrum, perturbed free-induction-decay sidebands appear \cite{0953-4075-49-6-062003}, which become more significant for increasing values of $|\tau|$ [see also the first line in Eq.~(\ref{eq:spectrumprpu})].

Two main features emerge for increasing values of $N$. Firstly, a pump--probe--pump region appears for positive time delays, where the periodic excitation due to the pump pulses, at a repetition period of $T_{\mathrm{p}} = 2\times 2\pi/\omega_{32} = 280\,\mathrm{fs}$, can be recognized in the time-delay dependence of the absorption spectral lines. Furthermore, the spectra in this positive-time-delay region also present perturbed free-induction-decay sidebands similar to the negative-time-delay case [see also the first line in Eq.~(\ref{eq:spectrumpuprpu})], which can be identified in Figs.~\ref{fig:TransitionGridPlot1}(d) and \ref{fig:TransitionGridPlot1}(f) for a finite number of pump pulses. Secondly, for $\tau<(N-1)T_{\mathrm{p}}$, the periodic excitation of the atomic dipole, resulting from the $N-\Mtau$ pump pulses which follow the probe pulse, induces the appearance of LISs. This becomes increasingly significant for larger values of $N$, up to the limit of infinitely many pulses shown in Fig.~\ref{fig:TransitionGridPlot1}(h). The onset and clear appearance of these additional lines is highlighted in the left column of Fig.~\ref{fig:TransitionGridPlot1}, which displays absorption spectral lines at a fixed value of the time delay.

The appearance of LISs is associated with the dynamics of the system following the probe pulse and periodically modified by a train of $N$ or $N - \Mtau$ pump pulses, for a probe--pump and pump--probe--pump setup, respectively. The larger the number of pump pulses following the probe pulse, the more defined and intense these additional lines will be. For this reason, at positive time delays and for a fixed total number $N$ of pump pulses, the additional spectral lines gradually fade out for increasing values of $\Mtau$, i.e., when one approaches the end of the pump-pulse train. This appears clearly in Fig.~\ref{fig:TransitionGridPlot1}(f) for large positive values of $\tau$.

% 
% As we further discuss in Appendix~\ref{Appendix:Central frequency of the spectral lines}, the central frequency of the spectral lines can be mathematically understood as resulting from the term $\hat{\mathcal{D}}_N(\bar{\omega})/[\uimm(\bar{\omega} - \omega_{k1})]$ in Eqs.~(\ref{eq:spectrumprpushort}) at negative time delays. While the denominator is responsible for the spectral lines centered on the transition energies $\omega_{k1}$, the appearance of additional lines in the spectrum for increasing values of $N$ is due to the operator $\hat{\mathcal{D}}_N(\bar{\omega})$ and in particular to its real poles for $N\rightarrow\infty$. For a pump--probe--pump setup at positive time delays, the central frequency of the spectral lines results from the operator $\hat{\mathcal{D}}_{N-\Mtau}(\bar{\omega})/[\uimm(\bar{\omega} - \omega_{k1})]$ in Eq.~(\ref{eq:spectrumpuprpushort}). The additional lines have therefore the same central frequency as for negative time delays, but their intensity in this case gradually fades out for increasing values of $\Mtau$, i.e., when one approaches the end of the pump-pulse train. This appears clearly in Fig.~\ref{fig:TransitionGridPlot1}(f) for large positive values of $\tau$.

\subsection{Dependence on laser control parameters for infinitely many pump pulses}
\label{Dependence on the pulse-to-pulse phase shift and pulse area for infinitely many pump pulses}
% \label{Dependence on the phase}

In this section, we explicitly focus on the case of infinitely many pump pulses, although it is apparent from the above discussion that the main spectral features exhibited by the spectra for $N\rightarrow \infty$ are already present for finite, sufficiently large numbers of pulses. We investigate the information encoded in the frequency of the LISs appearing in the spectrum as a function of control parameters such as the pulse area $\vartheta$ and pulse-to-pulse phase shift $\Delta\phi$. As discussed in Sec.~\ref{Theoretical Model}, LISs at $N\rightarrow \infty$ are due to the action of the infinite sequence of pump pulses following the probe pulse, reflected by the operator $\hat{\mathcal{D}}_{\infty}(\bar{\omega})$ in Eq.~(\ref{eq:actioninfinitelymanypumpafterprobe}). At the same time, we also investigate the time-delay-dependent features of the spectra, especially for $\tau>0$, focusing on the influence of the $\Mtau$ pump pulses preceding the probe pulse. Mathematical details are presented in the Appendixes~\ref{Appendix:Central frequency of the spectral lines}, \ref{Appendix:Spectral features in a pump--probe--pump setup determined by the pump pulses preceding the probe pulse}, and \ref{Appendix:Details on the time-delay-dependent features of the spectra}.

\subsubsection{Frequency-dependent features of the light-induced states}
\label{Frequency-dependent features and position of the additional spectral lines}

In order to gain an intuitive understanding of the origin of the LISs displayed in Fig.~\ref{fig:TransitionGridPlot1}, we can for instance focus on the probe--pump setup ($\tau<0$) and look at the time evolution of the atomic dipoles for $t>0$, i.e., following the first excitation from the pump-pulse train. Without loss of generality, we can then write
\begin{equation}
\begin{aligned}
\rho_{1k}(t,\tau) =\,& \sum_{j = 0}^{\infty}\rho_{1k}(j T_{\mathrm{p}}^+)\,\eu^{-\uimm \omega_{k1}(t - jT_{\mathrm{p}})}\\
&\times\,\{\theta(t-jT_{\mathrm{p}}) - \theta[t - (j+1)T_{\mathrm{p}}]\},
\end{aligned}
\end{equation}
where $\rho_{1k}(j T_{\mathrm{p}}^+)$ is the dipole immediately following the interaction with the $j$th pump pulse and where $\theta(x)$ is the Heaviside step function. The spectrum in Eq.~(\ref{eq:spdefinition}) will then be related to
\begin{equation}
\begin{aligned}
&\int_{0}^{\infty}\rho_{1k}(t,\tau)\,\eu^{-\uimm\bar{\omega}(t - \tau)}\,\diff t = \eu^{\uimm\bar{\omega}\tau}\,\eu^{-\uimm(\bar{\omega} - \omega_{k1})T_{\mathrm{p}}/2}\\
&\ \ \ \ \times\,T_{\mathrm{p}}\sinc{\left[(\bar{\omega} - \omega_{k1})\frac{T_{\mathrm{p}}}{2}\right]}\sum_{j = 0}^{\infty}\rho_{1k}(jT_{\mathrm{p}}^+)\,\eu^{-\uimm\bar{\omega} j T_{\mathrm{p}}}.
\end{aligned}
\end{equation}
Let us then suppose that the train of pulses, acting on the system with repetition frequency $\omega_{\mathrm{r}}$, will periodically generate the same atomic state with a frequency $\nu$, i.e., $\rho_{1k}(jT_{\mathrm{p}}^+)= \eu^{\uimm\nu j T_{\mathrm{p}}}$ is a periodic function. In such case,
\begin{equation}
\begin{aligned}
\sum_{j = 0}^{\infty}\rho_{1k}(jT_{\mathrm{p}}^+)\,\eu^{-\uimm\bar{\omega} j T_{\mathrm{p}}} &= \lim_{N\rightarrow\infty}\sum_{j = 0}^{N-1} \eu^{\uimm\nu j T_{\mathrm{p}}}\,\eu^{-\uimm(\bar{\omega} -\nu) j T_{\mathrm{p}}} \\
&= \lim_{N\rightarrow\infty}\frac{1-\eu^{-\uimm(\bar{\omega} -\nu)NT_{\mathrm{p}}}}{1-\eu^{-\uimm(\bar{\omega} -\nu)T_{\mathrm{p}}}}
\end{aligned}
\end{equation}
has a comb-like shape, with peaks centered on $\bar{\omega}_s = \nu + s \omega_{\mathrm{r}}$, $s\in\mathbb{Z}$ [see also Eq.~(\ref{eq:equality-peak-structure})]. Already from this  discussion, we can expect that the spectrum will consist of a series of lines, separated by the repetition frequency $\omega_{\mathrm{r}}$ and modulated by $\sinc{[(\bar{\omega} - \omega_{k1}){T_{\mathrm{p}}}/{2}]}$. If $\rho_{1k}(jT_{\mathrm{p}}^+)$ contains several frequency components $\eu^{\uimm \nu_n j T_{\mathrm{p}}}$ at frequencies $\nu_n$, then they will appear in the spectrum as groups of lines centered on the associated frequencies $\bar{\omega}_{ns} = \nu_n + s \omega_{\mathrm{r}}$. This is thoroughly discussed in Appendix~\ref{Appendix:Central frequency of the spectral lines}.

\begin{figure*}
\centering
\includegraphics[width=\textwidth]{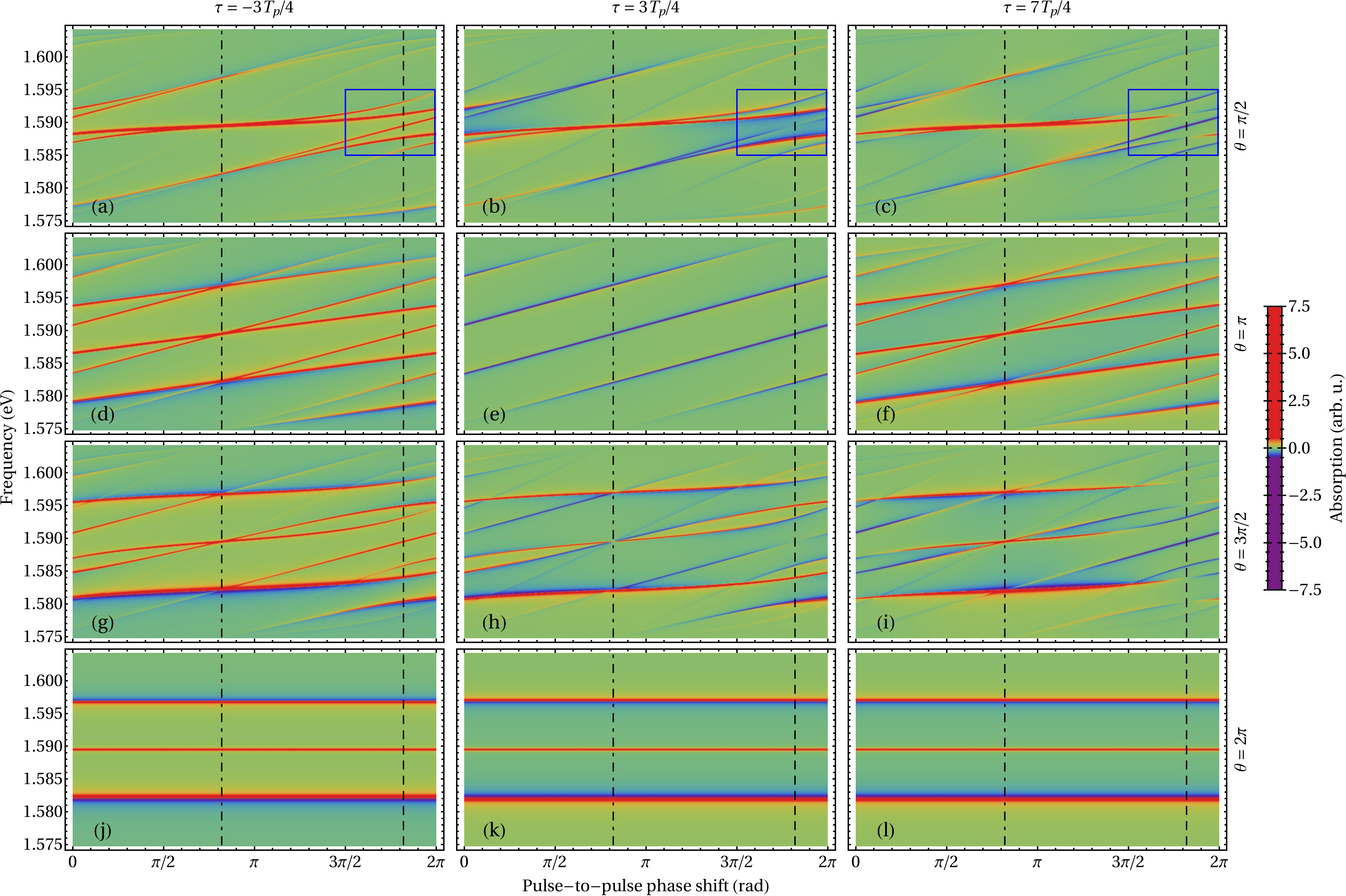}
\caption{Transient-absorption spectra for an infinite number of pump pulses as a function of frequency $\omega$ and pulse-to-pulse phase shift $\Delta\phi$, for pulse areas [(a)--(c)] $\vartheta = \pi/2$, [(d)--(f)] $\vartheta = \pi$, [(g)--(i)] $\vartheta = 3\pi/2$, and [(j)--(l)] $\vartheta = 2\pi$, and time delays [(a),(d),(g),(j)] $\tau = -3 T_{\mathrm{p}}/4$, [(b),(e),(h),(k)] $\tau = 3 T_{\mathrm{p}}/4$, and [(c),(f),(i),(l)] $\tau = 7 T_{\mathrm{p}}/4$. The dashed lines are centered at $\Delta\phi = \delta_{\mathrm{r}}T_{\mathrm{p}}$, the dot-dashed lines at $\Delta\phi = \delta_{\mathrm{r}}T_{\mathrm{p}} - \pi$. The blue boxes in panels (a)--(c) highlight the 5-level structures exhibited by the spectra.}
\label{fig:DetailsDeltaPhiDependenceGridPlot2}
\end{figure*}

For $N\rightarrow \infty$, Fig.~\ref{fig:DetailsDeltaPhiDependenceGridPlot2} displays the dependence of the central frequencies of the absorption lines on the phase shift $\Delta\phi$ for different values of $\tau$ and $\vartheta$. Here, in particular, we focus on the behavior around the transition energy $\omega_{31} = 1.59\,\mathrm{eV}$. Some general features can be recognized in the first row of Fig.~\ref{fig:DetailsDeltaPhiDependenceGridPlot2} [panels (a)--(c)] for $\vartheta = \pi/2$. Firstly, we notice that, for values of the phase shift $\Delta\phi>3\pi/2$, the absorption line present at $\omega_{31}$ is now shaped into five LISs, as highlighted in the blue boxes. As introduced in the above discussion, the frequencies of these five lines are associated with the frequency components of the evolution of $\rho_{13}(jT_{\mathrm{p}}^+)$. In particular, and as discussed thoroughly in Appendix~\ref{Appendix:omegao = deltar}, when $\omega_{\mathrm{o}} = \delta_{\mathrm{r}}$, i.e., at $\Delta\phi =\delta_{\mathrm{r}} T_{\mathrm{p}}$ (dashed lines), the five lines are equally spaced, separated by a frequency gap of $\Delta\omega = \vartheta/(2T_{\mathrm{p}})$ which here is equal to $\pi/(4T_{\mathrm{p}}) = 2\,\mathrm{meV}$. Several five-line structures appear in the spectrum, as expected from the above discussion: the structures are separated by the repetition frequency $\omega_{\mathrm{r}} = 2\pi/T_{\mathrm{p}} = 15\,\mathrm{meV}$, with the $s$th structure thus centered on $\omega = \omega_{31} + s \omega_{\mathrm{r}}$, $s\in\mathbb{Z}$. We notice that the pulse-to-pulse phase shift and the time delay both affect the shape of the lines, which turn from a Lorentzian to a Fano-like shape depending on the value of $\Delta\phi$ and $\tau$. %This is also influenced by the value of the time delay. For example, for $\tau = 3\pi/\omega_{32}$ [probe pulse arriving between the $0$th and the $1$st pump pulse, as shown in Fig. \ref{fig:DetailsDeltaPhiDependenceGridPlot2}(b)], the intensity of the central line in the {\color{red}blue box} vanishes for $\Delta\phi = \delta_{\mathrm{r}} T_{\mathrm{p}}$. In contrast, when $\tau = 5\pi/\omega_{32}$ [probe pulse arriving between the $1$st and the $2$nd pump pulse, as shown in Fig. \ref{fig:DetailsDeltaPhiDependenceGridPlot2}(c)] it is the intensity of the second and fourth lines which vanishes for the same value of $\Delta\phi$. %The disappearance of given lines for set values of the time delay is a general property of the spectra, related to the preparation of the system by the $\Mtau$ pulses preceding the probe pulse. We show this in one particular case

As a second general feature of the spectra, we notice that the spacing between the five lines changes with $\Delta\phi$, with the lines forming groups as shown in Figs.~\ref{fig:DetailsDeltaPhiDependenceGridPlot2}(a)--(c). In particular, when $\omega_{\mathrm{o}} = \delta_{\mathrm{r}} - \pi/T_{\mathrm{p}}$, such that $\Delta\phi =\delta_{\mathrm{r}} T_{\mathrm{p}}-\pi$ (dot-dashed line), the lines merge into single lines centered on $\omega = \omega_{31}$ or $\omega = \omega_{31}\pm \omega_{\mathrm{r}}/2$, as discussed in Appendix~\ref{Appendix:omegao = deltar - pi/T}. When decreasing $\Delta\phi$ even further, the lines ungroup again, to newly approach a five-line structure---the results are periodic in $\Delta\phi$ $\mathrm{mod}\,2\pi$. 

This line merging takes place also for higher values of the pulse area, as one can see by comparing Figs.~\ref{fig:DetailsDeltaPhiDependenceGridPlot2}(a), \ref{fig:DetailsDeltaPhiDependenceGridPlot2}(d), and \ref{fig:DetailsDeltaPhiDependenceGridPlot2}(g) [for the $2\pi$-area case of Fig.\ref{fig:DetailsDeltaPhiDependenceGridPlot2}(j), no merging takes place, as we will discuss afterwards]. In particular, the frequencies at which the lines merge, $\omega = \omega_{31}$ or $\omega = \omega_{31}\pm \omega_{\mathrm{r}}/2$, do not depend on $\vartheta$, as shown in Appendix~\ref{Appendix:omegao = deltar - pi/T}. Other lines tend to group towards single lines centered on $\omega = \omega_{31}\pm \omega_{\mathrm{r}}$, but their intensities decrease for $\Delta\phi\rightarrow \delta_{\mathrm{r}} T_{\mathrm{p}}-\pi$ so that no spectral line appears at $\omega_{31}\pm \omega_{\mathrm{r}}$ when $\Delta\phi$ is exactly equal to $\delta_{\mathrm{r}} T_{\mathrm{p}}-\pi$. 

With the increase in the pulse area, the frequency gap $\Delta\omega = \vartheta/(2T_{\mathrm{p}})$ between individual lines in each five-line structure also grows. This can lead to the intersection or merging of lines belonging to different structures. $\vartheta = \pi$ is the smallest pulse area for which such intersections take place: in this case and for $\Delta\phi = \delta_{\mathrm{r}} T_{\mathrm{p}}-\pi$, the central frequency of the top line in the $s$th structure, $\omega_{31} + s \omega_{\mathrm{r}} + 2\Delta\omega$, and that of the bottom line in the $(s+1)$th structure, $\omega_{31} + (s+1) \omega_{\mathrm{r}} - 2\Delta\omega$, coincide and are equal to $\omega_{31} + 2s\pi/T_{\mathrm{p}} + \pi/T_{\mathrm{p}}$. 

This can be recognized in Figs.~\ref{fig:DetailsDeltaPhiDependenceGridPlot2}(d)--(f) for $\vartheta = \pi$. The behavior of the spectrum and the position of the LISs for $\vartheta = \pi$ are described in detail in Appendix~\ref{Appendix:pi-area pulses}. Firstly, we notice that the position of all absorption lines depends linearly upon $\Delta\phi$ for this value of the pulse area. Furthermore, it is now more difficult than in the previous $\pi/2$-area case to identify groups consisting of five lines in the spectrum, because two lines belonging to different groups are here completely merged. It is interesting to see in Fig.~\ref{fig:DetailsDeltaPhiDependenceGridPlot2}(e) how some of the above lines do not appear at all when $t_0<\tau<t_1$. This dependence is a direct result of the action of the $\pi$-area pump pulses preceding the arrival of the probe pulse. These $\Mtau$ pulses prepare the system in the state which is then encountered by the probe pulse, and which determines the shapes of the lines in the spectrum, as explained in Appendix~\ref{Appendix:Spectral features in a pump--probe--pump setup determined by the pump pulses preceding the probe pulse}. This is a first example of the dependence of the spectra on time delay, which will be more clearly visible in Figs.~\ref{fig:OmegaTauGridPlot3} and \ref{fig:TauThetaGridPlot5}. 

While intersections of different lines appear only at $\Delta\phi = \delta_{\mathrm{r}} T_{\mathrm{p}}-\pi$ for $\vartheta \leq\pi$, lines will intersect also at additional values of $\Delta\phi$ for larger pulse areas. This is exhibited in Figs.~\ref{fig:DetailsDeltaPhiDependenceGridPlot2}(g)--(i) for $\vartheta = 3\pi/2$. However, these intersections render it also more difficult to distinguish five-line structures in the spectrum, although it would still be possible to formally group the lines as in the case of $\vartheta = \pi/2$. Finally, when $\vartheta = 2\pi$, as in Figs.~\ref{fig:DetailsDeltaPhiDependenceGridPlot2}(j)--(l), only three lines can be distinguished, whose positions and shapes do not depend on the pulse-to-pulse phase shift $\Delta\phi$. The lines are centered on $\omega_{31}$ and $\omega_{31}\pm\vartheta/(2T_{\mathrm{p}})$. For $\vartheta = 2\pi$, these frequencies are equal to $\omega = \omega_{31}$ and $\omega = \omega_{31}\pm \omega_{\mathrm{r}}/2$, and thus correspond to the above-mentioned $\vartheta$-independent frequencies at which the spectral lines are centered when $\Delta\phi =\delta_{\mathrm{r}} T_{\mathrm{p}}-\pi$. Appendix~\ref{Appendix:2pi-area pulses} presents the details of this $2\pi$-area case.

\begin{figure*}
\centering
\includegraphics[width=\textwidth]{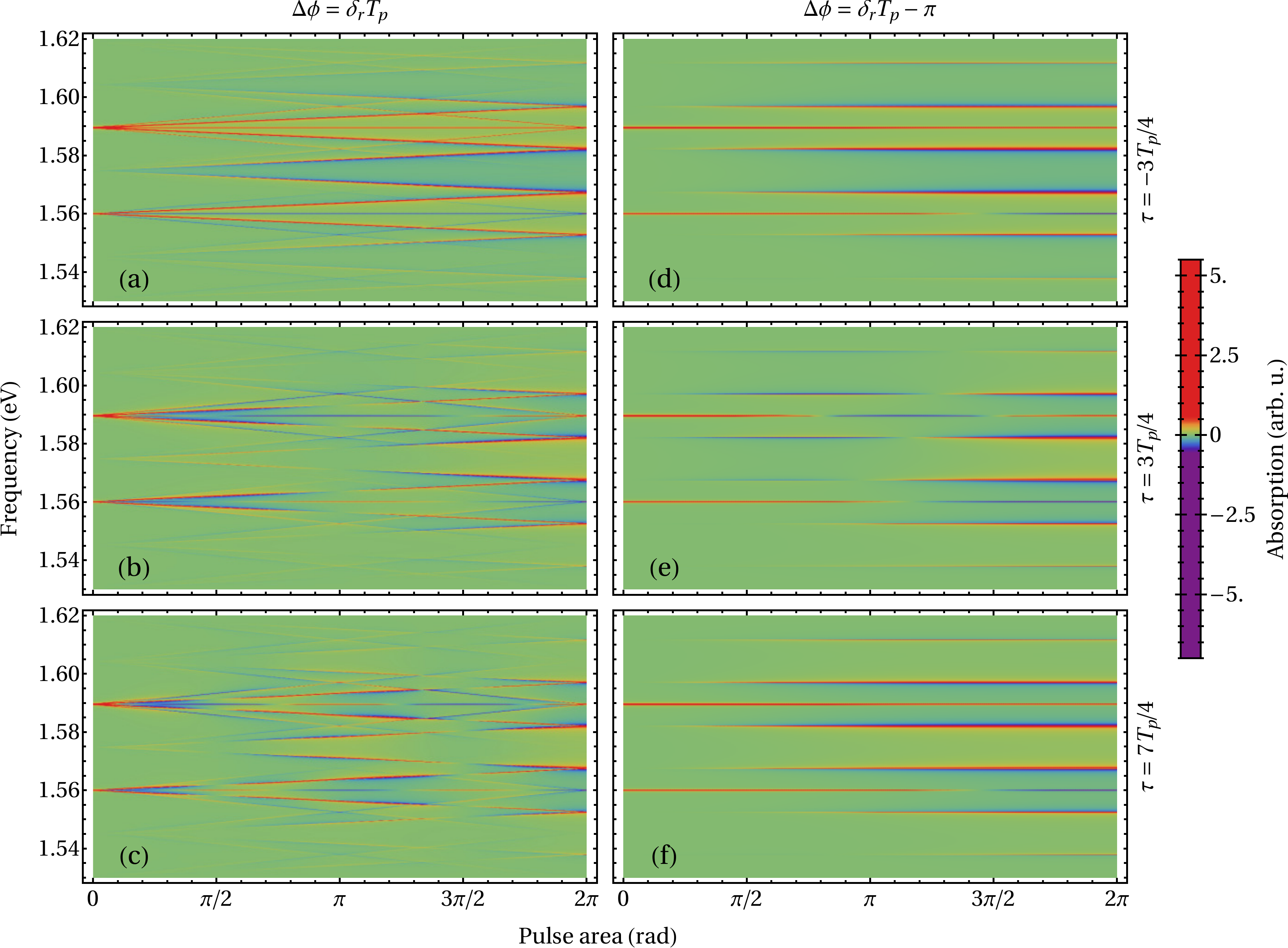}
\caption{Transient-absorption spectra for an infinite number of pump pulses as a function of frequency $\omega$ and pulse area $\vartheta$, for pulse-to-pulse phase shifts [(a)--(c)] $\Delta\phi = \delta_{\mathrm{r}}T_{\mathrm{p}}$ and [(d)--(f)] $\Delta\phi = \delta_{\mathrm{r}}T_{\mathrm{p}} - \pi$, and time delays [(a),(d)] $\tau = -3 T_{\mathrm{p}}/4$, [(b),(e)] $\tau = 3 T_{\mathrm{p}}/4$, and [(c),(f)] $\tau = 7 T_{\mathrm{p}}/4$.}
\label{fig:OmegaThetaGridPlot4}
\end{figure*}

Also the absorption spectral line centered at $\omega = \omega_{21}$ is shaped into several five-line structures when $N\rightarrow \infty$. In order to render this apparent, in Fig.~\ref{fig:OmegaThetaGridPlot4} we display transient-absorption spectra as a function of frequency and pulse area, evaluated at the two values of the pulse-to-pulse phase shift $\Delta\phi$ which were recognized to be important in the above discussion, and for the same discrete values of the time delay $\tau$ already used in Fig.~\ref{fig:DetailsDeltaPhiDependenceGridPlot2}. The left column [Figs.~\ref{fig:OmegaThetaGridPlot4}(a)--(c)] displays results evaluated at $\Delta\phi = \delta_{\mathrm{r}}T_{\mathrm{p}}$. Here, the expansion of the five-level structures as a function of $\vartheta$, with already described intersections for values of the pulse area larger than $\pi$, can be clearly recognized (see also Appendix~\ref{Appendix:omegao = deltar}). The right column [Figs.~\ref{fig:OmegaThetaGridPlot4}(d)--(f)], with the results evaluated at $\Delta\phi = \delta_{\mathrm{r}}T_{\mathrm{p}} - \pi$, shows once more that the position of the lines is not influenced by the value of $\vartheta$ for this particular choice of the pulse-to-pulse phase shift (see also Appendix~\ref{Appendix:omegao = deltar - pi/T}).

The central frequencies of the LISs appearing in the spectrum are related to the action of the intense pump pulses: this is immediate for $\Delta\phi = \delta_{\mathrm{r}}T_{\mathrm{p}}$, where the spacing between the lines in the same five-level structure is given by $\Delta\omega = \vartheta/(2T_{\mathrm{p}})$ and is thus due to the amplitude and phase action of each single pump pulse. For $N = 1$, a light-imposed amplitude and phase change would modify the dipole decay, and would therefore lead to a change of the absorption line shapes from Lorentzian to Fano-like. By acting on the system several times, however, the repeated amplitude and phase changes imposed by the pulses lead to the appearance of several LIS structures. Information on the action of the pulses can therefore be directly extracted from the central frequency of the LISs, complementing the information which could be obtained by a detailed analysis of the absorption line shapes. 

\subsubsection{Time-delay-dependent features \\and periodicity of the spectra}
\label{Time-delay-dependent features for given values of the pulse-to-pulse phase shift}

\begin{figure*}
\centering
\includegraphics[width=\textwidth]{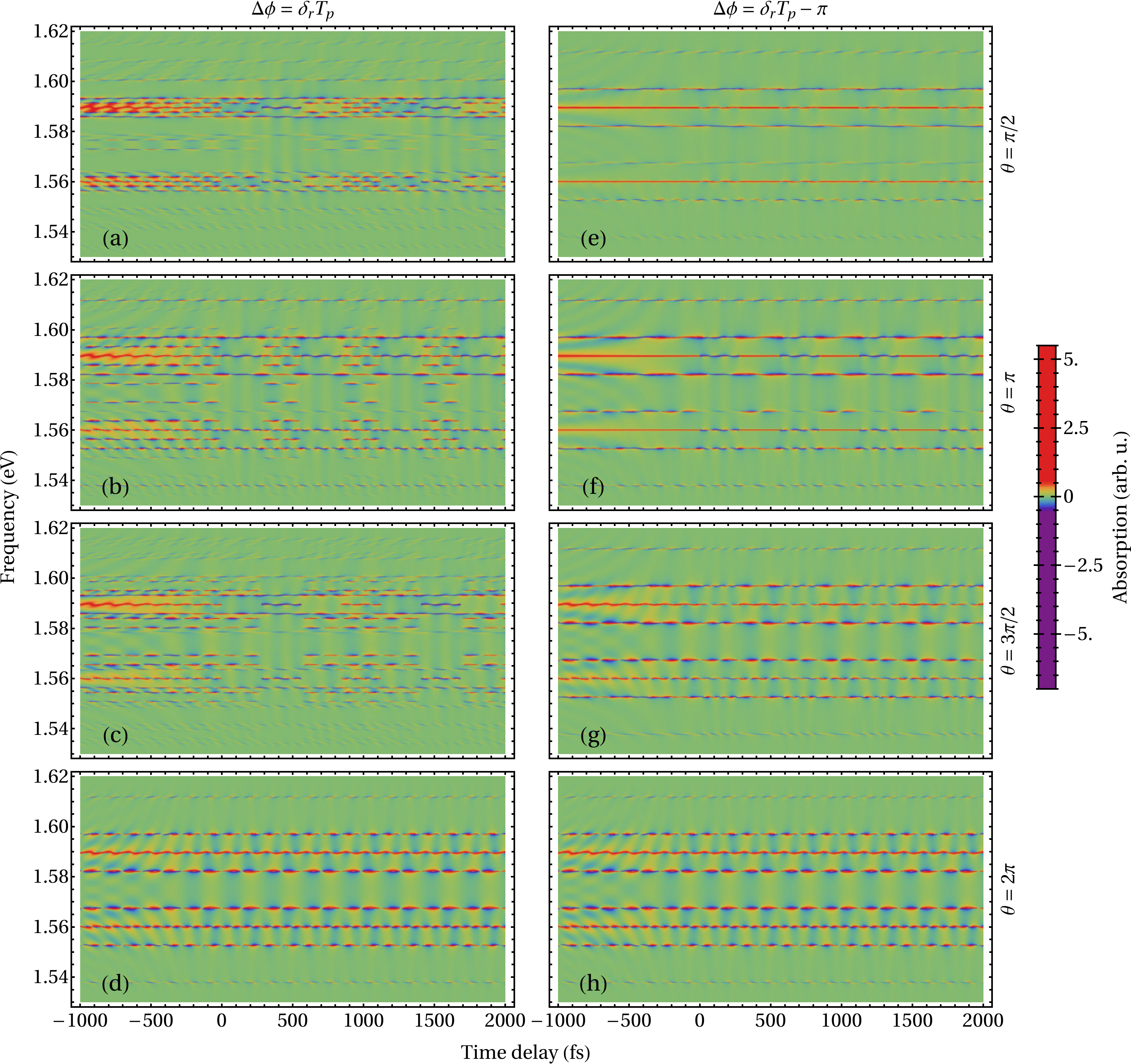}
\caption{Transient-absorption spectra for an infinite number of pump pulses as a function of frequency $\omega$ and time delay $\tau$, for pulse-to-pulse phase shifts [(a)--(d)] $\Delta\phi = \delta_{\mathrm{r}}T_{\mathrm{p}}$ and [(e)--(h)] $\Delta\phi = \delta_{\mathrm{r}}T_{\mathrm{p}} - \pi$, and pulse areas [(a),(e)] $\vartheta = \pi/2$, [(b),(f)] $\vartheta = \pi$, [(c),(g)] $\vartheta = 3\pi/2$, and [(d),(h)] $\vartheta = 2\pi$.}
\label{fig:OmegaTauGridPlot3}
\end{figure*}

In order to focus on the time-delay-dependent features of the spectrum, especially in the pump--probe--pump region at $\tau>0$, in Fig.~\ref{fig:OmegaTauGridPlot3} we display transient-absorption spectra as a function of frequency and time delay for given values of the pulse-to-pulse phase shift $\Delta\phi$ and pulse area $\vartheta$. The left column [Figs.~\ref{fig:OmegaTauGridPlot3}(a)--(d)] presents spectra at $\Delta\phi = \delta_{\mathrm{r}}T_{\mathrm{p}}$ for increasing values of the pulse area $\vartheta$. Several five-level structures are recognizable in Fig.~\ref{fig:OmegaTauGridPlot3}(a), separated by the repetition frequency $\omega_{\mathrm{r}}$. Furthermore, Figs.~\ref{fig:OmegaTauGridPlot3}(a)--(c) highlight the increase in the frequency spacing between lines belonging to the same structure for growing values of $\vartheta$, with the above-described intersections and merging for pulse areas $\vartheta\geq \pi$. For $2\pi$-area pulses, as shown in Fig.~\ref{fig:OmegaTauGridPlot3}(d) and already discussed for Figs.~\ref{fig:DetailsDeltaPhiDependenceGridPlot2}(j)--(l), a lower number of spectral lines appear. 

The results displayed in the right column [Figs.~\ref{fig:OmegaTauGridPlot3}(e)--(h)] are obtained for $\Delta\phi = \delta_{\mathrm{r}}T_{\mathrm{p}} - \pi$. In this case, the central frequencies of the lines appearing in the spectrum do not depend on $\vartheta$, and are the same in all four panels. They also coincide with the $\Delta\phi$-independent central frequencies of the spectra evaluated at $\vartheta = 2\pi$. We notice that the two spectra in Figs.~\ref{fig:OmegaTauGridPlot3}(d) and \ref{fig:OmegaTauGridPlot3}(h), evaluated at different values of $\Delta\phi$ and for $\vartheta = 2\pi$, show identical frequency- and time-delay-dependent features: this is a general feature of the spectra for $\vartheta = 2\pi$, which are independent of $\Delta\phi$ as we show in Appendix~\ref{Appendix:2pi-area pulses-time}.

Figure~\ref{fig:OmegaTauGridPlot3} also allows one to focus on the time-delay-dependent features of the spectrum in the pump--probe--pump region at positive delays. Figures~\ref{fig:OmegaTauGridPlot3}(a)--(c) show that certain lines, otherwise present in the spectrum, are suppressed for given time-delay intervals. While the position of the lines is, in general, determined by the periodic action of the pump-pulse sequence following the probe pulse [and in particular by the poles of the operator $\hat{\mathcal{D}}_{\infty}(\bar{\omega})$ in Eq.~(\ref{eq:actioninfinitelymanypumpafterprobe})], the shape of the spectral lines is determined by the state encountered by the probe pulse, resulting from the action of the sequence of $\Mtau$ pump pulses which precede it. A change of $\Mtau$ causes a modification in the resulting prepared state, and for given values of $\vartheta$ there exists a number of pulses $\Mtau$ for which some of the lines in the spectrum are suppressed. Although this is a general property of the time-delay-dependent spectra displayed here, in Appendix~\ref{Appendix:Spectral features in a pump--probe--pump setup determined by the pump pulses preceding the probe pulse} we explain the disappearance of the spectral lines in the particular case exhibited in Fig.~\ref{fig:OmegaTauGridPlot3}(b), i.e., for $\vartheta = \pi$ and for an odd number $\Mtau$ of pump pulses preceding the probe pulse. 

Figure~\ref{fig:OmegaTauGridPlot3} exhibits the periodic features of the spectrum as a function of time delay for $\tau>0$. For instance, for $\Delta\phi = \delta_{\mathrm{r}}T_{\mathrm{p}}$, one can recognize a periodicity of $4T_{\mathrm{p}}$ at $\vartheta = \pi/2$ and $\vartheta = 3\pi/2$ [Figs.~\ref{fig:OmegaTauGridPlot3}(a) and \ref{fig:OmegaTauGridPlot3}(c), respectively], $2T_{\mathrm{p}}$ at $\vartheta = \pi$ [Fig.~\ref{fig:OmegaTauGridPlot3}(c)], and $T_{\mathrm{p}}$ at $\vartheta = 2\pi$ [Fig.~\ref{fig:OmegaTauGridPlot3}(d)]. In contrast, all the spectra evaluated at $\Delta\phi = \delta_{\mathrm{r}}T_{\mathrm{p}} - \pi$ [Figs.~\ref{fig:OmegaTauGridPlot3}(e)--(h)] have a periodicity of $2T_{\mathrm{p}}$, including the particular case of the spectrum at $\vartheta = 2\pi$ with a periodicity of $T_{\mathrm{p}}$. This is discussed in detail in Appendixes~\ref{Appendix:omegao = deltar-time} and \ref{Appendix:omegao = deltar - pi/Tp-time}.

\begin{figure*}
\centering
\includegraphics[width=\textwidth]{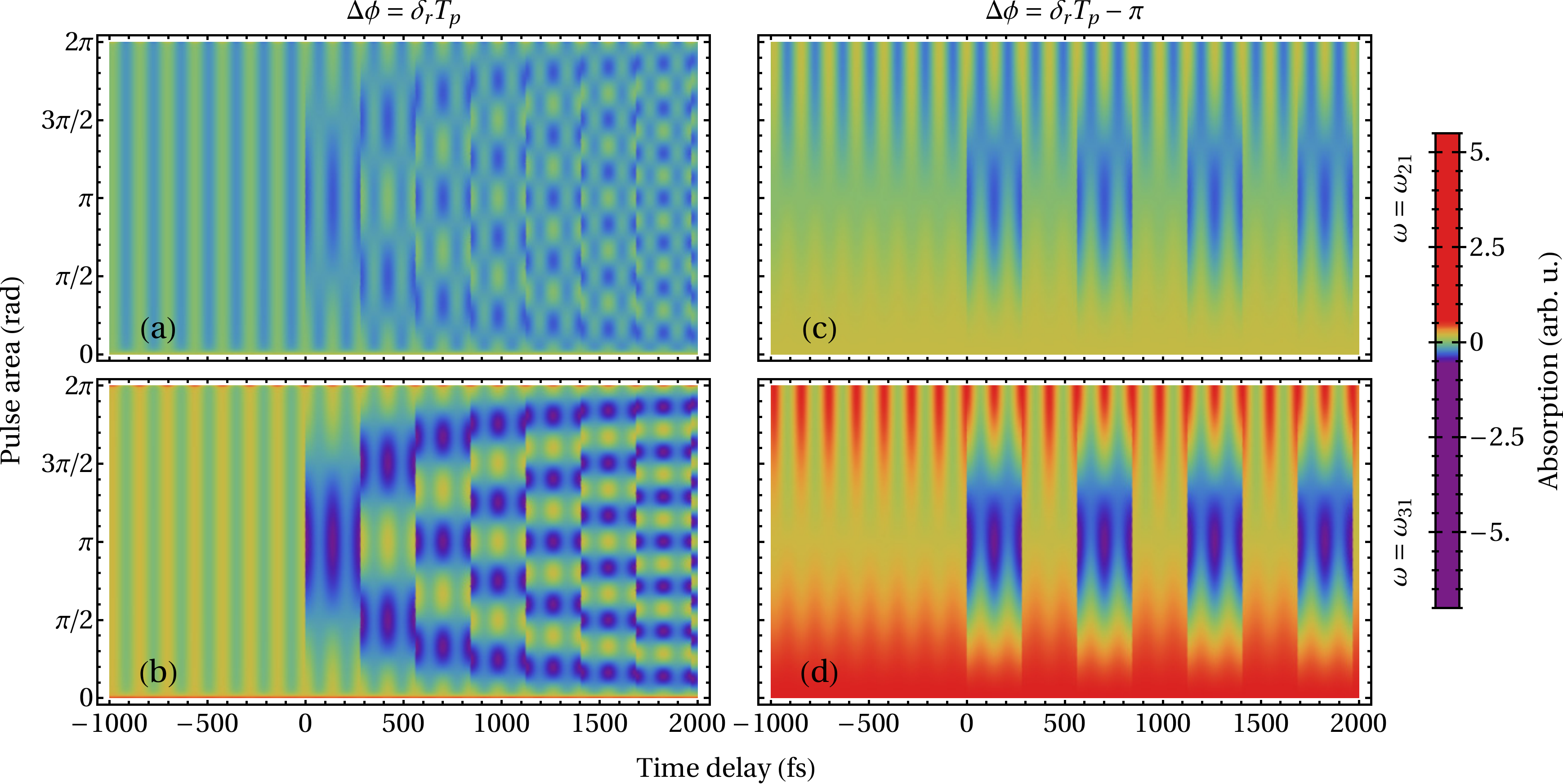}
\caption{Transient-absorption spectra for an infinite number of pump pulses as a function of pulse area $\vartheta$ and time delay $\tau$, for pulse-to-pulse phase shifts [(a),(b)] $\Delta\phi = \delta_{\mathrm{r}}T_{\mathrm{p}}$ and [(c),(d)] $\Delta\phi = \delta_{\mathrm{r}}T_{\mathrm{p}} - \pi$, and frequencies [(a),(c)] $\omega = \omega_{21}$ and [(b),(d)] $\omega = \omega_{31}$.}
\label{fig:TauThetaGridPlot5}
\end{figure*}

These periodic features are further highlighted in Fig.~\ref{fig:TauThetaGridPlot5}, showing time-delay-dependent spectra as a function of the pulse area $\vartheta$ for given values of pulse-to-pulse phase shift $\Delta\phi = \delta_{\mathrm{r}}T_{\mathrm{p}}$ and $\Delta\phi = \delta_{\mathrm{r}}T_{\mathrm{p}} - \pi$, and evaluated at frequencies equal to the transition energies $\omega_{21}$ and $\omega_{31}$. In previous works of transient-absorption spectroscopy in the presence of a single intense pump pulse \cite{PhysRevLett.115.033003}, it was shown that the line shapes encode amplitude and phase information about the action of the pulse on the atomic system. In particular, the spectra feature, both at positive and negative time delays, oscillations in $\tau$ at the beating frequency $\omega_{32}$, whose phases were shown to be directly related to the intensity-dependent atomic-phase change imposed the pump pulse. In the case investigated here for $\delta$~pulses, however, the phases of the matrix elements of the operator $\hat{A}(\vartheta,\alpha)$ in Eq.~(\ref{eq:pulseaction}) are not affected by the intensity of the pulse, i.e., by the value of $\vartheta$---only a change of amplitude is possible, including a change of sign. As a result, the phase of the time-delay-dependent oscillations exhibited by the spectrum at the beating frequency $\omega_{32}$ is independent of $\vartheta$. This clearly appears in Fig.~\ref{fig:TauThetaGridPlot5}. 

At positive time delays, the spectra display a modulation of their intensity as a function of $\vartheta$. This modulation reflects the action of the $\Mtau$ pump pulses preceding the probe pulse, and therefore strongly depends on $\tau$ as well. This is further discussed in Appendix~\ref{Appendix:Details on the time-delay-dependent features of the spectra}. The properties of this modulation can be more precisely investigated for the two values of $\Delta\phi$ used in Fig.~\ref{fig:TauThetaGridPlot5}, as discussed in Appendixes~\ref{Appendix:omegao = deltar-time} and \ref{Appendix:omegao = deltar - pi/Tp-time} and as shown below.

For $\Delta\phi = \delta_{\mathrm{r}}T_{\mathrm{p}}$ as in Figs.~\ref{fig:TauThetaGridPlot5}(a) and \ref{fig:TauThetaGridPlot5}(b), one can show that the dipoles generated by $M_{\tau_1}$ pulses of area $\vartheta_1$ and $M_{\tau_2}$ pulses of area $\vartheta_2$ are equal if there exists an integer $K$ for which
\begin{equation}
M_{\tau_1}\vartheta_1 = M_{\tau_2}\vartheta_2 + 2\pi K.
\label{eq:relationtautheta}
\end{equation}
When this condition is fulfilled and the generated state is the same, then also the associated spectra coincide. This can be recognized by inspecting the position of the minima in Figs.~\ref{fig:TauThetaGridPlot5}(a) and \ref{fig:TauThetaGridPlot5}(b) at positive time delays, which lie on the hyperbolic curves $\vartheta = (2K + 1)\pi/\Mtau$ in agreement with Eq.~(\ref{eq:relationtautheta}). For $0<\vartheta<2\pi$, as exhibited in the figure, there exist exactly $\Mtau$ possible integers $K$ for which the above condition is satisfied. This explains why the number of minima increases with $\tau$ and matches the associated value of $\Mtau = \lfloor \tau/T_{\mathrm{p}}\rfloor + 1$. Furthermore, by applying Eq.~(\ref{eq:relationtautheta}) with $\vartheta_1 = \vartheta_2 = \vartheta$, one obtains that two sequences of identically intense pulses prepare the system in the same state if $\Delta\Mtau = 2\pi K/\vartheta$, where $\Delta\Mtau$ and $K$ are both integers. This explains the periodicity of the spectra as a function of time delay, which we already noticed in Figs.~\ref{fig:OmegaTauGridPlot3}(a)--(d). For a given pulse area $\vartheta$ and at positive time delays, the spectra have namely a periodicity of $X_{\vartheta}T_{\mathrm{p}}$, where $X_{\vartheta}$ is the smallest integer which is also a multiple of $2\pi/\vartheta$. This agrees with the values we have already identified while discussing the spectra in Figs.~\ref{fig:OmegaTauGridPlot3}(a)--(d) for the pulse areas used therein. 

In contrast, when $\Delta\phi = \delta_{\mathrm{r}}T_{\mathrm{p}} - \pi$ as in Figs.~\ref{fig:TauThetaGridPlot5}(c) and \ref{fig:TauThetaGridPlot5}(d), the spectra have a periodicity of $2T_{\mathrm{p}}$, as already identified in Figs.~\ref{fig:OmegaTauGridPlot3}(e)--(h). Also in this case, this reflects the action of the $\Mtau$ preparatory pump pulses preceding the probe pulse, and in particular the fact that, for this value of the pulse-to-pulse phase shift, an even number of pulses acting on the ground state brings the system back to it, independent of the pulse area $\vartheta$. Consequently, any odd number of pulses will prepare the system in the same excited state. As a result, the ensuing spectra have a periodicity given by $2 T_{\mathrm{p}}$ for any value of the pulse area. 

By using a train of pump pulses, the evolution of the transient-absorption line shapes as a function of time delay thus exhibits periodic features, with a periodicity which can be directly related to the properties of the pump pulses used. The time-delay-dependent features of the spectra, as well as the frequency of the LISs, can therefore be used to access the intensity-dependent action of each pump pulse on the atomic system.

% \subsection{Spectra for particular phases $\Delta\phi2$}
% \label{Spectra for particular phases}
% 
% 
% 
% \subsection{Spectra as a function of pulse area and frequency}
% \label{Spectra as a function of pulse area and frequency}

% 
% \subsection{Spectra as a function of time delay and pulse area}
% \label{Spectra as a function of time delay and pulse area}

\section{Conclusion}
\label{Conclusion}

In conclusion, we have investigated the dynamics and the transient-absorption spectrum of a $V$-type three-level system excited by a train of $\delta$-like pulses and probed by a short pulse at different delays. We have shown that the periodic modification of the dipole response induces the appearance of LISs in the absorption spectrum of the probe pulse, in spite of the fact that each $\delta$-like pump pulse is as short as the probe pulse. We have also shown that the LIS frequencies are directly related to the action of each single intense pump pulse. Furthermore, we have shown that the spectrum exhibits periodic features as a function of time delay for $\tau>0$, which are related to the action of the pump pulses preceding the probe pulse. In the presence of a periodically pumped system, these frequency- and time-delay-dependent features provide further variables, in addition to the shape of the absorption lines, which can be experimentally measured in order to access and reconstruct the quantum dynamics of a strong-field-excited system. 

While the dynamics and spectra presented in this paper were calculated assuming a fixed ratio between the repetition frequency $\omega_{\mathrm{r}} = \omega_{32}/2$ and the beating frequency $\omega_{32}$, further studies could investigate the dependence of the transient-absorption spectra on $\omega_{\mathrm{r}}$. Furthermore, by considering pump and probe pulses of finite duration, instead of the $\delta$-like pulses assumed here, one would expect intensity-dependent phase effects analogous to those already reported in Refs.~\cite{PhysRevLett.115.033003, 0953-4075-51-3-035501}: understanding how these atomic phases are encoded in the spectrum of a periodically pumped system would be an interesting extension of the work presented here. 

Towards an experimental realization of the scheme with Rb atoms, an atomic-system description could be considered beyond the three-level model used here. Control schemes in Rb \cite{PhysRevLett.110.223601}, also with shaped optical-frequency combs \cite{PhysRevLett.100.203001}, have considered a closed-loop four-level model, including the coupling of the two excited states $|2\rangle$ and $|3\rangle$ to the more highly excited state $|4\rangle = 5d\,^2D_{3/2}$. However, this coupling is weaker than that to the ground state, and these studies explicitly aimed at shaping the pulses in order to optimize population transfer to this more highly excited state. This is not the case for the TAS experiments considered here, and studies of TAS with Rb atoms for a single pump pulse have already shown that a $V$-type three-level model well describes the frequency- and time-delay-dependent features of the absorption spectra for different pump-pulse intensities \cite{PhysRevLett.115.033003}. Finally, one could further study the influence of propagation effects on the resulting transient-absorption spectra beyond the single-atom response \cite{PhysRevLett.114.143002}, e.g., towards the experimental investigation of media which are not optically thin due to large densities or medium lengths.

\begin{acknowledgments}
The authors acknowledge valuable discussions with Christoph~H.~Keitel and Thomas~Pfeifer.
\end{acknowledgments}

\appendix
\section{Spectral features of a train of pump pulses}
\label{Appendix:Time--frequency description of the train of pump pulses}

In order to study the spectral features of the train of equally spaced pump pulses in Eq.~(\ref{eq:pumpfield}), we first introduce the positive-frequency part of the field
\begin{equation}
\mathcal{E}_{\mathrm{pu}}^{(+)}(t) = \frac{\eu^{\uimm\phi_{0,\mathrm{pu}}}}{2}\sum_{n=0}^{N-1}\mathcal{E}_{0,\mathrm{pu}}(t-nT_{\mathrm{p}})\,\eu^{\uimm\omega_{\mathrm{c}}(t - nT_{\mathrm{p}})}\,\eu^{\uimm n\Delta\phi},
\end{equation}
such that
\begin{equation}
\mathcal{E}_{\mathrm{pu}}(t) = \mathcal{E}_{\mathrm{pu}}^{(+)}(t) + \bigl[\mathcal{E}_{\mathrm{pu}}^{(+)}(t)\bigr]^*
\end{equation}
and
\begin{equation}
\tilde{\mathcal{E}}_{\mathrm{pu}}(\omega) = \tilde{\mathcal{E}}_{\mathrm{pu}}^{(+)}(\omega) + \bigl[\tilde{\mathcal{E}}_{\mathrm{pu}}^{(+)}(-\omega)\bigr]^*.
\end{equation}
By defining the convolution of two functions
\begin{equation}
f(t) * g(t) = \int f(t-t')\,g(t')\,\diff t',
\end{equation}
whose Fourier transform is given by
\begin{equation}
\int_{-\infty}^{\infty} f(t) * g(t)\,\eu^{-\uimm\omega t}\,\diff t = \tilde{f}(\omega)\,\tilde{g}(\omega),
\end{equation}
the positive-frequency part of the pump field can be written as
\begin{equation}
\mathcal{E}_{\mathrm{pu}}^{(+)}(t) = \frac{\eu^{\uimm\phi_{0,\mathrm{pu}}}}{2} \bigl[\mathcal{E}_{0,\mathrm{pu}}(t)\,\eu^{\uimm\omega_{\mathrm{c}} t}\bigr]*\sum_{n=0}^{N-1}\delta(t - nT_{\mathrm{p}})\,\eu^{\uimm n \Delta\phi},
\end{equation}
whose Fourier transform is given by
\begin{equation}
\begin{aligned}
\tilde{\mathcal{E}}_{\mathrm{pu}}^{(+)}(\omega) &= \frac{\eu^{\uimm\phi_{0,\mathrm{pu}}}}{2} \tilde{\mathcal{E}}_{0,\mathrm{pu}}(\omega - \omega_{\mathrm{c}})\,\sum_{n=0}^{N-1}\eu^{-\uimm(\omega - \omega_{\mathrm{o}}) n T_{\mathrm{p}}}\\
&= \frac{\eu^{\uimm\phi_{0,\mathrm{pu}}}}{2} \tilde{\mathcal{E}}_{0,\mathrm{pu}}(\omega - \omega_{\mathrm{c}})\,\frac{1-\eu^{-\uimm(\omega - \omega_{\mathrm{o}})NT_{\mathrm{p}}}}{1 - \eu^{-\uimm(\omega - \omega_{\mathrm{o}})T_{\mathrm{p}}}}.
\end{aligned}
\label{eq:Fouriertransformgeneralforcomb}
\end{equation}

In order to render the peak structure of $\tilde{\mathcal{E}}_{\mathrm{pu}}^{(+)}(\omega)$ more apparent, one can write $\mathcal{E}_{\mathrm{pu}}^{(+)}(t)$ as
\begin{equation}
\begin{aligned}
\mathcal{E}_{\mathrm{pu}}^{(+)}(t) &= \frac{\eu^{\uimm\phi_{0,\mathrm{pu}}}}{2} \, \bigl[\mathcal{E}_{0,\mathrm{pu}}(t)\,\eu^{\uimm\omega_{\mathrm{c}} t}\bigr]* \Biggl[\sum_{n=-\infty}^{\infty}\delta(t - nT_{\mathrm{p}})\,\eu^{\uimm n \Delta\phi}\\
&\ \ \ \ \times\, \{\theta(t-aT_{\mathrm{p}}) - \theta[t - (N-a)T_{\mathrm{p}}]\} \Biggr],
\end{aligned}
\end{equation}
with the Heaviside step function $\theta(x)$ and with $0<a<1$. Notice that the field is independent of the explicit value of $a$. Thereby, the field can be written in terms of an infinite train of $\delta$~pulses, whose Fourier transform is given by an infinite comb of $\delta$~peaks
\begin{equation}
\int_{-\infty}^{\infty} \sum_{n=-\infty}^{\infty}\delta(t - nT_{\mathrm{p}})\,\eu^{\uimm n \Delta\phi}\,\eu^{-\uimm\omega t}\,\diff t = \omega_{\mathrm{r}}\sum_{m = -\infty}^{\infty} \delta(\omega - \omega_m),
\end{equation}
where we have used the definitions in Eqs.~(\ref{eq:omegam}) and (\ref{eq:omegarandomegao}). By recalling that
\begin{equation}
\int_{-\infty}^{\infty} f(t) \, g(t)\,\eu^{-\uimm\omega t}\,\diff t = \frac{1}{2\pi}\,\tilde{f}(\omega) * \tilde{g}(\omega),
\end{equation}
the Fourier transform of $\mathcal{E}_{\mathrm{pu}}^{(+)}(t)$ is given by
\begin{equation}
\begin{aligned}
\tilde{\mathcal{E}}_{\mathrm{pu}}^{(+)}(\omega) &=\frac{\eu^{\uimm\phi_{0,\mathrm{pu}}}}{2}\,\tilde{\mathcal{E}}_{0,\mathrm{pu}}(\omega - \omega_{\mathrm{c}})\,\times\\
&\ \ \Biggl[ \frac{\omega_{\mathrm{r}}}{2\pi}\biggl(\eu^{\uimm\omega a T_{\mathrm{p}}}\,\frac{1-\eu^{-\uimm\omega N T_{\mathrm{p}}}}{\uimm\,\omega}\biggr)* \sum_{m = -\infty}^{\infty} \delta(\omega - \omega_m)\Biggr]\\
&= \frac{\eu^{\uimm\phi_{0,\mathrm{pu}}}}{2}\,\tilde{\mathcal{E}}_{0,\mathrm{pu}}(\omega - \omega_{\mathrm{c}})\,N\sum_{m = -\infty}^{\infty} \eu^{\uimm(\omega - \omega_m) aT_{\mathrm{p}}}\,\times\\
&\ \ \eu^{-\uimm(\omega - \omega_m) NT_{\mathrm{p}}/2}\,  \,\sinc\biggl[\frac{NT_{\mathrm{p}}}{2}(\omega - \omega_m)\biggr].
\end{aligned}
\label{eq:Fouriercomb}
\end{equation}
One can therefore recognize that the Fourier transform of a train of $N$ pulses is given by peaks centered at the frequency $\omega_m = m \omega_{\mathrm{r}} + \omega_{\mathrm{o}}$. The strength of the peaks is modulated by the Fourier transform $\tilde{\mathcal{E}}_{0,\mathrm{pu}}(\omega - \omega_{\mathrm{c}})$ of a single pulse, while the width of each peak is associated with the width of $\sinc[NT_{\mathrm{p}}(\omega - \omega_m)/2]$, which is much smaller than the separation frequency $\omega_{\mathrm{r}}$ if $N\gg 1$.

The Fourier transform $\tilde{\mathcal{E}}_{\mathrm{pu}}^{(+)}(\omega)$ is independent of $a$. The second line in Eq.~(\ref{eq:Fouriercomb}) can namely be written as
\begin{equation}
\begin{aligned}
&\frac{\omega_{\mathrm{r}}}{2\pi}\biggl(\eu^{\uimm\omega a T_{\mathrm{p}}}\,\frac{1-\eu^{-\uimm\omega N T_{\mathrm{p}}}}{\uimm\,\omega}\biggr)* \sum_{m = -\infty}^{\infty} \delta(\omega - \omega_m)\\
=\,& \frac{\omega_{\mathrm{r}}}{2\pi}\sum_{m = -\infty}^{\infty}\eu^{\uimm(\omega-\omega_m) a T_{\mathrm{p}}}\, \frac{1-\eu^{-\uimm(\omega-\omega_m) N T_{\mathrm{p}}}}{\uimm\,(\omega-\omega_m)}\\
=\,&\eu^{\uimm(\omega-\omega_{\mathrm{o}}) a T_{\mathrm{p}}}\,(1-\eu^{-\uimm(\omega-\omega_{\mathrm{o}}) N T_{\mathrm{p}}})\,\frac{\omega_{\mathrm{r}}}{2\pi}\sum_{m = -\infty}^{\infty}\frac{\eu^{-\uimm 2\pi m a}}{\uimm\,(\omega-\omega_m)},
\end{aligned}
\end{equation}
due to the fact that
\begin{equation}
1-\eu^{-\uimm(\omega-\omega_m) N T_{\mathrm{p}}} = 1-\eu^{-\uimm(\omega-\omega_{\mathrm{o}}) N T_{\mathrm{p}}}
\end{equation}
is independent of $m$. By recognizing that \cite{Gradshteyn}
\begin{equation}
\begin{aligned}
&\frac{\omega_{\mathrm{r}}}{2\pi}\sum_{m = -\infty}^{\infty}\frac{\eu^{-\uimm 2\pi m a}}{\uimm\,(\omega-\omega_m)} = 
\frac{1}{2\pi\uimm}\sum_{m = -\infty}^{\infty}\frac{\eu^{-\uimm 2\pi m a}}{\frac{\omega - \omega_{\mathrm{o}}}{\omega_{\mathrm{r}}}-m} \\
=\,&\frac{1}{2\pi\uimm}\Biggl(\frac{1}{\frac{\omega - \omega_{\mathrm{o}}}{\omega_{\mathrm{r}}}} + 2\uimm   \sum_{m=1}^{\infty}\frac{m\sin(2\pi a m)}{m^2 - \left(\frac{\omega - \omega_{\mathrm{o}}}{\omega_{\mathrm{r}}}\right)^2}\\
& - 2\,\frac{\omega - \omega_{\mathrm{o}}}{\omega_{\mathrm{r}}}\,\sum_{m=1}^{\infty}\frac{\cos(2\pi a m)}{m^2 - \left(\frac{\omega - \omega_{\mathrm{o}}}{\omega_{\mathrm{r}}}\right)^2}\Biggr) = \frac{\eu^{-\uimm(\omega - \omega_{\mathrm{o}})aT_{\mathrm{p}}}}{1-\eu^{-\uimm(\omega - \omega_{\mathrm{o}})T_{\mathrm{p}}}}
\end{aligned}
\end{equation}
for $0<a<1$, one can conclude that
\begin{equation}
\begin{aligned}
&\frac{\omega_{\mathrm{r}}}{2\pi}\biggl(\eu^{\uimm\omega a T_{\mathrm{p}}}\,\frac{1-\eu^{-\uimm\omega N T_{\mathrm{p}}}}{\uimm\,\omega}\biggr)* \sum_{m = -\infty}^{\infty} \delta(\omega - \omega_m) \\
=\,& \frac{1-\eu^{-\uimm(\omega-\omega_{\mathrm{o}}) N T_{\mathrm{p}}}}{1-\eu^{-\uimm(\omega - \omega_{\mathrm{o}})T_{\mathrm{p}}}},
\end{aligned}
\label{eq:equality-peak-structure}
\end{equation}
which is independent of $a$ and coincides with the result in Eq.~(\ref{eq:Fouriertransformgeneralforcomb}).

\section{Evolution of the system between two generic pump pulses $a$ and $b$}
\label{Appendix:Position-dependent action of the pump pulse}

The interaction with two or more consecutive pump pulses explicitly depends on their position in the train of pulses as a result of the phase-dependent term $[\hat{\mathcal{F}}(\Delta\phi)]^n$. We will show this here explicitly, by considering the evolution of $\vec{R}(t)$ between $t_a^-$ and $t_b^+$, where $a$ and $b$ are two integers, $0\leq a \leq b\leq N-1$, associated with the $a$th and $b$th pump pulses, respectively, and where $t_n^{-}$ ($t_n^+$) denotes the time $t_n$ approached from the left (right), preceding (following) the interaction with the $n$th pump pulse. We assume that $\tau\not\in[t_a,\,t_b]$, such that the evolution of the system results from the interaction with $(b-a+1)$ pump pulses, separated by $(b-a)$ intervals of free evolution. The state reached by the system is then given by
\begin{equation}
\begin{aligned}
\vec{R}(t_b^+) &= \hat{\mathcal{U}}_{\mathrm{pu},b}\,\hat{\mathcal{V}}_{\mathrm{p}}\,\hat{\mathcal{U}}_{\mathrm{pu},b-1} \cdots\hat{\mathcal{V}}_{\mathrm{p}}\,\hat{\mathcal{U}}_{\mathrm{pu},a} \,\vec{R}(t_a^-)\\
&= \hat{\mathcal{F}}_{0,\mathrm{pu}}\daga\,(\hat{\mathcal{F}}_{\Delta}\daga)^b\,\hat{\mathcal{A}}_{\mathrm{pu}}\\
&\times\,(\hat{\mathcal{F}}_{\Delta})^b\,\hat{\mathcal{F}}_{0,\mathrm{pu}}\,\hat{\mathcal{V}}_{\mathrm{p}}\,\hat{\mathcal{F}}_{0,\mathrm{pu}}\daga\,(\hat{\mathcal{F}}_{\Delta}\daga)^{b-1}\,\hat{\mathcal{A}}_{\mathrm{pu}}\\
&\times\,(\hat{\mathcal{F}}_{\Delta})^{b-1}\,\hat{\mathcal{F}}_{0,\mathrm{pu}} \cdots\hat{\mathcal{V}}_{\mathrm{p}}\,\hat{\mathcal{F}}_{0,\mathrm{pu}}\daga\,(\hat{\mathcal{F}}_{\Delta}\daga)^a\,\hat{\mathcal{A}}_{\mathrm{pu}}\\
&\times\,(\hat{\mathcal{F}}_{\Delta})^a\,\hat{\mathcal{F}}_{0,\mathrm{pu}}\,\vec{R}(t_a^-)\\
&= \hat{\mathcal{F}}_{0,\mathrm{pu}}\daga\,(\hat{\mathcal{F}}_{\Delta}\daga)^b\,\hat{\mathcal{A}}_{\mathrm{pu}}\,(\hat{\mathcal{F}}_{\Delta}\,\hat{\mathcal{V}}_{\mathrm{p}}\,\hat{\mathcal{A}}_{\mathrm{pu}})^{b-a}\\
&\times\,(\hat{\mathcal{F}}_{\Delta})^a\,\hat{\mathcal{F}}_{0,\mathrm{pu}}\,\vec{R}(t_a^-),
\end{aligned}
\label{eq:finitesequence}
\end{equation}
where we have used the fact that the diagonal matrices $\hat{\mathcal{V}}(t)$ and $\hat{\mathcal{F}}(\phi)$ commute.

\section{Evolution of the dipole response $\rho_{1k}(t)$}
\label{Appendix:Evolution of the dipole response}

The off-diagonal matrix elements $\rho_{1k} = R_{k}(t) = \vec{v}_k \vec{R}(t)$ used for the calculation of the absorption spectrum are displayed below for the probe--pump [Eq.~(\ref{eq:evolutionprpu-k})], pump--probe [Eq.~(\ref{eq:evolutionpupr-k})], and pump--probe--pump setup [Eq.~(\ref{eq:evolutionpuprpu-k})].
\begin{widetext}
\begin{equation}
% \begin{aligned}
R_k(t) =
\left\{
\begin{aligned}
&0, & t<\tau, \\
&\eu^{\uimm\omega_{k1}(t-\tau)}\,\vec{v}_k\,\hat{\mathcal{U}}_{\mathrm{pr}}\,\vec{R}_0, & \tau<t<0,\\ 
&\eu^{\uimm\omega_{k1}(t - lT_{\mathrm{p}})}\,\eu^{\uimm\phi_{0,\mathrm{pu}}}\,\vec{v}_k\,\hat{\mathcal{A}}_{\mathrm{pu}}\,(\eu^{\uimm \Delta\phi}\,\hat{\mathcal{F}}_{\Delta}\,\hat{\mathcal{V}}_{\mathrm{p}}\,\hat{\mathcal{A}}_{\mathrm{pu}})^l\,\hat{\mathcal{F}}_{0,\mathrm{pu}}\,\hat{\mathcal{V}}(-\tau)\,\hat{\mathcal{U}}_{\mathrm{pr}}\,\vec{R}_0, & lT_{\mathrm{p}}<t<(l+1)T_{\mathrm{p}},\\
&\eu^{\uimm\omega_{k1}[t - (N-1)T_{\mathrm{p}}]}\,\eu^{\uimm\phi_{0,\mathrm{pu}}}\,\vec{v}_k\,\hat{\mathcal{A}}_{\mathrm{pu}}\,(\eu^{\uimm \Delta\phi}\,\hat{\mathcal{F}}_{\Delta}\,\hat{\mathcal{V}}_{\mathrm{p}}\,\hat{\mathcal{A}}_{\mathrm{pu}})^{N-1}\,\hat{\mathcal{F}}_{0,\mathrm{pu}}\,\hat{\mathcal{V}}(-\tau)\,\hat{\mathcal{U}}_{\mathrm{pr}}\,\vec{R}_0,& t>(N-1)T_{\mathrm{p}}.
\end{aligned}
\right.
% \end{aligned}
\label{eq:evolutionprpu-k}
\end{equation}

\begin{equation}
% \begin{aligned}
R_k(t) =
\left\{
\begin{aligned}
&0, & t<0, \\
&\eu^{\uimm\omega_{k1}(t - lT_{\mathrm{p}})}\,\eu^{\uimm\phi_{0,\mathrm{pu}}}\,\,\eu^{\uimm l\Delta\phi}\,\vec{v}_k\,\hat{\mathcal{A}}_{\mathrm{pu}}\,(\hat{\mathcal{F}}_{\Delta}\,\hat{\mathcal{V}}_{\mathrm{p}}\,\hat{\mathcal{A}}_{\mathrm{pu}})^l\,\vec{R}_0, & lT_{\mathrm{p}}<t<(l+1)T_{\mathrm{p}},\\
&\eu^{\uimm\omega_{k1}[t - (N-1)T_{\mathrm{p}}]}\,\eu^{\uimm\phi_{0,\mathrm{pu}}}\,\,\eu^{\uimm (N-1)\Delta\phi}\,\vec{v}_k\,\hat{\mathcal{A}}_{\mathrm{pu}}\,(\hat{\mathcal{F}}_{\Delta}\,\hat{\mathcal{V}}_{\mathrm{p}}\,\hat{\mathcal{A}}_{\mathrm{pu}})^{N-1}\,\vec{R}_0, & (N-1)T_{\mathrm{p}}<t<\tau,\\
&\eu^{\uimm\omega_{k1}(t - \tau)}\,\vec{v}_k\,\hat{\mathcal{U}}_{\mathrm{pr}}\,\hat{\mathcal{V}}(\tau - (N-1)T_{\mathrm{p}})\,\hat{\mathcal{F}}_{0,\mathrm{pu}}\daga\,(\hat{\mathcal{F}}_{\Delta}\daga)^{N-1}\,\hat{\mathcal{A}}_{\mathrm{pu}}\,(\hat{\mathcal{F}}_{\Delta}\,\hat{\mathcal{V}}_{\mathrm{p}}\,\hat{\mathcal{A}}_{\mathrm{pu}})^{N-1}\,\vec{R}_0, & t>\tau,\\ 
\end{aligned}
\right.
% \end{aligned}
\label{eq:evolutionpupr-k}
\end{equation}

\begin{equation}
% \begin{aligned}
R_k(t) =
\left\{
\begin{aligned}
&0, & t<0, \\
&\eu^{\uimm\omega_{k1}(t - pT_{\mathrm{p}})}\,\eu^{\uimm\phi_{0,\mathrm{pu}}}\,\vec{v}_k\,\hat{\mathcal{A}}_{\mathrm{pu}}\,(\eu^{\uimm \Delta\phi}\,\hat{\mathcal{F}}_{\Delta}\,\hat{\mathcal{V}}_{\mathrm{p}}\,\hat{\mathcal{A}}_{\mathrm{pu}})^p\,\vec{R}_0, & pT_{\mathrm{p}}<t<(p+1)T_{\mathrm{p}},\\
&\eu^{\uimm\omega_{k1}[t - (\Mtau-1)T_{\mathrm{p}}]}\,\eu^{\uimm\phi_{0,\mathrm{pu}}}\,\vec{v}_k\,\hat{\mathcal{A}}_{\mathrm{pu}}\,(\eu^{\uimm \Delta\phi}\,\hat{\mathcal{F}}_{\Delta}\,\hat{\mathcal{V}}_{\mathrm{p}}\,\hat{\mathcal{A}}_{\mathrm{pu}})^{\Mtau-1}\,\vec{R}_0, & (\Mtau-1)T_{\mathrm{p}}<t<\tau,\\
&\eu^{\uimm\omega_{k1}(t - \tau)}\,\vec{v}_k\,\hat{\mathcal{U}}_{\mathrm{pr}}\,\hat{\mathcal{V}}(\tau - (\Mtau-1)T_{\mathrm{p}})\,\hat{\mathcal{F}}_{0,\mathrm{pu}}\daga\,(\hat{\mathcal{F}}_{\Delta}\daga)^{\Mtau-1}\,\hat{\mathcal{A}}_{\mathrm{pu}}\,(\hat{\mathcal{F}}_{\Delta}\,\hat{\mathcal{V}}_{\mathrm{p}}\,\hat{\mathcal{A}}_{\mathrm{pu}})^{\Mtau-1}\,\vec{R}_0, & \tau<t<\Mtau T_{\mathrm{p}},\\ 
&\left.
\hspace{-1.5 mm}\begin{aligned}
&\eu^{\uimm\omega_{k1}(t - qT_{\mathrm{p}})}\,\eu^{\uimm\phi_{0,\mathrm{pu}}}\,\vec{v}_k\,\hat{\mathcal{A}}_{\mathrm{pu}}\,(\eu^{\uimm \Delta\phi}\,\hat{\mathcal{F}}_{\Delta}\,\hat{\mathcal{V}}_{\mathrm{p}}\,\hat{\mathcal{A}}_{\mathrm{pu}})^{q-\Mtau}\,\eu^{\uimm \Mtau\Delta\phi}\,(\hat{\mathcal{F}}_{\Delta})^{\Mtau}\,\hat{\mathcal{F}}_{0,\mathrm{pu}}\\
&\times \hat{\mathcal{V}}(\Mtau T_{\mathrm{p}}-\tau)\,\hat{\mathcal{U}}_{\mathrm{pr}}\,\hat{\mathcal{V}}(\tau - (\Mtau-1)T_{\mathrm{p}})\,\hat{\mathcal{F}}_{0,\mathrm{pu}}\daga\,(\hat{\mathcal{F}}_{\Delta}\daga)^{\Mtau-1}\,\hat{\mathcal{A}}_{\mathrm{pu}}\,(\hat{\mathcal{F}}_{\Delta}\,\hat{\mathcal{V}}_{\mathrm{p}}\,\hat{\mathcal{A}}_{\mathrm{pu}})^{\Mtau-1}\,\vec{R}_0, 
\end{aligned}
\right\}
& qT_{\mathrm{p}}<t<(q+1)T_{\mathrm{p}},\\
&\left.
\hspace{-1.5 mm}\begin{aligned}
&\eu^{\uimm\omega_{k1}[t - (N-1)T_{\mathrm{p}}]}\,\eu^{\uimm\phi_{0,\mathrm{pu}}}\,\vec{v}_k\,\hat{\mathcal{A}}_{\mathrm{pu}}\,(\eu^{\uimm \Delta\phi}\,\hat{\mathcal{F}}_{\Delta}\,\hat{\mathcal{V}}_{\mathrm{p}}\,\hat{\mathcal{A}}_{\mathrm{pu}})^{N-\Mtau-1}\,\eu^{\uimm \Mtau\Delta\phi}\,(\hat{\mathcal{F}}_{\Delta})^{\Mtau}\,\hat{\mathcal{F}}_{0,\mathrm{pu}}\\
&\times \hat{\mathcal{V}}(\Mtau T_{\mathrm{p}}-\tau)\,\hat{\mathcal{U}}_{\mathrm{pr}}\,\hat{\mathcal{V}}(\tau - (\Mtau-1)T_{\mathrm{p}})\,\hat{\mathcal{F}}_{0,\mathrm{pu}}\daga\,(\hat{\mathcal{F}}_{\Delta}\daga)^{\Mtau-1}\,\hat{\mathcal{A}}_{\mathrm{pu}}\,(\hat{\mathcal{F}}_{\Delta}\,\hat{\mathcal{V}}_{\mathrm{p}}\,\hat{\mathcal{A}}_{\mathrm{pu}})^{\Mtau-1}\,\vec{R}_0, 
\end{aligned}
\right\}
& t>(N-1) T_{\mathrm{p}},
\end{aligned}
\right.
% \end{aligned}
\label{eq:evolutionpuprpu-k}
\end{equation}
% \end{widetext}

\section{The operator $\hat{\mathcal{Z}}(\tau)$}
\label{Appendix:The operator Z}

By averaging over the fast time-delay-dependent oscillations in Eq.~(\ref{eq:Z}), several matrix elements of $\hat{\mathcal{Z}}(\tau')$ vanish. The $9\times 9$ matrix $\hat{\mathcal{Z}}(\tau')$ can then be written in terms of a sum of Kronecker products 
\begin{equation}
\begin{aligned}
\hat{\mathcal{Z}} =&\,
\begin{pmatrix}
0 &0 &0 \\
0 &1 &0 \\
0 &0 &1
\end{pmatrix}
\hat{Z}
\begin{pmatrix}
0 &0 &0 \\
0 &1 &0 \\
0 &0 &1
\end{pmatrix}\,\otimes\,
\begin{pmatrix}
0 &0 &0 \\
0 &1 &0 \\
0 &0 &1
\end{pmatrix}
\hat{Z}^*
\begin{pmatrix}
1 &0 &0 \\
0 &0 &0 \\
0 &0 &0
\end{pmatrix} + 
\begin{pmatrix}
1 &0 &0 \\
0 &0 &0 \\
0 &0 &0
\end{pmatrix}
\hat{Z}
\begin{pmatrix}
1 &0 &0 \\
0 &0 &0 \\
0 &0 &0
\end{pmatrix}\,\otimes\,
\begin{pmatrix}
0 &0 &0 \\
0 &1 &0 \\
0 &0 &1
\end{pmatrix}
\hat{Z}^*
\begin{pmatrix}
1 &0 &0 \\
0 &0 &0 \\
0 &0 &0
\end{pmatrix} \\
+& 
\begin{pmatrix}
1 &0 &0 \\
0 &0 &0 \\
0 &0 &0
\end{pmatrix}
\hat{Z}
\begin{pmatrix}
0 &0 &0 \\
0 &1 &0 \\
0 &0 &1
\end{pmatrix}\,\otimes\,
\begin{pmatrix}
1 &0 &0 \\
0 &0 &0 \\
0 &0 &0
\end{pmatrix}
\hat{Z}^*
\begin{pmatrix}
1 &0 &0 \\
0 &0 &0 \\
0 &0 &0
\end{pmatrix} +
\begin{pmatrix}
1 &0 &0 \\
0 &0 &0 \\
0 &0 &0
\end{pmatrix}
\hat{Z}
\begin{pmatrix}
0 &0 &0 \\
0 &1 &0 \\
0 &0 &1
\end{pmatrix}\,\otimes\,
\begin{pmatrix}
0 &0 &0 \\
0 &1 &0 \\
0 &0 &1
\end{pmatrix}
\hat{Z}^*
\begin{pmatrix}
0 &0 &0 \\
0 &1 &0 \\
0 &0 &1
\end{pmatrix} ,
\end{aligned}
\label{eq:Zdetail}
\end{equation}
% 
% \begin{equation}
% \hat{\mathcal{Z}}=
% \begin{pmatrix}
% 0 &0 &0 &\mathcal{Z}_{14} &0 &0 &\mathcal{Z}_{17} &0 &0\\
% \mathcal{Z}_{21} &0 &0 &0 &\mathcal{Z}_{25} &\mathcal{Z}_{26} &0 &\mathcal{Z}_{28} &\mathcal{Z}_{29}\\
% \mathcal{Z}_{31} &0 &0 &0 &\mathcal{Z}_{35} &\mathcal{Z}_{36} &0 &\mathcal{Z}_{38} &\mathcal{Z}_{39}\\
% 0 &0 &0 &0 &0 &0 &0 &0 &0\\
% 0 &0 &0 &\mathcal{Z}_{54} &0 &0 &\mathcal{Z}_{57} &0 &0\\
% 0 &0 &0 &\mathcal{Z}_{64} &0 &0 &\mathcal{Z}_{67} &0 &0\\
% 0 &0 &0 &0 &0 &0 &0 &0 &0\\
% 0 &0 &0 &\mathcal{Z}_{84} &0 &0 &\mathcal{Z}_{87} &0 &0\\
% 0 &0 &0 &\mathcal{Z}_{94} &0 &0 &\mathcal{Z}_{97} &0 &0
% \end{pmatrix},
% \end{equation}
\end{widetext}
involving the operator
\begin{equation}
\hat{Z}(\tau') = \hat{V}(T_{\mathrm{p}} - \tau')\,\hat{U}_{\mathrm{pr}}(\vartheta_{\mathrm{pr}},\alpha,\phi_{0,\mathrm{pr}})\,\hat{V}(\tau').
\end{equation}
Written explicitly, the operator reads
\begin{equation}
\hat{\mathcal{Z}}=
\begin{pmatrix}
0 &0 &0 &\mathcal{Z}_{14} &0 &0 &\mathcal{Z}_{17} &0 &0\\
\mathcal{Z}_{21} &0 &0 &0 &\mathcal{Z}_{25} &\mathcal{Z}_{26} &0 &\mathcal{Z}_{28} &\mathcal{Z}_{29}\\
\mathcal{Z}_{31} &0 &0 &0 &\mathcal{Z}_{35} &\mathcal{Z}_{36} &0 &\mathcal{Z}_{38} &\mathcal{Z}_{39}\\
0 &0 &0 &0 &0 &0 &0 &0 &0\\
0 &0 &0 &\mathcal{Z}_{54} &0 &0 &\mathcal{Z}_{57} &0 &0\\
0 &0 &0 &\mathcal{Z}_{64} &0 &0 &\mathcal{Z}_{67} &0 &0\\
0 &0 &0 &0 &0 &0 &0 &0 &0\\
0 &0 &0 &\mathcal{Z}_{84} &0 &0 &\mathcal{Z}_{87} &0 &0\\
0 &0 &0 &\mathcal{Z}_{94} &0 &0 &\mathcal{Z}_{97} &0 &0
\end{pmatrix},
\end{equation}
with nonvanishing elements equal to $\mathcal{Z}_{ij}(\tau') = \mathcal{V}_{ii}(T_{\mathrm{p}}-\tau')\,\mathcal{U}_{\mathrm{pr},ij}\,\mathcal{V}_{jj}(\tau')$. Notice that some of the nonvanishing matrix elements of $\hat{\mathcal{Z}}(\tau')$ may be negligibly small compared to others for small intensities of the probe pulse, since they are of different orders in $\vartheta_{\mathrm{pr}}$, and may thus vanish if we use the probe-pulse interaction operator given in Eq.~(\ref{eq:probepulseapprox}). 

\section{Central frequencies of the light-induced states appearing in the spectrum}
\label{Appendix:Central frequency of the spectral lines}

In order to quantify the central frequencies of the LISs appearing in the spectrum, we show that they are determined by the poles of the operator $[\vec{v}_k\,\hat{\mathcal{D}}_N(\bar{\omega})]/[\uimm(\bar{\omega} - \omega_{k1})]$ in Eq.~(\ref{eq:spectrumprpushort}). The same can be used to explain the spectra at positive time delays, determined by the term $[\vec{v}_k\,\hat{\mathcal{D}}_{N-\Mtau}(\bar{\omega})]/[\uimm(\bar{\omega} - \omega_{k1})]$ in Eq.~(\ref{eq:spectrumpuprpushort}). It is important to notice that the poles are real, so that a divergence in the spectrum would appear if $\bar{\omega}$ were real. Since we evaluate the spectrum at the complex frequency $\bar{\omega} = \omega-\uimm\gamma/2$, no divergences appear in the spectrum, as these reduce to peaks with a width of $\gamma/2$ and centered on the corresponding real poles.

For $N=1$, $[\vec{v}_k\,\hat{\mathcal{D}}_N(\bar{\omega})]/[\uimm(\bar{\omega} - \omega_{k1})]$ reduces to $[\vec{v}_k\,\hat{\mathcal{A}}_{\mathrm{pu}}]/[\uimm(\bar{\omega} - \omega_{k1})]$, whose only poles are $\bar{\omega} = \omega_{k1}$. However, when $N\rightarrow \infty$, this operator reads
\begin{equation}
\begin{aligned}
\frac{\vec{v}_k\,\hat{\mathcal{D}}_{\infty}(\bar{\omega})}{\uimm(\bar{\omega} - \omega_{k1})} &=  -\uimm\,\frac{T_{\mathrm{p}}}{2}\,\eu^{-\uimm(\bar{\omega} - \omega_{k1})\frac{T_{\mathrm{p}}}{2}}\,\sinc{\left[(\bar{\omega} - \omega_{k1})\frac{T_{\mathrm{p}}}{2}\right]}\\
&\ \times\,\vec{v}_k\,\hat{\mathcal{A}}_{\mathrm{pu}}\,\bigl(\hat{\mathcal{I}} -\eu^{-\uimm (\bar{\omega}T_{\mathrm{p}}- \Delta\phi)}\,\hat{\mathcal{F}}_{\Delta}\,\hat{\mathcal{V}}_{\mathrm{p}}\,\hat{\mathcal{A}}_{\mathrm{pu}}\bigr)^{-1}.
\end{aligned}
\label{eq:responsibledivergence}
\end{equation}
Firstly, due to the presence of $\sinc{[(\bar{\omega} - \omega_{k1})T_{\mathrm{p}}/2]}$ in the first line, the pole at $\omega_{k1}$ present for a finite number of pump pulses is here removed, unless it appears explicitly as a pole of the inverse operator in the second line. We also notice that this operator has zeros at
\begin{equation}
\bar{\omega} =\bar{\omega}^{\mathrm{zero}}_r = \omega_{k1} + r\omega_{\mathrm{r}},\ \ r\neq 0
\label{eq:zeros}
\end{equation}
for any $r\in\mathbb{Z}$ other than 0. In order to identify the poles of Eq.~(\ref{eq:responsibledivergence}), we need to focus on the inverse operator in the second line. In particular, we notice that 
$$\hat{\mathcal{F}}_{\Delta}\,\hat{\mathcal{V}}_{\mathrm{p}}\,\hat{\mathcal{A}}_{\mathrm{pu}} = [\hat{F}_{\Delta}\,\hat{V}_{\mathrm{p}}\,\hat{A}_{\mathrm{pu}}]\otimes[\hat{F}_{\Delta}\,\hat{V}_{\mathrm{p}}\,\hat{A}_{\mathrm{pu}}]^*,$$ 
where we have introduced $\hat{F}_{\Delta}=\hat{F}(\Delta\phi)$, $\hat{V}_{\mathrm{p}} = \hat{V}(T_{\mathrm{p}})$, and $\hat{A}_{\mathrm{pu}}=\hat{A}(\vartheta,\alpha)$. The product
\begin{equation} 
\hat{F}_{\Delta}\,\hat{V}_{\mathrm{p}} = 
\begin{pmatrix}
1 &0 &0\\
0 &\eu^{-\uimm(\delta_{\mathrm{r}} - \omega_{\mathrm{o}})T_{\mathrm{p}}} &0\\
0 &0 &\eu^{-\uimm(\delta_{\mathrm{r}} - \omega_{\mathrm{o}})T_{\mathrm{p}}}\\
\end{pmatrix}
\end{equation}
is a diagonal matrix describing the change in the atomic phases of the two excited states during one period. Since $\hat{F}_{\Delta}\,\hat{V}_{\mathrm{p}}\,\hat{A}_{\mathrm{pu}}$ is a unitary operator, its eigenvalues $\eu^{\uimm\lambda_j}$, $j\in\{1,\,2,\,3\}$, lie on the unit circle. After introducing the phases
\begin{equation}
\beta = (\delta_{\mathrm{r}} - \omega_{\mathrm{o}})T_{\mathrm{p}}
\end{equation}
and
\begin{equation}
\varepsilon = \arccos\left[\cos\left(\frac{\vartheta}{2}\right)\,\cos\left(\frac{\beta}{2}\right)\right] ,
\end{equation}
the eigenvalues $\eu^{\uimm\lambda_j}$ and associated eigenvectors $\vec{P}_j$ can be calculated exactly as
\begin{equation}
\begin{aligned}
\eu^{\uimm\lambda_1} &= \eu^{-\uimm\beta},\\
\eu^{\uimm\lambda_2} &= \eu^{-\uimm\beta/2 + \uimm \varepsilon},\\
\eu^{\uimm\lambda_3} &= \eu^{-\uimm\beta/2 - \uimm \varepsilon},
\end{aligned}
\end{equation}
and
\begin{equation}
\begin{aligned}
\vec{P}_1 &=
\begin{pmatrix}
0\\
-\cos(\alpha)\\
\sin(\alpha)
\end{pmatrix},\\
\vec{P}_2 &=
\begin{pmatrix}
\frac{\sin\left(\vartheta/2\right)}{\sqrt{\sin^2\left(\vartheta/2\right) + \left|\cos\left(\vartheta/2\right) - \eu^{\uimm\beta/2-\uimm\varepsilon}\right|^2}}\\
\uimm\frac{\cos\left(\vartheta/2\right) - \eu^{-\uimm\beta/2+\uimm\varepsilon}}{\sqrt{\sin^2\left(\vartheta/2\right) + \left|\cos\left(\vartheta/2\right) - \eu^{\uimm\beta/2-\uimm\varepsilon}\right|^2}}\,\sin(\alpha)\\
\uimm\frac{\cos\left(\vartheta/2\right) - \eu^{-\uimm\beta/2+\uimm\varepsilon}}{\sqrt{\sin^2\left(\vartheta/2\right) + \left|\cos\left(\vartheta/2\right) - \eu^{\uimm\beta/2-\uimm\varepsilon}\right|^2}}\,\cos(\alpha)
\end{pmatrix},\\
\vec{P}_3 &=
\begin{pmatrix}
\uimm\frac{\cos\left(\vartheta/2\right) - \eu^{\uimm\beta/2-\uimm\varepsilon}}{\sqrt{\sin^2\left(\vartheta/2\right) + \left|\cos\left(\vartheta/2\right) - \eu^{\uimm\beta/2-\uimm\varepsilon}\right|^2}}\\
\frac{\sin\left(\vartheta/2\right)}{\sqrt{\sin^2\left(\vartheta/2\right) + \left|\cos\left(\vartheta/2\right) - \eu^{\uimm\beta/2-\uimm\varepsilon}\right|^2}}\,\sin(\alpha)\\
\frac{\sin\left(\vartheta/2\right)}{\sqrt{\sin^2\left(\vartheta/2\right) + \left|\cos\left(\vartheta/2\right) - \eu^{\uimm\beta/2-\uimm\varepsilon}\right|^2}}\,\cos(\alpha)
\end{pmatrix}.
\end{aligned}
\label{eq:eigenvectors}
\end{equation}
By introducing the diagonal matrix $\hat{\varLambda} = \diag(\eu^{\uimm\lambda_1},\,\eu^{\uimm\lambda_2},\,\eu^{\uimm\lambda_3})$, and the matrix $\hat{P} = (\vec{P}_1,\,\vec{P}_2,\,\vec{P}_3)$, whose $j$th column is the eigenvectors $\vec{P}_j$ of $\hat{F}_{\Delta}\,\hat{V}_{\mathrm{p}}\,\hat{A}_{\mathrm{pu}}$, we obtain that
\begin{equation}
\hat{F}_{\Delta}\,\hat{V}_{\mathrm{p}}\,\hat{A}_{\mathrm{pu}}= \hat{P}\,\hat{\varLambda}\,\hat{P}^{-1}.
\label{eq:diagonalization}
\end{equation}
% and therefore
% \begin{equation}
% \hat{\mathcal{F}}_{\Delta}\,\hat{\mathcal{V}}_{\mathrm{p}}\,\hat{\mathcal{A}}_{\mathrm{pu}}= \hat{\mathcal{P}}\,\hat{\mathcal{L}}\,\hat{\mathcal{P}}^{-1},
% \end{equation}
% with $\hat{\mathcal{L}} = \hat{L}\otimes\hat{L}^*$ and $\hat{\mathcal{P}}= \hat{P}\otimes\hat{P}^*$. 
Notice that the eigenvectors in Eq.~(\ref{eq:eigenvectors}) have been determined such that $\hat{P}^{-1} = \hat{P}\daga$. As a result, the inverse operator in the second line in Eq.~(\ref{eq:responsibledivergence}) reduces to
$$
(\hat{P}\otimes\hat{P}^*)\,\bigl[\hat{\mathcal{I}} -\eu^{-\uimm (\bar{\omega}T_{\mathrm{p}}- \Delta\phi)}\,(\hat{\varLambda}\otimes\hat{\varLambda}^*)\bigr]^{-1}\,(\hat{P}\otimes\hat{P}^*)^{-1},
$$
where $[\hat{\mathcal{I}} -\eu^{-\uimm (\bar{\omega}T_{\mathrm{p}}- \Delta\phi)}\,(\hat{\varLambda}\otimes\hat{\varLambda}^*)]^{-1}$ is a diagonal matrix of elements $(\hat{\mathcal{I}} -\eu^{-\uimm (\bar{\omega}T_{\mathrm{p}}- \Delta\phi)}\,\eu^{\uimm(\lambda_j - \lambda_{j'})}))^{-1}$, $j,\,j'\in\{1,\,2,\,3\}$, with poles at $\bar{\omega} = \omega_{\mathrm{o}} + (\lambda_j - \lambda_{j'})/T_{\mathrm{p}} + s'\omega_{\mathrm{r}}$, for any $s'\in\mathbb{Z}$. However, from Eq.~(\ref{eq:diagonalization}), we also notice that
\begin{equation}
\hat{A}_{\mathrm{pu}}\,\hat{P} = (\hat{F}_{\Delta}\,\hat{V}_{\mathrm{p}})^{-1}\,\hat{P}\,\hat{\varLambda},
\end{equation}
such that
\begin{equation}
\begin{aligned}
&\vec{v}_k\,\hat{\mathcal{A}}_{\mathrm{pu}}\,(\hat{P}\otimes\hat{P}^*) \\
=\,&[(1,\,0,\,0)\,\hat{A}_{\mathrm{pu}}\,\hat{P}]\otimes[(0,\,\delta_{k2},\,\delta_{k3})\,\hat{A}_{\mathrm{pu}}^*\,\hat{P}^*] \\
% =\,&[(1,\,0,\,0)\,\hat{P}\,\hat{\varLambda}]\otimes[(0,\,\delta_{k2},\,\delta_{k3})\,\eu^{-\uimm\beta}\,\hat{P}^*\,\hat{\varLambda}^*]\\
=\,&\sum_{k',k''=2}^3\sum_{j'=1}^3 \delta_{kk''}\,P_{1k'}\,P_{k''j}^*\,\\
&\times\,[(0,\,\delta_{k'2},\,\delta_{k'3})\otimes(\delta_{j'1},\,\delta_{j'2},\,\delta_{j'3})]\,[\eu^{-\uimm\beta}\hat{\varLambda}\otimes\hat{\varLambda}^*],
\end{aligned}
\end{equation}
where we have explicitly used the fact that $P_{11} = 0$. As a result, the second line in Eq.~(\ref{eq:responsibledivergence}) can be written as
\begin{equation}
\begin{aligned}
&\vec{v}_k\,\hat{\mathcal{A}}_{\mathrm{pu}}\,\bigl(\hat{\mathcal{I}} -\eu^{-\uimm (\bar{\omega}T_{\mathrm{p}}- \Delta\phi)}\,\hat{\mathcal{F}}_{\Delta}\,\hat{\mathcal{V}}_{\mathrm{p}}\,\hat{\mathcal{A}}_{\mathrm{pu}}\bigr)^{-1}\\
=\,&\sum_{k',k''=2}^3\sum_{j'=1}^3 \delta_{kk''}\,P_{1k'}\,P_{k''j}^*\,\\
&\times\,[(0,\,\delta_{k'2},\,\delta_{k'3})\otimes(\delta_{j'1},\,\delta_{j'2},\,\delta_{j'3})]\\
&\times\,\bigl[\hat{\mathcal{I}} -\eu^{-\uimm (\bar{\omega}T_{\mathrm{p}}- \Delta\phi)}\,(\hat{\varLambda}\otimes\hat{\varLambda}^*)\bigr]^{-1}\\
&\times\,[\eu^{-\uimm\beta}\hat{\varLambda}\otimes\hat{\varLambda}^*]\,(\hat{P}\otimes\hat{P}^*)^{-1}.
\end{aligned}
\label{eq:whynotall}
\end{equation}
Due to the term $[(0,\,\delta_{k'2},\,\delta_{k'3})\otimes(\delta_{j1},\,\delta_{j2},\,\delta_{j3})]$, not all matrix elements of the diagonal operator $[\hat{\mathcal{I}} -\eu^{-\uimm (\bar{\omega}T_{\mathrm{p}}- \Delta\phi)}\,(\hat{\varLambda}\otimes\hat{\varLambda}^*)]^{-1}$ contribute to the spectrum, but only $(\hat{\mathcal{I}} -\eu^{-\uimm (\bar{\omega}T_{\mathrm{p}}- \Delta\phi)}\,\eu^{\uimm(\lambda_{k'} - \lambda_{j'})}))^{-1}$, with $k'\in\{2,\,3\}$ and $j'\in\{1,\,2,\,3\}$. The only poles determining the peaks in the spectrum are thus
\begin{equation}
\begin{aligned}
\bar{\omega} = \bar{\omega}^{\mathrm{pole}}_{k'j's_k} &= \omega_{\mathrm{o}} + \frac{\lambda_{k'} - \lambda_{j'}}{T_{\mathrm{p}}} + s'\omega_{\mathrm{r}},\\
&=\omega_{k1} + \frac{\lambda_{k'} - \lambda_{j'} - \beta}{T_{\mathrm{p}}} + s_k\omega_{\mathrm{r}}
\end{aligned}
\end{equation}
for $k,\,k'\in\{2,\,3\}$, $j'\in\{1,\,2,\,3\}$, and for any $s_k\in\mathbb{Z}$ (in the above equality, $s_k = s' - \lfloor \omega_{k1}/\omega_{\mathrm{r}}\rfloor$, with $s_3 = s_2 - 2$ since $\omega_{32} = 2\omega_{\mathrm{r}}$).

For a fixed value of $s_k$, this provides the central frequencies of the five-line structures appearing in the spectrum for $N\rightarrow \infty$ and discussed in Sec.~\ref{Results and discussion}:
\begin{equation}
\begin{aligned}
\bar{\omega}^{\mathrm{pole}}_{k'k's_k} &= \omega_{k1} -\frac{\beta}{T_{\mathrm{p}}} + s_k\,\omega_{\mathrm{r}},\\
\bar{\omega}^{\mathrm{pole}}_{k'1s_k}  &=\omega_{k1} -\frac{\beta}{2T_{\mathrm{p}}} \pm \frac{\varepsilon}{T_{\mathrm{p}}}+s_k\,\omega_{\mathrm{r}},\\
\bar{\omega}^{\mathrm{pole}}_{k'k''s_k} &=\omega_{k1} -\frac{\beta}{T_{\mathrm{p}}}  \pm\frac{2\varepsilon}{T_{\mathrm{p}}} +s_k\,\omega_{\mathrm{r}},
\end{aligned}
\label{eq:poles}
\end{equation}
with $k'\neq k''$. Different values of the index $s_k$ are associated with different five-level structures. The term $\sinc{[(\bar{\omega} - \omega_{k1})T_{\mathrm{p}}/2]}$ in the first line of Eq.~(\ref{eq:responsibledivergence}) modulates the intensity of the lines, such that structures in proximity of the transition energies $\omega_{k1}$ are stronger than the remaining ones. Furthermore, whenever the frequencies $\bar{\omega}^{\mathrm{pole}}_{k'j's_k}$ in Eq.~(\ref{eq:poles}) coincide with the frequencies $\bar{\omega}^{\mathrm{zero}}_{r}$ in Eq.~(\ref{eq:zeros}), the corresponding lines are suppressed in the spectrum.

The dependence of the poles upon the pulse-to-pulse phase shift $\Delta\phi \doteq \omega_{\mathrm{o}}T_{\mathrm{p}}$ is in general complex due to the presence of $\varepsilon$ in Eq.~(\ref{eq:poles}). We notice, however, that one of the spectral peaks is always centered on $\bar{\omega}^{\mathrm{pole}}_{k'k's_k}$, independent of the pulse area $\vartheta$. This central frequency has a linear dependence on $\Delta\phi$, and the corresponding peak can be recognized in Fig.~\ref{fig:DetailsDeltaPhiDependenceGridPlot2} for all values of $\vartheta$ except $2\pi$. As we will discuss later, the contribution to the spectrum due to this line is suppressed for $\vartheta = 2\pi$ . 

In the following, we investigate in detail a few particular cases on which we have focused during the discussion of the results in Sec.~\ref{Results and discussion}.

\subsection{$\omega_{\mathrm{o}} = \delta_{\mathrm{r}}$}
\label{Appendix:omegao = deltar}

Whenever the offset frequency $\omega_{\mathrm{o}}$ is equal to the effective detuning $\delta_{\mathrm{r}}$ (see also Fig.~\ref{fig:MoreLevelSystem}), then $\hat{F}_{\Delta}\,\hat{V}_{\mathrm{p}} = \hat{I}$, i.e., the pulse-to-pulse phase shift $\Delta\phi$ perfectly balances the difference in the phase of the two excited states $\omega_{32}T_{\mathrm{p}}$ accumulated during the interval $T_{\mathrm{p}}$ in between the two pump pulses. The operator $\hat{F}_{\Delta}\,\hat{V}_{\mathrm{p}}\,\hat{A}_{\mathrm{pu}}$ then reduces to the symmetric operator $\hat{A}_{\mathrm{pu}}$ in Eq.~(\ref{eq:pulseaction}), $\beta = 0$, $\varepsilon = \vartheta/2$, such that Eq.~(\ref{eq:poles}) gives
\begin{equation}
\begin{aligned}
\hat{\varLambda} &= \diag(1,\,\eu^{\uimm\vartheta/2},\,\eu^{-\uimm\vartheta/2}),\\
\hat{P}&=
\begin{pmatrix}
0 &\frac{1}{\sqrt{2}} &-\frac{1}{\sqrt{2}} \\
-\cos(\alpha) &\frac{\sin(\alpha)}{\sqrt{2}} & \frac{\sin(\alpha)}{\sqrt{2}} \\
\sin(\alpha) &\frac{\cos(\alpha)}{\sqrt{2}} & \frac{\cos(\alpha)}{\sqrt{2}} 
\end{pmatrix},
\end{aligned}
\end{equation}
with $\vartheta$-independent eigenvectors. The central frequencies of the spectral lines from Eq.~(\ref{eq:poles}) can therefore be written as 
\begin{equation}
\begin{aligned}
\bar{\omega}^{\mathrm{pole}}_{k'k's_k} &= \omega_{k1} + s_k\,\omega_{\mathrm{r}},\\
\bar{\omega}^{\mathrm{pole}}_{k'1s_k} &=\omega_{k1} \pm \frac{\vartheta}{2T_{\mathrm{p}}}+s_k\,\omega_{\mathrm{r}},\\
\bar{\omega}^{\mathrm{pole}}_{k'k''s_k} &=\omega_{k1} \pm \frac{\vartheta}{T_{\mathrm{p}}}+s_k\,\omega_{\mathrm{r}}.
\end{aligned}
\label{eq:positionpoles1}
\end{equation}
These frequencies correspond to the central frequencies of the five-level structures identified in Sec.~\ref{Results and discussion} for $\Delta\phi = \delta_{\mathrm{r}}T_{\mathrm{p}}$, separated by the frequency gap $\Delta\omega = \vartheta/(2T_{\mathrm{p}})$. Notice that $\bar{\omega}^{\mathrm{pole}}_{k'k's_k}$ in Eq.~(\ref{eq:positionpoles1}) corresponds to the position $\bar{\omega}^{\mathrm{zero}}_r$ of the zeros of $\sinc{[(\bar{\omega} - \omega_{k1})T_{\mathrm{p}}/2]}$ except when $s_k = 0$. Hence, while the five-level structures centered on $\omega_{k1}$ do show the associated central line, this is suppressed in the additional structures appearing above and below, as apparent in Figs.~\ref{fig:DetailsDeltaPhiDependenceGridPlot2}, \ref{fig:OmegaThetaGridPlot4}(a)--(c), and \ref{fig:OmegaTauGridPlot3}(a)--(d) in Sec.~\ref{Results and discussion}.

\subsection{$\omega_{\mathrm{o}} = \delta_{\mathrm{r}} - \pi/T_{\mathrm{p}}$}
\label{Appendix:omegao = deltar - pi/T}

When offset frequency and effective detuning differ by $\pi/T_{\mathrm{p}}$, it follows that $\hat{F}_{\Delta}\,\hat{V}_{\mathrm{p}} = \diag(1,\,-1,\,-1)$, $\beta = \pi$, and $\varepsilon = \pi/2$. As a result, the Hermitian operator $\hat{F}_{\Delta}\,\hat{V}_{\mathrm{p}}\,\hat{A}_{\mathrm{pu}}$ has eigenvalues and eigenvectors given by
\begin{equation}
\begin{aligned}
\hat{\varLambda} &= \diag(-1,\,1,\,-1),\\
\hat{P}&=
\begin{pmatrix}
0 &\cos{\left(\frac{\vartheta}{4}\right)} &-\uimm\sin{\left(\frac{\vartheta}{4}\right)} \\
-\cos(\alpha) &-\uimm\,\sin{\left(\frac{\vartheta}{4}\right)}\,\sin(\alpha) & \cos{\left(\frac{\vartheta}{4}\right)}\,\sin(\alpha) \\
\sin(\alpha) &-\uimm\,\sin{\left(\frac{\vartheta}{4}\right)}\,\cos(\alpha) & \cos{\left(\frac{\vartheta}{4}\right)}\,\cos(\alpha)
\end{pmatrix},
\end{aligned}
\end{equation}
such that all poles in Eq.~(\ref{eq:poles}) are given by
\begin{equation}
\begin{aligned}
\bar{\omega}^{\mathrm{pole}}_{22s_k} = \bar{\omega}^{\mathrm{pole}}_{33s_k} =\bar{\omega}^{\mathrm{pole}}_{31s_k}  &= \omega_{k1} -\frac{\pi}{T_{\mathrm{p}}} + s_k\,\omega_{\mathrm{r}},\\
\bar{\omega}^{\mathrm{pole}}_{21s_k} =\bar{\omega}^{\mathrm{pole}}_{23s_k} &=\omega_{k1} -\frac{2\pi}{T_{\mathrm{p}}}+s_k\,\omega_{\mathrm{r}},\\
\bar{\omega}^{\mathrm{pole}}_{32s_k} &=\omega_{k1} +s_k\,\omega_{\mathrm{r}},
\end{aligned}
\end{equation}
which can be summarized as the $\vartheta$-independent frequencies
\begin{equation}
\bar{\omega}^{\mathrm{pole}}_{s} = \omega_{k1} + s\,\frac{\omega_{\mathrm{r}}}{2},
\label{eq:positionpoles2}
\end{equation}
with $s\in\mathbb{Z}$. Notice that the corresponding spectral lines will be suppressed whenever their central frequencies are equal to the zeros in Eq.~(\ref{eq:zeros}). This is apparent in Figs.~\ref{fig:DetailsDeltaPhiDependenceGridPlot2}, \ref{fig:OmegaThetaGridPlot4}(d)--(f), and \ref{fig:OmegaTauGridPlot3}(e)--(h) in Sec.~\ref{Results and discussion}.

\subsection{$\pi$-area pulses}
\label{Appendix:pi-area pulses}

When $\vartheta =\pi$, such that $\varepsilon = \pi/2$, then the diagonalization of the operator $\hat{F}_{\Delta}\,\hat{V}_{\mathrm{p}}\,\hat{A}_{\mathrm{pu}} $
% \begin{equation}
% \begin{aligned}
% \hat{F}_{\Delta}\,\hat{V}_{\mathrm{p}}\,\hat{A}_{\mathrm{pu}} &= 
% \begin{pmatrix}
% 1 &0 &0\\
% 0 &\eu^{-\uimm(\delta_{\mathrm{r}} - \omega_{\mathrm{o}})T_{\mathrm{p}}} &0\\
% 0 &0 &\eu^{-\uimm(\delta_{\mathrm{r}} - \omega_{\mathrm{o}})T_{\mathrm{p}}}\\
% \end{pmatrix}\\
% &\times 
% \begin{pmatrix}
% 0 & \uimm\,\sin(\alpha) &\uimm\,\cos(\alpha)\\
% \uimm\sin(\alpha) & \cos^2(\alpha ) &-\frac{1}{2}\sin(2\alpha)\\
% \uimm\cos(\alpha) & -\frac{1}{2}\sin(2\alpha) &\sin^2(\alpha)
% \end{pmatrix}
% \end{aligned}
% \end{equation}
leads to
\begin{equation}
\begin{aligned}
\hat{\varLambda} &= \diag(\eu^{-\uimm(\delta_{\mathrm{r}} - \omega_{\mathrm{o}})T_{\mathrm{p}}} ,\,\uimm\,\eu^{-\uimm(\delta_{\mathrm{r}} - \omega_{\mathrm{o}})T_{\mathrm{p}}/2} ,\,-\uimm\,\eu^{-\uimm(\delta_{\mathrm{r}} - \omega_{\mathrm{o}})T_{\mathrm{p}}/2} ),\\
\hat{P}&=
\begin{pmatrix}
0 &\frac{1}{\sqrt{2}} &-\frac{1}{\sqrt{2}}\eu^{\uimm(\delta_{\mathrm{r}} - \omega_{\mathrm{o}})T_{\mathrm{p}}/2} \\
-\cos(\alpha) & \eu^{-\uimm(\delta_{\mathrm{r}} - \omega_{\mathrm{o}})T_{\mathrm{p}}/2}\,\frac{\sin(\alpha) }{\sqrt{2}}& \frac{\sin(\alpha)}{\sqrt{2}}\\
\sin(\alpha) & \eu^{-\uimm(\delta_{\mathrm{r}} - \omega_{\mathrm{o}})T_{\mathrm{p}}/2}\,\frac{\cos(\alpha) }{\sqrt{2}} &\frac{\cos(\alpha)}{\sqrt{2}}
\end{pmatrix}.
\end{aligned}
\end{equation}
Equation~(\ref{eq:poles}) then provides the equations for the central frequencies of the peaks as a function of both offset frequency and effective detuning
\begin{equation}
\begin{aligned}
\bar{\omega}^{\mathrm{pole}}_{k'k's_k} &=\omega_{k1} + (\omega_{\mathrm{o}} - \delta_{\mathrm{r}}) + s_k\,\omega_{\mathrm{r}},\\
\bar{\omega}^{\mathrm{pole}}_{k'k''s_k} &=\omega_{k1} + \left[\omega_{\mathrm{o}} - \left(\delta_{\mathrm{r}} \pm\frac{\pi}{T_{\mathrm{p}}}\right)\right] +s_k\,\omega_{\mathrm{r}},\\
\bar{\omega}^{\mathrm{pole}}_{k'1s_k} &=\omega_{k1} +\frac{\omega_{\mathrm{o}} - \delta_{\mathrm{r}}}{2} \pm \frac{\omega_{\mathrm{r}}}{4}+s_k\,\omega_{\mathrm{r}}.
\end{aligned}
\label{eq:positionpolespi}
\end{equation}
Notice that the $\pm$ sign in $\bar{\omega}^{\mathrm{pole}}_{k'k''s_k}$ is superfluous, since the $+$ solution associated with the index $s_k$ coincides with the $-$ solution for the index $(s_{k} +1)$. This leads to the level structures shown in Figs.~\ref{fig:DetailsDeltaPhiDependenceGridPlot2}(d)--(f), explaining the linear dependence of the position of the absorption lines upon the pulse-to-pulse phase shift. Two parallel lines, given by $\bar{\omega}^{\mathrm{pole}}_{k'k's_k} $ and $\bar{\omega}^{\mathrm{pole}}_{k'k''s_k} $, have the same unitary slope and are spaced by $\pi/T_{\mathrm{p}} = \omega_{\mathrm{r}}/2$. The remaining two lines, given by $\bar{\omega}^{\mathrm{pole}}_{k'1s_k}$, are also parallel and separated by $\omega_{\mathrm{r}}/2$, but with a slope equal to $1/2$. These two couples of lines intersect at $\omega_{\mathrm{o}} = \delta_{\mathrm{r}} - \pi/T_{\mathrm{p}}$, as confirmed in Figs.~\ref{fig:DetailsDeltaPhiDependenceGridPlot2}(d)--(f).

\subsection{$2\pi$-area pulses}
\label{Appendix:2pi-area pulses}

A pulse with area $\vartheta = 2\pi$ will not mix the subspace formed by the ground state with that associated with the two excited states, since its action is given by the block-diagonal operator
\begin{equation}
\hat{A}_{\mathrm{pu}} =
\begin{pmatrix}
-1 &0 &0 \\
0 &\cos(2\alpha) & -\sin(2\alpha) \\
0 &-\sin(2\alpha) & -\cos(2\alpha)
\end{pmatrix}.
\end{equation}
Multiplying it by $\hat{F}_{\Delta}\,\hat{V}_{\mathrm{p}}$ still preserves its block-diagonal form. In this case, $\varepsilon = \pi - \beta/2$, such that eigenvalues and eigenvectors of $\hat{F}_{\Delta}\,\hat{V}_{\mathrm{p}}\,\hat{A}_{\mathrm{pu}} $ can be written as
\begin{equation}
\begin{aligned}
\hat{\varLambda} &= \diag(\eu^{-\uimm(\delta_{\mathrm{r}} - \omega_{\mathrm{o}})T_{\mathrm{p}}} ,\,-\eu^{-\uimm(\delta_{\mathrm{r}} - \omega_{\mathrm{o}})T_{\mathrm{p}}} ,\,-1),\\
\hat{P}&=
\begin{pmatrix}
0 &0 &1 \\
-\cos(\alpha) & \sin(\alpha) & 0\\
\sin(\alpha) & \cos(\alpha) & 0
\end{pmatrix}.
\label{eq:eigenvectors2pi}
\end{aligned}
\end{equation}
Owing to the many vanishing elements of $\hat{P}$, not all 5 peaks in Eq.~(\ref{eq:poles}) contribute to the spectrum. To see this, one can refer to Eq.~(\ref{eq:whynotall}), which for a $2\pi$-area pulse reads
\begin{equation}
\begin{aligned}
&\vec{v}_k\,\hat{\mathcal{A}}_{\mathrm{pu}}\,\bigl(\hat{\mathcal{I}} -\eu^{-\uimm (\bar{\omega}T_{\mathrm{p}}- \Delta\phi)}\,\hat{\mathcal{F}}_{\Delta}\,\hat{\mathcal{V}}_{\mathrm{p}}\,\hat{\mathcal{A}}_{\mathrm{pu}}\bigr)^{-1}\\
=\,&\sum_{k''=2}^3\sum_{j'=1}^2 \delta_{kk''}\,P_{k''j}^*\,[(0,\,0,\,1)\otimes(\delta_{j'1},\,\delta_{j'2},\,0)]\\
&\times\,[\eu^{-\uimm\beta}\hat{\varLambda}\otimes\hat{\varLambda}^*]\,\bigl[\hat{\mathcal{I}} -\eu^{-\uimm (\bar{\omega}T_{\mathrm{p}}- \Delta\phi)}\,(\hat{\varLambda}\otimes\hat{\varLambda}^*)\bigr]^{-1}\\
&\times\,(\hat{P}\otimes\hat{P}^*)^{-1},
\end{aligned}
\label{eq:whynotall2pi}
\end{equation}
where we have used the fact that $P_{1k'} = \delta_{k'3}$ and that $P_{k''j'}$ vanishes for $j' = 3$. As a result, the only lines appearing in the spectrum are due to the poles of
\begin{equation}
\begin{aligned}
&\frac{\eu^{-\uimm\beta}\,\eu^{\uimm(\lambda_3- \lambda_{j'})}}{1-\eu^{-\uimm(\bar{\omega} - \omega_{k1})T_{\mathrm{p}}})\,\eu^{-\uimm\beta}\,\eu^{\uimm(\lambda_3- \lambda_{j'})}}\\
=\,&\frac{(-1)^{j'+1}}{1-\eu^{-\uimm(\bar{\omega} - \omega_{k1})T_{\mathrm{p}}})\,(-1)^{j'+1}},
\end{aligned}
\label{eq:blabla}
\end{equation}
which are given by
\begin{equation}
\begin{aligned}
\bar{\omega}^{\mathrm{pole}}_{31s_k} &=\omega_{k1} + \frac{\omega_{\mathrm{r}}}{2} + s_k\,\omega_{\mathrm{r}},\\
\bar{\omega}^{\mathrm{pole}}_{32s_k} &=\omega_{k1} + (s_k + 1)\,\omega_{\mathrm{r}},
\end{aligned}
\label{eq:positionpoles2pi}
\end{equation}
are spaced by $\omega_{\mathrm{r}}/2$, and independent of $\omega_{\mathrm{o}}$, as shown in Figs.~\ref{fig:DetailsDeltaPhiDependenceGridPlot2}(j)--(l). They are equal to the $\vartheta$-independent frequencies in Eq.~(\ref{eq:positionpoles2}) for $\omega_{\mathrm{o}} = \delta_{\mathrm{r}} - \pi/T_{\mathrm{p}}$. Also here, if these central frequencies are equal to the zeros in Eq.~(\ref{eq:zeros}), then the corresponding spectral lines are suppressed, as shown in Figs.~\ref{fig:OmegaTauGridPlot3}(d) and \ref{fig:OmegaTauGridPlot3}(h). We finally notice that Eq.~(\ref{eq:whynotall2pi}) is independent of $\beta$ as a consequence of Eqs.~(\ref{eq:eigenvectors2pi}) and (\ref{eq:blabla}). This will be used in Appendix~\ref{Appendix:2pi-area pulses-time}.

\section{Spectral features in a pump--probe--pump setup determined by the pump pulses preceding the probe pulse}
\label{Appendix:Spectral features in a pump--probe--pump setup determined by the pump pulses preceding the probe pulse}

The area of the pump pulses preceding the probe pulse determines the state in which the system is prepared and encountered by the probe pulse. This influences the frequency-dependent features of the spectrum in a pump--probe--pump setup, causing, e.g., the disappearance of some of the spectral lines identified in Appendix~\ref{Appendix:Central frequency of the spectral lines}. This is clearly visible in Figs.~\ref{fig:OmegaThetaGridPlot4} and \ref{fig:OmegaTauGridPlot3}, displaying the dependence of the spectral lines upon pulse area and time delay: one can see that lines otherwise present in the spectrum are suppressed for given values of $\vartheta$ and $\tau$. 

This feature is a result of the state in which the system is prepared by the $\Mtau$ pump pulses preceding the probe pulse. In order to provide an example for this general property, we focus on the case of $\vartheta = \pi$, and show how the preparation of the system determines the disappearance of given lines. This is clearly apparent in Fig.~\ref{fig:DetailsDeltaPhiDependenceGridPlot2}(e) for $\Mtau = 1$: in this figure, half of the spectral lines identified in Appendix~\ref{Appendix:pi-area pulses} for $\vartheta = \pi$ are suppressed, whereas they appear in  Fig.~\ref{fig:DetailsDeltaPhiDependenceGridPlot2}(f) for $\Mtau = 2$.

To show this, we notice that, for $\vartheta = \pi$, a train of $\Mtau$ pulses prepares the system in the state
\begin{equation}
\begin{aligned}
&\hat{A}_{\mathrm{pu}}\,(\hat{F}_{\Delta}\,\hat{V}_{\mathrm{p}}\,\hat{A}_{\mathrm{pu}})^{\Mtau-1}\,(1,\,0,\,0)^{\mathrm{T}} \\
=\, & \frac{(\uimm\,\eu^{-\uimm\beta/2})^{\Mtau}}{2}\,
\begin{pmatrix}
1+(-1)^{\Mtau}\\
\eu^{\uimm\beta/2}\,\sin(\alpha)\,[1-(-1)^{\Mtau}]\\
\eu^{\uimm\beta/2}\,\cos(\alpha)\,[1-(-1)^{\Mtau}]
\end{pmatrix},
\end{aligned}
\label{eq:initialstate-pi}
\end{equation}
see also Eq.~(\ref{eq:initialstate}). Therefore, whenever $\Mtau$ is odd, only the two excited states are occupied. In such case, the spectrum from Eq.~(\ref{eq:spectrumpuprpushort}) contains only the last addend appearing in Eq.~(\ref{eq:Zdetail}),
\begin{equation}
\begin{aligned}
&\eu^{-\uimm\bar{\omega}(T_{\mathrm{p}} - \tau')}\,\hat{\mathcal{Z}}(\tau')\,\hat{\mathcal{F}}_{\Delta}\,\hat{\mathcal{A}}_{\mathrm{pu}}\,(\hat{\mathcal{F}}_{\Delta}\,\hat{\mathcal{V}}_{\mathrm{p}}\,\hat{\mathcal{A}}_{\mathrm{pu}})^{\Mtau-1}\,\vec{R}_0 \\
=\, & \eu^{-\uimm\bar{\omega}(T_{\mathrm{p}} - \tau')}
\begin{pmatrix}
1 &0 &0 \\
0 &0 &0 \\
0 &0 &0
\end{pmatrix}
\hat{Z}
\begin{pmatrix}
0 \\
\sin(\alpha)\\
\cos(\alpha)
\end{pmatrix}\\
&\otimes
\begin{pmatrix}
0 &0 &0 \\
0 &1 &0 \\
0 &0 &1
\end{pmatrix}
\hat{Z}^*
\begin{pmatrix}
0 \\
\sin(\alpha)\\
\cos(\alpha)
\end{pmatrix},
\end{aligned}
\end{equation}
and the central frequencies of the lines appearing in the spectrum can be determined by inspecting
\begin{equation}
\begin{aligned}
&\bigl[\hat{\mathcal{I}} -\eu^{-\uimm (\bar{\omega}T_{\mathrm{p}}- \Delta\phi)}\,(\hat{\varLambda}\otimes\hat{\varLambda}^*)\bigr]^{-1}\,(\hat{P}\otimes\hat{P}^*)^{-1}\\
&\times\,\eu^{-\uimm\bar{\omega}(T_{\mathrm{p}} - \tau')}\,\hat{\mathcal{Z}}(\tau')\,\hat{\mathcal{F}}_{\Delta}\,\hat{\mathcal{A}}_{\mathrm{pu}}\,(\hat{\mathcal{F}}_{\Delta}\,\hat{\mathcal{V}}_{\mathrm{p}}\,\hat{\mathcal{A}}_{\mathrm{pu}})^{\Mtau-1}\,\vec{R}_0 \\
=\, &\eu^{-\uimm\bar{\omega}(T_{\mathrm{p}} - \tau')}\,\bigl[\hat{\mathcal{I}} -\eu^{-\uimm (\bar{\omega}T_{\mathrm{p}}- \Delta\phi)}\,(\hat{\varLambda}\otimes\hat{\varLambda}^*)\bigr]^{-1}\,\vec{x}\otimes\vec{y},
\end{aligned}
\label{eq:influenceofpreparationforpi}
\end{equation}
with the 3-dimensional vectors
\begin{equation}
\vec{x} = \begin{pmatrix}
x_1 \\
x_2 \\
x_3
\end{pmatrix} = \hat{P}\daga
\begin{pmatrix}
1 &0 &0 \\
0 &0 &0 \\
0 &0 &0
\end{pmatrix}
\hat{Z}
\begin{pmatrix}
0 \\
\sin(\alpha)\\
\cos(\alpha)
\end{pmatrix}
\end{equation}
and
\begin{equation}
\vec{y} = \begin{pmatrix}
y_1 \\
y_2 \\
y_3
\end{pmatrix} = 
\hat{P}^{\mathrm{T}}
\begin{pmatrix}
0 &0 &0 \\
0 &1 &0 \\
0 &0 &1
\end{pmatrix}
\hat{Z}^*
\begin{pmatrix}
0 \\
\sin(\alpha)\\
\cos(\alpha)
\end{pmatrix}.
\end{equation}
By noticing that the components $x_1$ and $y_1$ vanish for $\vartheta = \pi$ and for the weak probe pulses ($\vartheta_{\mathrm{pr}}\ll 1$) described by Eq.~(\ref{eq:probepulseapprox}), then one can conclude from Eq.~(\ref{eq:influenceofpreparationforpi}) that the poles $\bar{\omega}_{k'1s_k}^{\mathrm{pole}}$ identified in Eq.~(\ref{eq:positionpolespi}) do not correspond to peaks in the pump--probe--pump spectrum for $\vartheta = \pi$ and for an odd number $\Mtau$ of pulses preceding the weak probe pulse. This is in agreement with the results exhibited in Fig.~\ref{fig:DetailsDeltaPhiDependenceGridPlot2}(e).

\section{Periodicity of the spectra as a function of time delay}
\label{Appendix:Details on the time-delay-dependent features of the spectra}

The periodicity of the pump--probe--pump spectrum in Eq.~(\ref{eq:spectrumpuprpushort}) is exclusively determined by the operator $\hat{\mathcal{A}}_{\mathrm{pu}}\,(\hat{\mathcal{F}}_{\Delta}\,\hat{\mathcal{V}}_{\mathrm{p}}\,\hat{\mathcal{A}}_{\mathrm{pu}})^{\Mtau-1}\,\vec{R}_0$, which prepares the system in the state encountered by the probe pulse. All remaining terms in the spectrum depend on $T_{\mathrm{p}} - \tau' = \Mtau T_{\mathrm{p}} - \tau$ and are thus periodic in $\tau$ with period $T_{\mathrm{p}}$. Whenever two sequences of pump pulses $M_{\tau_1}$ and $M_{\tau_2}$ prepare the system in the same state, also the associated spectra will exhibit the same features. 

In order to investigate the properties of the state prepared by the pump pulses preceding the probe pulse, we observe that
\begin{equation}
\begin{aligned}
&\ \ \ \ \hat{A}_{\mathrm{pu}}\,(\hat{F}_{\Delta}\,\hat{V}_{\mathrm{p}}\,\hat{A}_{\mathrm{pu}})^{\Mtau-1}\,(1,\,0,\,0)^{\mathrm{T}} \\
&= (\hat{F}_{\Delta}\,\hat{V}_{\mathrm{p}})^{-1}\,\hat{P}\,\hat{\varLambda}^{\Mtau}\,\hat{P}^{-1}\,(1,\,0,\,0)^{\mathrm{T}}\\
% =\, & (\hat{F}_{\Delta}\,\hat{V}_{\mathrm{p}})^{-1}\,\hat{P}\,(0,\,P_{12}^*\,\eu^{\uimm\lambda_2 \Mtau},\,P_{13}^*\,\eu^{\uimm\lambda_3 \Mtau})^{\mathrm{T}}\\
&=\frac{\eu^{-\uimm\beta\Mtau/2}}{\sin^2\left(\vartheta/2\right) + \left|\cos\left(\vartheta/2\right) - \eu^{\uimm\beta/2-\uimm\varepsilon}\right|^2}\,\times\\
&
\begin{pmatrix}
\sin^2\left(\vartheta/2\right)\eu^{\uimm\varepsilon\Mtau} + \left|\cos\left(\vartheta/2\right) - \eu^{\uimm\beta/2-\uimm\varepsilon}\right|^2\eu^{-\uimm\varepsilon\Mtau} \\
-2\sin\left(\vartheta/2\right)\left[\cos\left(\vartheta/2\right) - \eu^{-\uimm\beta/2+\uimm\varepsilon}\right]\eu^{\uimm\beta}\sin(\varepsilon\Mtau)\sin(\alpha)\\
-2\sin\left(\vartheta/2\right)\left[\cos\left(\vartheta/2\right) - \eu^{-\uimm\beta/2+\uimm\varepsilon}\right]\eu^{\uimm\beta}\sin(\varepsilon\Mtau)\cos(\alpha)
\end{pmatrix}.
\end{aligned}
\label{eq:initialstate}
\end{equation}
Since the spectrum depends on
\begin{equation}
\begin{aligned}
&\hat{\mathcal{A}}_{\mathrm{pu}}\,(\hat{\mathcal{F}}_{\Delta}\,\hat{\mathcal{V}}_{\mathrm{p}}\,\hat{\mathcal{A}}_{\mathrm{pu}})^{\Mtau-1}\,\vec{R}_0 \\
=\,&\hat{A}_{\mathrm{pu}}\,(\hat{F}_{\Delta}\,\hat{V}_{\mathrm{p}}\,\hat{A}_{\mathrm{pu}})^{\Mtau-1}\,(1,\,0,\,0)^{\mathrm{T}} \\
&\otimes\,\hat{A}_{\mathrm{pu}}\,(\hat{F}_{\Delta}\,\hat{V}_{\mathrm{p}}\,\hat{A}_{\mathrm{pu}})^{\Mtau-1}\,(1,\,0,\,0)^{\mathrm{T}} ,
\end{aligned}
\end{equation}
we observe that (i) it does not depend on the common phase term $\eu^{-\uimm\beta\Mtau/2}$ in Eq.~(\ref{eq:initialstate}), and (ii) its dependence upon $\Mtau$ is only via terms of the form $\eu^{\pm\uimm 2\varepsilon\Mtau}$. In other words, the dipoles generated by $M_{\tau_1}$ pulses associated with $\varepsilon_1$ and $M_{\tau_2}$ pulses associated with $\varepsilon_2$ are equal---and the corresponding spectra coincide---if there exists an integer $K$ for which
\begin{equation}
M_{\tau_1}\varepsilon_1 = M_{\tau_2}\varepsilon_2 + \pi K.
\label{eq:relationtautheta-supp}
\end{equation}
For fixed pulse parameters $\vartheta$ and $\beta$, the spectrum is periodic with respect to the number of preparatory pump pulses, with period $\Delta \Mtau = \pi K/\varepsilon$, where $\Delta \Mtau$ and $K$ are both integers.

We analyze this in depth for the same particular cases already discussed in Appendix~\ref{Appendix:Central frequency of the spectral lines}.

\subsection{$\omega_{\mathrm{o}} = \delta_{\mathrm{r}}$}
\label{Appendix:omegao = deltar-time}

In this case, with $\beta = 0$ and $\varepsilon = \vartheta/2$, the state prepared by the initial $\Mtau$ pump pulses is given by
\begin{equation}
\hat{A}_{\mathrm{pu}}\,(\hat{F}_{\Delta}\,\hat{V}_{\mathrm{p}}\,\hat{A}_{\mathrm{pu}})^{\Mtau-1}\,(1,\,0,\,0)^{\mathrm{T}} = 
\begin{pmatrix}
\cos\left(\frac{\vartheta\Mtau}{2}\right)\\
\uimm\,\sin\left(\frac{\vartheta\Mtau}{2}\right)\,\sin(\alpha)\\
\uimm\,\sin\left(\frac{\vartheta\Mtau}{2}\right)\,\cos(\alpha)
\end{pmatrix}
\end{equation}
and Eq.~(\ref{eq:relationtautheta-supp}) leads to Eq.~(\ref{eq:relationtautheta}), thus explaining the periodic features in Figs.~\ref{fig:TauThetaGridPlot5}(a) and \ref{fig:TauThetaGridPlot5}(b) and their dependence on $\vartheta$.

\subsection{$\omega_{\mathrm{o}} = \delta_{\mathrm{r}} - \pi/T_{\mathrm{p}}$}
\label{Appendix:omegao = deltar - pi/Tp-time}

With $\beta = \pi$ and $\varepsilon = \pi/2$, the state encountered by the probe pulse is given by
\begin{equation}
\begin{aligned}
&\hat{A}_{\mathrm{pu}}\,(\hat{F}_{\Delta}\,\hat{V}_{\mathrm{p}}\,\hat{A}_{\mathrm{pu}})^{\Mtau-1}\,(1,\,0,\,0)^{\mathrm{T}} \\
=\,& 
\begin{pmatrix}
\frac{1+(-1)^{\Mtau}}{2} + \frac{1-(-1)^{\Mtau}}{2}\,\cos\left(\frac{\vartheta}{2}\right)\\
\uimm\,\frac{1-(-1)^{\Mtau}}{2}\,\sin\left(\frac{\vartheta}{2}\right)\,\sin(\alpha)\\
\uimm\,\frac{1-(-1)^{\Mtau}}{2}\,\sin\left(\frac{\vartheta}{2}\right)\,\cos(\alpha)
\end{pmatrix}\\
=\,& \left\{
\begin{aligned}
&\begin{pmatrix}
\cos\left(\frac{\vartheta}{2}\right)\\
\uimm\,\sin\left(\frac{\vartheta}{2}\right)\,\sin(\alpha)\\
\uimm\,\sin\left(\frac{\vartheta}{2}\right)\,\cos(\alpha)
\end{pmatrix}, 
&\text{if $\Mtau$ odd,  }\\
&\ \ \ \ \ \ (1,\,0,\,0)^{\mathrm{T}},
&\text{if $\Mtau$ even,}
\end{aligned} \right. 
\end{aligned}
\end{equation}
explaining the results in Figs.~\ref{fig:TauThetaGridPlot5}(c) and \ref{fig:TauThetaGridPlot5}(d) and the periodicity of the spectra as a function of $\tau$, with period $2T_{\mathrm{p}}$.

\subsection{$\pi$-area pulses}
\label{Appendix:pi-area pulses-time}
As shown in Eq.~(\ref{eq:initialstate-pi}), a sequence of $\Mtau$ $\pi$-area pulses prepares the system in the state
\begin{equation}
\begin{aligned}
&\hat{A}_{\mathrm{pu}}\,(\hat{F}_{\Delta}\,\hat{V}_{\mathrm{p}}\,\hat{A}_{\mathrm{pu}})^{\Mtau-1}\,(1,\,0,\,0)^{\mathrm{T}} \\
=\,& \left\{
\begin{aligned}
&(\uimm\,\eu^{-\uimm\beta/2})^{\Mtau}\,\eu^{\uimm\beta/2}\,
(0,\,\sin(\alpha),\,\cos(\alpha))^{\mathrm{T}},
&\text{if $\Mtau$ odd,  }\\
&(\uimm\,\eu^{-\uimm\beta/2})^{\Mtau}\,
(1,\,0,\,0)^{\mathrm{T}},
&\text{if $\Mtau$ even,}
\end{aligned} \right. 
\end{aligned}
\end{equation}
so that the associated spectra are periodic in $\tau$, with period $2T_{\mathrm{p}}$ for any $\beta$.

\subsection{$2\pi$-area pulses}
\label{Appendix:2pi-area pulses-time}

A sequence of $\Mtau$ $2\pi$-area pulses prepares the system in the state
\begin{equation}
\hat{A}_{\mathrm{pu}}\,(\hat{F}_{\Delta}\,\hat{V}_{\mathrm{p}}\,\hat{A}_{\mathrm{pu}})^{\Mtau-1}\,(1,\,0,\,0)^{\mathrm{T}} \\
= ((-1)^{\Mtau},\,0,\,0)^{\mathrm{T}},
\label{eq:initialstate2pi}
\end{equation}
and the time-delay-dependent spectra have period $T_{\mathrm{p}}$---the spectra are not sensitive to the absolute phase of the state associated with $(-1)^{\Mtau}$. In Appendix~\ref{Appendix:2pi-area pulses}, we already noticed that Eq.~(\ref{eq:whynotall2pi}) is independent of $\beta$. Due to Eq.~(\ref{eq:initialstate2pi}) and therefore as a result of
\begin{equation}
\hat{\mathcal{F}}_{\Delta}\,\hat{\mathcal{A}}_{\mathrm{pu}}\,(\hat{\mathcal{F}}_{\Delta}\,\hat{\mathcal{V}}_{\mathrm{p}}\,\hat{\mathcal{A}}_{\mathrm{pu}})^{\Mtau-1}\,\vec{R}_0 = \vec{R}_0,
\end{equation}
the spectrum in Eq.~(\ref{eq:spectrumpuprpushort}) contains only the second addend appearing in Eq.~(\ref{eq:Zdetail}), leading to
\begin{equation}
\begin{aligned}
&\eu^{-\uimm\bar{\omega}(T_{\mathrm{p}} - \tau')}\,\hat{\mathcal{Z}}(\tau')\,\hat{\mathcal{F}}_{\Delta}\,\hat{\mathcal{A}}_{\mathrm{pu}}\,(\hat{\mathcal{F}}_{\Delta}\,\hat{\mathcal{V}}_{\mathrm{p}}\,\hat{\mathcal{A}}_{\mathrm{pu}})^{\Mtau-1}\,\vec{R}_0 \\
=\, & \eu^{-\uimm\bar{\omega}(T_{\mathrm{p}} - \tau')}\,(Z_{11},\,0,\,0)^{\mathrm{T}}\otimes (0,\,Z^*_{21},\,Z^*_{31})^{\mathrm{T}}\\
=\, & -\uimm\,\frac{\vartheta}{2}\,
\begin{pmatrix}
1\\
0\\
0
\end{pmatrix}
\otimes 
\begin{pmatrix}
0\\ 
\eu^{-\uimm(\bar{\omega}-\omega_{21})(T_{\mathrm{p}} - \tau')}\,\sin(\alpha)\\
\eu^{-\uimm(\bar{\omega}-\omega_{31})(T_{\mathrm{p}} - \tau')}\,\cos(\alpha)
\end{pmatrix}
\end{aligned}
\end{equation}
for the weak probe pulses described by Eq.~(\ref{eq:probepulseapprox}). Hence, the spectra in Eq.~(\ref{eq:spectrumpuprpushort}) at $\vartheta = 2\pi$ are independent of the pulse-to-pulse phase shift. This explains why the spectra displayed in Figs.~\ref{fig:OmegaTauGridPlot3}(d) and \ref{fig:OmegaTauGridPlot3}(h), evaluated at $\vartheta = 2\pi$ for two different values of $\beta$, are identical.

% \bibliography{biblio}

\begin{thebibliography}{53}%
\makeatletter
\providecommand \@ifxundefined [1]{%
 \@ifx{#1\undefined}
}%
\providecommand \@ifnum [1]{%
 \ifnum #1\expandafter \@firstoftwo
 \else \expandafter \@secondoftwo
 \fi
}%
\providecommand \@ifx [1]{%
 \ifx #1\expandafter \@firstoftwo
 \else \expandafter \@secondoftwo
 \fi
}%
\providecommand \natexlab [1]{#1}%
\providecommand \enquote  [1]{``#1''}%
\providecommand \bibnamefont  [1]{#1}%
\providecommand \bibfnamefont [1]{#1}%
\providecommand \citenamefont [1]{#1}%
\providecommand \href@noop [0]{\@secondoftwo}%
\providecommand \href [0]{\begingroup \@sanitize@url \@href}%
\providecommand \@href[1]{\@@startlink{#1}\@@href}%
\providecommand \@@href[1]{\endgroup#1\@@endlink}%
\providecommand \@sanitize@url [0]{\catcode `\\12\catcode `\$12\catcode
  `\&12\catcode `\#12\catcode `\^12\catcode `\_12\catcode `\%12\relax}%
\providecommand \@@startlink[1]{}%
\providecommand \@@endlink[0]{}%
\providecommand \url  [0]{\begingroup\@sanitize@url \@url }%
\providecommand \@url [1]{\endgroup\@href {#1}{\urlprefix }}%
\providecommand \urlprefix  [0]{URL }%
\providecommand \Eprint [0]{\href }%
\providecommand \doibase [0]{http://dx.doi.org/}%
\providecommand \selectlanguage [0]{\@gobble}%
\providecommand \bibinfo  [0]{\@secondoftwo}%
\providecommand \bibfield  [0]{\@secondoftwo}%
\providecommand \translation [1]{[#1]}%
\providecommand \BibitemOpen [0]{}%
\providecommand \bibitemStop [0]{}%
\providecommand \bibitemNoStop [0]{.\EOS\space}%
\providecommand \EOS [0]{\spacefactor3000\relax}%
\providecommand \BibitemShut  [1]{\csname bibitem#1\endcsname}%
\let\auto@bib@innerbib\@empty
%</preamble>
\bibitem [{\citenamefont {Pollard}\ and\ \citenamefont
  {Mathies}(1992)}]{doi:10.1146/annurev.pc.43.100192.002433}%
  \BibitemOpen
  \bibfield  {author} {\bibinfo {author} {\bibfnamefont {W.~T.}\ \bibnamefont
  {Pollard}}\ and\ \bibinfo {author} {\bibfnamefont {R.~A.}\ \bibnamefont
  {Mathies}},\ }\bibfield  {title} {\enquote {\bibinfo {title} {Analysis of
  femtosecond dynamic absorption spectra of nonstationary states},}\ }\href
  {\doibase 10.1146/annurev.pc.43.100192.002433} {\bibfield  {journal}
  {\bibinfo  {journal} {Annu. Rev. Phys. Chem.}\ }\textbf {\bibinfo {volume}
  {43}},\ \bibinfo {pages} {497--523} (\bibinfo {year} {1992})}\BibitemShut
  {NoStop}%
\bibitem [{\citenamefont {Wu}\ \emph {et~al.}(2016)\citenamefont {Wu},
  \citenamefont {Chen}, \citenamefont {Camp}, \citenamefont {Schafer},\ and\
  \citenamefont {Gaarde}}]{0953-4075-49-6-062003}%
  \BibitemOpen
  \bibfield  {author} {\bibinfo {author} {\bibfnamefont {M.}~\bibnamefont
  {Wu}}, \bibinfo {author} {\bibfnamefont {S.}~\bibnamefont {Chen}}, \bibinfo
  {author} {\bibfnamefont {S.}~\bibnamefont {Camp}}, \bibinfo {author}
  {\bibfnamefont {K.~J.}\ \bibnamefont {Schafer}}, \ and\ \bibinfo {author}
  {\bibfnamefont {M.~B.}\ \bibnamefont {Gaarde}},\ }\bibfield  {title}
  {\enquote {\bibinfo {title} {Theory of strong-field attosecond transient
  absorption},}\ }\href {http://stacks.iop.org/0953-4075/49/i=6/a=062003}
  {\bibfield  {journal} {\bibinfo  {journal} {J. Phys. B}\ }\textbf {\bibinfo
  {volume} {49}},\ \bibinfo {pages} {062003} (\bibinfo {year}
  {2016})}\BibitemShut {NoStop}%
\bibitem [{\citenamefont {Loh}\ \emph {et~al.}(2007)\citenamefont {Loh},
  \citenamefont {Khalil}, \citenamefont {Correa}, \citenamefont {Santra},
  \citenamefont {Buth},\ and\ \citenamefont {Leone}}]{PhysRevLett.98.143601}%
  \BibitemOpen
  \bibfield  {author} {\bibinfo {author} {\bibfnamefont {Z.-H.}\ \bibnamefont
  {Loh}}, \bibinfo {author} {\bibfnamefont {M.}~\bibnamefont {Khalil}},
  \bibinfo {author} {\bibfnamefont {R.~E.}\ \bibnamefont {Correa}}, \bibinfo
  {author} {\bibfnamefont {R.}~\bibnamefont {Santra}}, \bibinfo {author}
  {\bibfnamefont {C.}~\bibnamefont {Buth}}, \ and\ \bibinfo {author}
  {\bibfnamefont {S.~R.}\ \bibnamefont {Leone}},\ }\bibfield  {title} {\enquote
  {\bibinfo {title} {Quantum state-resolved probing of strong-field-ionized
  xenon atoms using femtosecond high-order harmonic transient absorption
  spectroscopy},}\ }\href {\doibase 10.1103/PhysRevLett.98.143601} {\bibfield
  {journal} {\bibinfo  {journal} {Phys. Rev. Lett.}\ }\textbf {\bibinfo
  {volume} {98}},\ \bibinfo {pages} {143601} (\bibinfo {year}
  {2007})}\BibitemShut {NoStop}%
\bibitem [{\citenamefont {Goulielmakis}\ \emph {et~al.}(2010)\citenamefont
  {Goulielmakis}, \citenamefont {Loh}, \citenamefont {Wirth}, \citenamefont
  {Santra}, \citenamefont {Rohringer}, \citenamefont {Yakovlev}, \citenamefont
  {Zherebtsov}, \citenamefont {Pfeifer}, \citenamefont {Azzeer}, \citenamefont
  {Kling}, \citenamefont {Leone},\ and\ \citenamefont
  {Krausz}}]{GoulielmakisNature466}%
  \BibitemOpen
  \bibfield  {author} {\bibinfo {author} {\bibfnamefont {E.}~\bibnamefont
  {Goulielmakis}}, \bibinfo {author} {\bibfnamefont {Z.-H.}\ \bibnamefont
  {Loh}}, \bibinfo {author} {\bibfnamefont {A.}~\bibnamefont {Wirth}}, \bibinfo
  {author} {\bibfnamefont {R.}~\bibnamefont {Santra}}, \bibinfo {author}
  {\bibfnamefont {N.}~\bibnamefont {Rohringer}}, \bibinfo {author}
  {\bibfnamefont {V.~S.}\ \bibnamefont {Yakovlev}}, \bibinfo {author}
  {\bibfnamefont {S.}~\bibnamefont {Zherebtsov}}, \bibinfo {author}
  {\bibfnamefont {T.}~\bibnamefont {Pfeifer}}, \bibinfo {author} {\bibfnamefont
  {A.~M.}\ \bibnamefont {Azzeer}}, \bibinfo {author} {\bibfnamefont {M.~F.}\
  \bibnamefont {Kling}}, \bibinfo {author} {\bibfnamefont {S.~R.}\ \bibnamefont
  {Leone}}, \ and\ \bibinfo {author} {\bibfnamefont {F.}~\bibnamefont
  {Krausz}},\ }\bibfield  {title} {\enquote {\bibinfo {title} {Real-time
  observation of valence electron motion},}\ }\href
  {http://dx.doi.org/10.1038/nature09212} {\bibfield  {journal} {\bibinfo
  {journal} {Nature (London)}\ }\textbf {\bibinfo {volume} {466}},\ \bibinfo
  {pages} {739--743} (\bibinfo {year} {2010})}\BibitemShut {NoStop}%
\bibitem [{\citenamefont {Wang}\ \emph {et~al.}(2010)\citenamefont {Wang},
  \citenamefont {Chini}, \citenamefont {Chen}, \citenamefont {Zhang},
  \citenamefont {He}, \citenamefont {Cheng}, \citenamefont {Wu}, \citenamefont
  {Thumm},\ and\ \citenamefont {Chang}}]{PhysRevLett.105.143002}%
  \BibitemOpen
  \bibfield  {author} {\bibinfo {author} {\bibfnamefont {H.}~\bibnamefont
  {Wang}}, \bibinfo {author} {\bibfnamefont {M.}~\bibnamefont {Chini}},
  \bibinfo {author} {\bibfnamefont {S.}~\bibnamefont {Chen}}, \bibinfo {author}
  {\bibfnamefont {C.-H.}\ \bibnamefont {Zhang}}, \bibinfo {author}
  {\bibfnamefont {F.}~\bibnamefont {He}}, \bibinfo {author} {\bibfnamefont
  {Y.}~\bibnamefont {Cheng}}, \bibinfo {author} {\bibfnamefont
  {Y.}~\bibnamefont {Wu}}, \bibinfo {author} {\bibfnamefont {U.}~\bibnamefont
  {Thumm}}, \ and\ \bibinfo {author} {\bibfnamefont {Z.}~\bibnamefont
  {Chang}},\ }\bibfield  {title} {\enquote {\bibinfo {title} {Attosecond
  time-resolved autoionization of argon},}\ }\href {\doibase
  10.1103/PhysRevLett.105.143002} {\bibfield  {journal} {\bibinfo  {journal}
  {Phys. Rev. Lett.}\ }\textbf {\bibinfo {volume} {105}},\ \bibinfo {pages}
  {143002} (\bibinfo {year} {2010})}\BibitemShut {NoStop}%
\bibitem [{\citenamefont {Wirth}\ \emph {et~al.}(2011)\citenamefont {Wirth},
  \citenamefont {Hassan}, \citenamefont {Grgura{\v s}}, \citenamefont {Gagnon},
  \citenamefont {Moulet}, \citenamefont {Luu}, \citenamefont {Pabst},
  \citenamefont {Santra}, \citenamefont {Alahmed}, \citenamefont {Azzeer},
  \citenamefont {Yakovlev}, \citenamefont {Pervak}, \citenamefont {Krausz},\
  and\ \citenamefont {Goulielmakis}}]{Wirth195}%
  \BibitemOpen
  \bibfield  {author} {\bibinfo {author} {\bibfnamefont {A.}~\bibnamefont
  {Wirth}}, \bibinfo {author} {\bibfnamefont {M.~Th.}\ \bibnamefont {Hassan}},
  \bibinfo {author} {\bibfnamefont {I.}~\bibnamefont {Grgura{\v s}}}, \bibinfo
  {author} {\bibfnamefont {J.}~\bibnamefont {Gagnon}}, \bibinfo {author}
  {\bibfnamefont {A.}~\bibnamefont {Moulet}}, \bibinfo {author} {\bibfnamefont
  {T.~T.}\ \bibnamefont {Luu}}, \bibinfo {author} {\bibfnamefont
  {S.}~\bibnamefont {Pabst}}, \bibinfo {author} {\bibfnamefont
  {R.}~\bibnamefont {Santra}}, \bibinfo {author} {\bibfnamefont {Z.~A.}\
  \bibnamefont {Alahmed}}, \bibinfo {author} {\bibfnamefont {A.~M.}\
  \bibnamefont {Azzeer}}, \bibinfo {author} {\bibfnamefont {V.~S.}\
  \bibnamefont {Yakovlev}}, \bibinfo {author} {\bibfnamefont {V.}~\bibnamefont
  {Pervak}}, \bibinfo {author} {\bibfnamefont {F.}~\bibnamefont {Krausz}}, \
  and\ \bibinfo {author} {\bibfnamefont {E.}~\bibnamefont {Goulielmakis}},\
  }\bibfield  {title} {\enquote {\bibinfo {title} {Synthesized light
  transients},}\ }\href {\doibase 10.1126/science.1210268} {\bibfield
  {journal} {\bibinfo  {journal} {Science}\ }\textbf {\bibinfo {volume}
  {334}},\ \bibinfo {pages} {195--200} (\bibinfo {year} {2011})}\BibitemShut
  {NoStop}%
\bibitem [{\citenamefont {Holler}\ \emph {et~al.}(2011)\citenamefont {Holler},
  \citenamefont {Schapper}, \citenamefont {Gallmann},\ and\ \citenamefont
  {Keller}}]{PhysRevLett.106.123601}%
  \BibitemOpen
  \bibfield  {author} {\bibinfo {author} {\bibfnamefont {M.}~\bibnamefont
  {Holler}}, \bibinfo {author} {\bibfnamefont {F.}~\bibnamefont {Schapper}},
  \bibinfo {author} {\bibfnamefont {L.}~\bibnamefont {Gallmann}}, \ and\
  \bibinfo {author} {\bibfnamefont {U.}~\bibnamefont {Keller}},\ }\bibfield
  {title} {\enquote {\bibinfo {title} {Attosecond electron wave-packet
  interference observed by transient absorption},}\ }\href {\doibase
  10.1103/PhysRevLett.106.123601} {\bibfield  {journal} {\bibinfo  {journal}
  {Phys. Rev. Lett.}\ }\textbf {\bibinfo {volume} {106}},\ \bibinfo {pages}
  {123601} (\bibinfo {year} {2011})}\BibitemShut {NoStop}%
\bibitem [{\citenamefont {Sabbar}\ \emph {et~al.}(2017)\citenamefont {Sabbar},
  \citenamefont {Timmers}, \citenamefont {Chen}, \citenamefont {Pymer},
  \citenamefont {Loh}, \citenamefont {Sayres}, \citenamefont {Pabst},
  \citenamefont {Santra},\ and\ \citenamefont {Leone}}]{Sabbar-NaturePhys}%
  \BibitemOpen
  \bibfield  {author} {\bibinfo {author} {\bibfnamefont {M.}~\bibnamefont
  {Sabbar}}, \bibinfo {author} {\bibfnamefont {H.}~\bibnamefont {Timmers}},
  \bibinfo {author} {\bibfnamefont {Y.-J.}\ \bibnamefont {Chen}}, \bibinfo
  {author} {\bibfnamefont {A.~K.}\ \bibnamefont {Pymer}}, \bibinfo {author}
  {\bibfnamefont {Z.-H.}\ \bibnamefont {Loh}}, \bibinfo {author} {\bibfnamefont
  {S.~G.}\ \bibnamefont {Sayres}}, \bibinfo {author} {\bibfnamefont
  {S.}~\bibnamefont {Pabst}}, \bibinfo {author} {\bibfnamefont
  {R.}~\bibnamefont {Santra}}, \ and\ \bibinfo {author} {\bibfnamefont {S.~R.}\
  \bibnamefont {Leone}},\ }\bibfield  {title} {\enquote {\bibinfo {title}
  {State-resolved attosecond reversible and irreversible dynamics in strong
  optical fields},}\ }\href {\doibase 10.1038/nphys4027} {\bibfield  {journal}
  {\bibinfo  {journal} {Nat. Physics}\ }\textbf {\bibinfo {volume} {13}},\
  \bibinfo {pages} {472--478} (\bibinfo {year} {2017})}\BibitemShut {NoStop}%
\bibitem [{\citenamefont {Warrick}\ \emph {et~al.}(2017)\citenamefont
  {Warrick}, \citenamefont {B\ae{}kh\o{}j}, \citenamefont {Cao}, \citenamefont
  {Fidler}, \citenamefont {Jensen}, \citenamefont {Madsen}, \citenamefont
  {Leone},\ and\ \citenamefont {Neumark}}]{WARRICK2017408}%
  \BibitemOpen
  \bibfield  {author} {\bibinfo {author} {\bibfnamefont {E.~R.}\ \bibnamefont
  {Warrick}}, \bibinfo {author} {\bibfnamefont {J.~E.}\ \bibnamefont
  {B\ae{}kh\o{}j}}, \bibinfo {author} {\bibfnamefont {W.}~\bibnamefont {Cao}},
  \bibinfo {author} {\bibfnamefont {A.~P.}\ \bibnamefont {Fidler}}, \bibinfo
  {author} {\bibfnamefont {F.}~\bibnamefont {Jensen}}, \bibinfo {author}
  {\bibfnamefont {L.~B.}\ \bibnamefont {Madsen}}, \bibinfo {author}
  {\bibfnamefont {S.~R.}\ \bibnamefont {Leone}}, \ and\ \bibinfo {author}
  {\bibfnamefont {D.~M.}\ \bibnamefont {Neumark}},\ }\bibfield  {title}
  {\enquote {\bibinfo {title} {Attosecond transient absorption spectroscopy of
  molecular nitrogen: Vibrational coherences in the $b'
  \,^1\sigma^+_{\mathrm{u}}$ state},}\ }\href {\doibase
  https://doi.org/10.1016/j.cplett.2017.02.013} {\bibfield  {journal} {\bibinfo
   {journal} {Chem. Phys. Lett.}\ }\textbf {\bibinfo {volume} {683}},\ \bibinfo
  {pages} {408--415} (\bibinfo {year} {2017})}\BibitemShut {NoStop}%
\bibitem [{\citenamefont {Reduzzi}\ \emph {et~al.}(2016)\citenamefont
  {Reduzzi}, \citenamefont {Chu}, \citenamefont {Feng}, \citenamefont
  {Dubrouil}, \citenamefont {Hummert}, \citenamefont {Calegari}, \citenamefont
  {Frassetto}, \citenamefont {Poletto}, \citenamefont {Kornilov}, \citenamefont
  {M.}, \citenamefont {Lin},\ and\ \citenamefont {Sansone}}]{ReduzziJPB}%
  \BibitemOpen
  \bibfield  {author} {\bibinfo {author} {\bibfnamefont {M.}~\bibnamefont
  {Reduzzi}}, \bibinfo {author} {\bibfnamefont {W.-C.}\ \bibnamefont {Chu}},
  \bibinfo {author} {\bibfnamefont {C.}~\bibnamefont {Feng}}, \bibinfo {author}
  {\bibfnamefont {A.}~\bibnamefont {Dubrouil}}, \bibinfo {author}
  {\bibfnamefont {J.}~\bibnamefont {Hummert}}, \bibinfo {author} {\bibfnamefont
  {F.}~\bibnamefont {Calegari}}, \bibinfo {author} {\bibfnamefont
  {F.}~\bibnamefont {Frassetto}}, \bibinfo {author} {\bibfnamefont
  {L.}~\bibnamefont {Poletto}}, \bibinfo {author} {\bibfnamefont
  {O.}~\bibnamefont {Kornilov}}, \bibinfo {author} {\bibfnamefont {Nisoli}\
  \bibnamefont {M.}}, \bibinfo {author} {\bibfnamefont {C.-D.}\ \bibnamefont
  {Lin}}, \ and\ \bibinfo {author} {\bibfnamefont {G.}~\bibnamefont
  {Sansone}},\ }\bibfield  {title} {\enquote {\bibinfo {title} {Observation of
  autoionization dynamics and sub-cycle quantum beating in electronic molecular
  wave packets},}\ }\href {https://doi.org/10.1088/0953-4075/49/6/065102}
  {\bibfield  {journal} {\bibinfo  {journal} {J. Phys. B}\ }\textbf {\bibinfo
  {volume} {49}},\ \bibinfo {pages} {065102} (\bibinfo {year}
  {2016})}\BibitemShut {NoStop}%
\bibitem [{\citenamefont {Cheng}\ \emph {et~al.}(2016)\citenamefont {Cheng},
  \citenamefont {Chini}, \citenamefont {Wang}, \citenamefont
  {Gonz\'alez-Castrillo}, \citenamefont {Palacios}, \citenamefont {Argenti},
  \citenamefont {Mart\'{\i}n},\ and\ \citenamefont
  {Chang}}]{PhysRevA.94.023403}%
  \BibitemOpen
  \bibfield  {author} {\bibinfo {author} {\bibfnamefont {Y.}~\bibnamefont
  {Cheng}}, \bibinfo {author} {\bibfnamefont {M.}~\bibnamefont {Chini}},
  \bibinfo {author} {\bibfnamefont {X.}~\bibnamefont {Wang}}, \bibinfo {author}
  {\bibfnamefont {A.}~\bibnamefont {Gonz\'alez-Castrillo}}, \bibinfo {author}
  {\bibfnamefont {A.}~\bibnamefont {Palacios}}, \bibinfo {author}
  {\bibfnamefont {L.}~\bibnamefont {Argenti}}, \bibinfo {author} {\bibfnamefont
  {F.}~\bibnamefont {Mart\'{\i}n}}, \ and\ \bibinfo {author} {\bibfnamefont
  {Z.}~\bibnamefont {Chang}},\ }\bibfield  {title} {\enquote {\bibinfo {title}
  {Reconstruction of an excited-state molecular wave packet with attosecond
  transient absorption spectroscopy},}\ }\href {\doibase
  10.1103/PhysRevA.94.023403} {\bibfield  {journal} {\bibinfo  {journal} {Phys.
  Rev. A}\ }\textbf {\bibinfo {volume} {94}},\ \bibinfo {pages} {023403}
  (\bibinfo {year} {2016})}\BibitemShut {NoStop}%
\bibitem [{\citenamefont {Schultze}\ \emph {et~al.}(2013)\citenamefont
  {Schultze}, \citenamefont {Bothschafter}, \citenamefont {Sommer},
  \citenamefont {Holzner}, \citenamefont {Schweinberger}, \citenamefont
  {Fiess}, \citenamefont {Hofstetter}, \citenamefont {Kienberger},
  \citenamefont {Apalkov}, \citenamefont {Yakovlev}, \citenamefont {Stockman},\
  and\ \citenamefont {Krausz}}]{Schultze-Nature}%
  \BibitemOpen
  \bibfield  {author} {\bibinfo {author} {\bibfnamefont {M.}~\bibnamefont
  {Schultze}}, \bibinfo {author} {\bibfnamefont {E.~M.}\ \bibnamefont
  {Bothschafter}}, \bibinfo {author} {\bibfnamefont {A.}~\bibnamefont
  {Sommer}}, \bibinfo {author} {\bibfnamefont {S.}~\bibnamefont {Holzner}},
  \bibinfo {author} {\bibfnamefont {W.}~\bibnamefont {Schweinberger}}, \bibinfo
  {author} {\bibfnamefont {M.}~\bibnamefont {Fiess}}, \bibinfo {author}
  {\bibfnamefont {M.}~\bibnamefont {Hofstetter}}, \bibinfo {author}
  {\bibfnamefont {R.}~\bibnamefont {Kienberger}}, \bibinfo {author}
  {\bibfnamefont {V.}~\bibnamefont {Apalkov}}, \bibinfo {author} {\bibfnamefont
  {V.~S.}\ \bibnamefont {Yakovlev}}, \bibinfo {author} {\bibfnamefont {M.~I.}\
  \bibnamefont {Stockman}}, \ and\ \bibinfo {author} {\bibfnamefont
  {F.}~\bibnamefont {Krausz}},\ }\bibfield  {title} {\enquote {\bibinfo {title}
  {Controlling dielectrics with the electric field of light},}\ }\href
  {\doibase 10.1038/nature11720} {\bibfield  {journal} {\bibinfo  {journal}
  {Nature (London)}\ }\textbf {\bibinfo {volume} {493}},\ \bibinfo {pages} {75}
  (\bibinfo {year} {2013})}\BibitemShut {NoStop}%
\bibitem [{\citenamefont {Schultze}\ \emph {et~al.}(2014)\citenamefont
  {Schultze}, \citenamefont {Ramasesha}, \citenamefont {Pemmaraju},
  \citenamefont {Sato}, \citenamefont {Whitmore}, \citenamefont {Gandman},
  \citenamefont {Prell}, \citenamefont {Borja}, \citenamefont {Prendergast},
  \citenamefont {Yabana}, \citenamefont {Neumark},\ and\ \citenamefont
  {Leone}}]{Schultze1348}%
  \BibitemOpen
  \bibfield  {author} {\bibinfo {author} {\bibfnamefont {M.}~\bibnamefont
  {Schultze}}, \bibinfo {author} {\bibfnamefont {K.}~\bibnamefont {Ramasesha}},
  \bibinfo {author} {\bibfnamefont {C.~D.}\ \bibnamefont {Pemmaraju}}, \bibinfo
  {author} {\bibfnamefont {S.~A.}\ \bibnamefont {Sato}}, \bibinfo {author}
  {\bibfnamefont {D.}~\bibnamefont {Whitmore}}, \bibinfo {author}
  {\bibfnamefont {A.}~\bibnamefont {Gandman}}, \bibinfo {author} {\bibfnamefont
  {J.~S.}\ \bibnamefont {Prell}}, \bibinfo {author} {\bibfnamefont {L.~J.}\
  \bibnamefont {Borja}}, \bibinfo {author} {\bibfnamefont {D.}~\bibnamefont
  {Prendergast}}, \bibinfo {author} {\bibfnamefont {K.}~\bibnamefont {Yabana}},
  \bibinfo {author} {\bibfnamefont {D.~M.}\ \bibnamefont {Neumark}}, \ and\
  \bibinfo {author} {\bibfnamefont {S.~R.}\ \bibnamefont {Leone}},\ }\bibfield
  {title} {\enquote {\bibinfo {title} {Attosecond band-gap dynamics in
  silicon},}\ }\href {\doibase 10.1126/science.1260311} {\bibfield  {journal}
  {\bibinfo  {journal} {Science}\ }\textbf {\bibinfo {volume} {346}},\ \bibinfo
  {pages} {1348--1352} (\bibinfo {year} {2014})}\BibitemShut {NoStop}%
\bibitem [{\citenamefont {Lucchini}\ \emph {et~al.}(2016)\citenamefont
  {Lucchini}, \citenamefont {Sato}, \citenamefont {Ludwig}, \citenamefont
  {Herrmann}, \citenamefont {Volkov}, \citenamefont {Kasmi}, \citenamefont
  {Shinohara}, \citenamefont {Yabana}, \citenamefont {Gallmann},\ and\
  \citenamefont {Keller}}]{Lucchini916}%
  \BibitemOpen
  \bibfield  {author} {\bibinfo {author} {\bibfnamefont {M.}~\bibnamefont
  {Lucchini}}, \bibinfo {author} {\bibfnamefont {S.~A.}\ \bibnamefont {Sato}},
  \bibinfo {author} {\bibfnamefont {A.}~\bibnamefont {Ludwig}}, \bibinfo
  {author} {\bibfnamefont {J.}~\bibnamefont {Herrmann}}, \bibinfo {author}
  {\bibfnamefont {M.}~\bibnamefont {Volkov}}, \bibinfo {author} {\bibfnamefont
  {L.}~\bibnamefont {Kasmi}}, \bibinfo {author} {\bibfnamefont
  {Y.}~\bibnamefont {Shinohara}}, \bibinfo {author} {\bibfnamefont
  {K.}~\bibnamefont {Yabana}}, \bibinfo {author} {\bibfnamefont
  {L.}~\bibnamefont {Gallmann}}, \ and\ \bibinfo {author} {\bibfnamefont
  {U.}~\bibnamefont {Keller}},\ }\bibfield  {title} {\enquote {\bibinfo {title}
  {Attosecond dynamical franz-keldysh effect in polycrystalline diamond},}\
  }\href {\doibase 10.1126/science.aag1268} {\bibfield  {journal} {\bibinfo
  {journal} {Science}\ }\textbf {\bibinfo {volume} {353}},\ \bibinfo {pages}
  {916--919} (\bibinfo {year} {2016})}\BibitemShut {NoStop}%
\bibitem [{\citenamefont {Moulet}\ \emph {et~al.}(2017)\citenamefont {Moulet},
  \citenamefont {Bertrand}, \citenamefont {Klostermann}, \citenamefont
  {Guggenmos}, \citenamefont {Karpowicz},\ and\ \citenamefont
  {Goulielmakis}}]{Moulet1134}%
  \BibitemOpen
  \bibfield  {author} {\bibinfo {author} {\bibfnamefont {A.}~\bibnamefont
  {Moulet}}, \bibinfo {author} {\bibfnamefont {J.~B.}\ \bibnamefont
  {Bertrand}}, \bibinfo {author} {\bibfnamefont {T.}~\bibnamefont
  {Klostermann}}, \bibinfo {author} {\bibfnamefont {A.}~\bibnamefont
  {Guggenmos}}, \bibinfo {author} {\bibfnamefont {N.}~\bibnamefont
  {Karpowicz}}, \ and\ \bibinfo {author} {\bibfnamefont {E.}~\bibnamefont
  {Goulielmakis}},\ }\bibfield  {title} {\enquote {\bibinfo {title} {Soft x-ray
  excitonics},}\ }\href {\doibase 10.1126/science.aan4737} {\bibfield
  {journal} {\bibinfo  {journal} {Science}\ }\textbf {\bibinfo {volume}
  {357}},\ \bibinfo {pages} {1134--1138} (\bibinfo {year} {2017})}\BibitemShut
  {NoStop}%
\bibitem [{\citenamefont {Mathies}\ \emph {et~al.}(1988)\citenamefont
  {Mathies}, \citenamefont {Brito~Cruz}, \citenamefont {Pollard},\ and\
  \citenamefont {Shank}}]{Mathies06051988}%
  \BibitemOpen
  \bibfield  {author} {\bibinfo {author} {\bibfnamefont {R.~A.}\ \bibnamefont
  {Mathies}}, \bibinfo {author} {\bibfnamefont {C.~H.}\ \bibnamefont
  {Brito~Cruz}}, \bibinfo {author} {\bibfnamefont {W.~T.}\ \bibnamefont
  {Pollard}}, \ and\ \bibinfo {author} {\bibfnamefont {C.~V.}\ \bibnamefont
  {Shank}},\ }\bibfield  {title} {\enquote {\bibinfo {title} {Direct
  observation of the femtosecond excited-state cis-trans isomerization in
  bacteriorhodopsin},}\ }\href {\doibase 10.1126/science.3363359} {\bibfield
  {journal} {\bibinfo  {journal} {Science}\ }\textbf {\bibinfo {volume}
  {240}},\ \bibinfo {pages} {777--779} (\bibinfo {year} {1988})}\BibitemShut
  {NoStop}%
\bibitem [{\citenamefont {Chen}\ \emph {et~al.}(2012)\citenamefont {Chen},
  \citenamefont {Bell}, \citenamefont {Beck}, \citenamefont {Mashiko},
  \citenamefont {Wu}, \citenamefont {Pfeiffer}, \citenamefont {Gaarde},
  \citenamefont {Neumark}, \citenamefont {Leone},\ and\ \citenamefont
  {Schafer}}]{PhysRevA.86.063408}%
  \BibitemOpen
  \bibfield  {author} {\bibinfo {author} {\bibfnamefont {S.}~\bibnamefont
  {Chen}}, \bibinfo {author} {\bibfnamefont {M.~J.}\ \bibnamefont {Bell}},
  \bibinfo {author} {\bibfnamefont {A.~R.}\ \bibnamefont {Beck}}, \bibinfo
  {author} {\bibfnamefont {H.}~\bibnamefont {Mashiko}}, \bibinfo {author}
  {\bibfnamefont {M.}~\bibnamefont {Wu}}, \bibinfo {author} {\bibfnamefont
  {A.~N.}\ \bibnamefont {Pfeiffer}}, \bibinfo {author} {\bibfnamefont {M.~B.}\
  \bibnamefont {Gaarde}}, \bibinfo {author} {\bibfnamefont {D.~M.}\
  \bibnamefont {Neumark}}, \bibinfo {author} {\bibfnamefont {S.~R.}\
  \bibnamefont {Leone}}, \ and\ \bibinfo {author} {\bibfnamefont {K.~J.}\
  \bibnamefont {Schafer}},\ }\bibfield  {title} {\enquote {\bibinfo {title}
  {Light-induced states in attosecond transient absorption spectra of
  laser-dressed helium},}\ }\href {\doibase 10.1103/PhysRevA.86.063408}
  {\bibfield  {journal} {\bibinfo  {journal} {Phys. Rev. A}\ }\textbf {\bibinfo
  {volume} {86}},\ \bibinfo {pages} {063408} (\bibinfo {year}
  {2012})}\BibitemShut {NoStop}%
\bibitem [{\citenamefont {Chini}\ \emph {et~al.}(2013)\citenamefont {Chini},
  \citenamefont {Wang}, \citenamefont {Cheng}, \citenamefont {Wu},
  \citenamefont {Zhao}, \citenamefont {Telnov}, \citenamefont {Chu},\ and\
  \citenamefont {Chang}}]{Chini-SciRep}%
  \BibitemOpen
  \bibfield  {author} {\bibinfo {author} {\bibfnamefont {M.}~\bibnamefont
  {Chini}}, \bibinfo {author} {\bibfnamefont {X.}~\bibnamefont {Wang}},
  \bibinfo {author} {\bibfnamefont {Y.}~\bibnamefont {Cheng}}, \bibinfo
  {author} {\bibfnamefont {Y.}~\bibnamefont {Wu}}, \bibinfo {author}
  {\bibfnamefont {D.}~\bibnamefont {Zhao}}, \bibinfo {author} {\bibfnamefont
  {D.~A.}\ \bibnamefont {Telnov}}, \bibinfo {author} {\bibfnamefont {S.-I.}\
  \bibnamefont {Chu}}, \ and\ \bibinfo {author} {\bibfnamefont
  {Z.}~\bibnamefont {Chang}},\ }\bibfield  {title} {\enquote {\bibinfo {title}
  {Sub-cycle oscillations in virtual states brought to light},}\ }\href
  {http://dx.doi.org/10.1038/srep01105} {\bibfield  {journal} {\bibinfo
  {journal} {Sci. Rep.}\ }\textbf {\bibinfo {volume} {3}},\ \bibinfo {pages}
  {1105--} (\bibinfo {year} {2013})}\BibitemShut {NoStop}%
\bibitem [{\citenamefont {R\o{}rstad}\ \emph {et~al.}(2017)\citenamefont
  {R\o{}rstad}, \citenamefont {B\ae{}kh\o{}j},\ and\ \citenamefont
  {Madsen}}]{PhysRevA.96.013430}%
  \BibitemOpen
  \bibfield  {author} {\bibinfo {author} {\bibfnamefont {J.~J.}\ \bibnamefont
  {R\o{}rstad}}, \bibinfo {author} {\bibfnamefont {J.~E.}\ \bibnamefont
  {B\ae{}kh\o{}j}}, \ and\ \bibinfo {author} {\bibfnamefont {L.~B.}\
  \bibnamefont {Madsen}},\ }\bibfield  {title} {\enquote {\bibinfo {title}
  {Analytic modeling of structures in attosecond transient-absorption
  spectra},}\ }\href {\doibase 10.1103/PhysRevA.96.013430} {\bibfield
  {journal} {\bibinfo  {journal} {Phys. Rev. A}\ }\textbf {\bibinfo {volume}
  {96}},\ \bibinfo {pages} {013430} (\bibinfo {year} {2017})}\BibitemShut
  {NoStop}%
\bibitem [{\citenamefont {Stoo{\ss}}\ \emph {et~al.}(2018)\citenamefont
  {Stoo{\ss}}, \citenamefont {Cavaletto}, \citenamefont {Donsa}, \citenamefont
  {Bl\"{a}ttermann}, \citenamefont {Birk}, \citenamefont {Keitel},
  \citenamefont {Brezinov\'{a}}, \citenamefont {Burgd\"{o}rfer}, \citenamefont
  {Ott},\ and\ \citenamefont {Pfeifer}}]{stooss_reconstructing}%
  \BibitemOpen
  \bibfield  {author} {\bibinfo {author} {\bibfnamefont {V.}~\bibnamefont
  {Stoo{\ss}}}, \bibinfo {author} {\bibfnamefont {S.~M.}\ \bibnamefont
  {Cavaletto}}, \bibinfo {author} {\bibfnamefont {S.}~\bibnamefont {Donsa}},
  \bibinfo {author} {\bibfnamefont {A.}~\bibnamefont {Bl\"{a}ttermann}},
  \bibinfo {author} {\bibfnamefont {P.}~\bibnamefont {Birk}}, \bibinfo {author}
  {\bibfnamefont {C.~H.}\ \bibnamefont {Keitel}}, \bibinfo {author}
  {\bibfnamefont {I.}~\bibnamefont {Brezinov\'{a}}}, \bibinfo {author}
  {\bibfnamefont {J.}~\bibnamefont {Burgd\"{o}rfer}}, \bibinfo {author}
  {\bibfnamefont {C.}~\bibnamefont {Ott}}, \ and\ \bibinfo {author}
  {\bibfnamefont {T.}~\bibnamefont {Pfeifer}},\ }\href
  {https://journals.aps.org/prl/accepted/0d073Y9fIdb1de5ca5a46586320491899104df693}
  {\bibfield  {journal} {\bibinfo  {journal} {Phys. Rev. Lett., in print}\ }
  (\bibinfo {year} {2018})}\BibitemShut {NoStop}%
\bibitem [{\citenamefont {Liu}\ \emph {et~al.}(2015)\citenamefont {Liu},
  \citenamefont {Cavaletto}, \citenamefont {Ott}, \citenamefont {Meyer},
  \citenamefont {Mi}, \citenamefont {Harman}, \citenamefont {Keitel},\ and\
  \citenamefont {Pfeifer}}]{PhysRevLett.115.033003}%
  \BibitemOpen
  \bibfield  {author} {\bibinfo {author} {\bibfnamefont {Z.}~\bibnamefont
  {Liu}}, \bibinfo {author} {\bibfnamefont {S.~M.}\ \bibnamefont {Cavaletto}},
  \bibinfo {author} {\bibfnamefont {C.}~\bibnamefont {Ott}}, \bibinfo {author}
  {\bibfnamefont {K.}~\bibnamefont {Meyer}}, \bibinfo {author} {\bibfnamefont
  {Y.}~\bibnamefont {Mi}}, \bibinfo {author} {\bibfnamefont {Z.}~\bibnamefont
  {Harman}}, \bibinfo {author} {\bibfnamefont {C.~H.}\ \bibnamefont {Keitel}},
  \ and\ \bibinfo {author} {\bibfnamefont {T.}~\bibnamefont {Pfeifer}},\
  }\bibfield  {title} {\enquote {\bibinfo {title} {Phase reconstruction of
  strong-field excited systems by transient-absorption spectroscopy},}\ }\href
  {\doibase 10.1103/PhysRevLett.115.033003} {\bibfield  {journal} {\bibinfo
  {journal} {Phys. Rev. Lett.}\ }\textbf {\bibinfo {volume} {115}},\ \bibinfo
  {pages} {033003} (\bibinfo {year} {2015})}\BibitemShut {NoStop}%
\bibitem [{\citenamefont {Liu}\ \emph {et~al.}(2017)\citenamefont {Liu},
  \citenamefont {Wang}, \citenamefont {Ding}, \citenamefont {Cavaletto},
  \citenamefont {Pfeifer},\ and\ \citenamefont {Hu}}]{Liu-SciRep}%
  \BibitemOpen
  \bibfield  {author} {\bibinfo {author} {\bibfnamefont {Z.}~\bibnamefont
  {Liu}}, \bibinfo {author} {\bibfnamefont {Q.}~\bibnamefont {Wang}}, \bibinfo
  {author} {\bibfnamefont {J.}~\bibnamefont {Ding}}, \bibinfo {author}
  {\bibfnamefont {S.~M.}\ \bibnamefont {Cavaletto}}, \bibinfo {author}
  {\bibfnamefont {T.}~\bibnamefont {Pfeifer}}, \ and\ \bibinfo {author}
  {\bibfnamefont {B.}~\bibnamefont {Hu}},\ }\bibfield  {title} {\enquote
  {\bibinfo {title} {Observation and quantification of the quantum dynamics of
  a strong-field excited multi-level system},}\ }\href
  {http://dx.doi.org/10.1038/srep39993} {\bibfield  {journal} {\bibinfo
  {journal} {Sci. Rep.}\ }\textbf {\bibinfo {volume} {7}},\ \bibinfo {pages}
  {39993--} (\bibinfo {year} {2017})}\BibitemShut {NoStop}%
\bibitem [{\citenamefont {Cavaletto}\ \emph {et~al.}(2017)\citenamefont
  {Cavaletto}, \citenamefont {Harman}, \citenamefont {Pfeifer},\ and\
  \citenamefont {Keitel}}]{PhysRevA.95.043413}%
  \BibitemOpen
  \bibfield  {author} {\bibinfo {author} {\bibfnamefont {S.~M.}\ \bibnamefont
  {Cavaletto}}, \bibinfo {author} {\bibfnamefont {Z.}~\bibnamefont {Harman}},
  \bibinfo {author} {\bibfnamefont {T.}~\bibnamefont {Pfeifer}}, \ and\
  \bibinfo {author} {\bibfnamefont {C.~H.}\ \bibnamefont {Keitel}},\ }\bibfield
   {title} {\enquote {\bibinfo {title} {Deterministic strong-field quantum
  control},}\ }\href {\doibase 10.1103/PhysRevA.95.043413} {\bibfield
  {journal} {\bibinfo  {journal} {Phys. Rev. A}\ }\textbf {\bibinfo {volume}
  {95}},\ \bibinfo {pages} {043413} (\bibinfo {year} {2017})}\BibitemShut
  {NoStop}%
\bibitem [{\citenamefont {Becquet}\ and\ \citenamefont
  {Cavaletto}(2018)}]{0953-4075-51-3-035501}%
  \BibitemOpen
  \bibfield  {author} {\bibinfo {author} {\bibfnamefont {V.}~\bibnamefont
  {Becquet}}\ and\ \bibinfo {author} {\bibfnamefont {S.~M.}\ \bibnamefont
  {Cavaletto}},\ }\bibfield  {title} {\enquote {\bibinfo {title}
  {Transient-absorption phases with strong probe and pump pulses},}\ }\href
  {http://stacks.iop.org/0953-4075/51/i=3/a=035501} {\bibfield  {journal}
  {\bibinfo  {journal} {J. Phys. B.}\ }\textbf {\bibinfo {volume} {51}},\
  \bibinfo {pages} {035501} (\bibinfo {year} {2018})}\BibitemShut {NoStop}%
\bibitem [{\citenamefont {Chini}\ \emph {et~al.}(2012)\citenamefont {Chini},
  \citenamefont {Zhao}, \citenamefont {Wang}, \citenamefont {Cheng},
  \citenamefont {Hu},\ and\ \citenamefont {Chang}}]{PhysRevLett.109.073601}%
  \BibitemOpen
  \bibfield  {author} {\bibinfo {author} {\bibfnamefont {M.}~\bibnamefont
  {Chini}}, \bibinfo {author} {\bibfnamefont {B.}~\bibnamefont {Zhao}},
  \bibinfo {author} {\bibfnamefont {H.}~\bibnamefont {Wang}}, \bibinfo {author}
  {\bibfnamefont {Y.}~\bibnamefont {Cheng}}, \bibinfo {author} {\bibfnamefont
  {S.~X.}\ \bibnamefont {Hu}}, \ and\ \bibinfo {author} {\bibfnamefont
  {Z.}~\bibnamefont {Chang}},\ }\bibfield  {title} {\enquote {\bibinfo {title}
  {Subcycle ac stark shift of helium excited states probed with isolated
  attosecond pulses},}\ }\href {\doibase 10.1103/PhysRevLett.109.073601}
  {\bibfield  {journal} {\bibinfo  {journal} {Phys. Rev. Lett.}\ }\textbf
  {\bibinfo {volume} {109}},\ \bibinfo {pages} {073601} (\bibinfo {year}
  {2012})}\BibitemShut {NoStop}%
\bibitem [{\citenamefont {Chen}\ \emph {et~al.}(2013)\citenamefont {Chen},
  \citenamefont {Wu}, \citenamefont {Gaarde},\ and\ \citenamefont
  {Schafer}}]{PhysRevA.88.033409}%
  \BibitemOpen
  \bibfield  {author} {\bibinfo {author} {\bibfnamefont {S.}~\bibnamefont
  {Chen}}, \bibinfo {author} {\bibfnamefont {M.}~\bibnamefont {Wu}}, \bibinfo
  {author} {\bibfnamefont {M.~B.}\ \bibnamefont {Gaarde}}, \ and\ \bibinfo
  {author} {\bibfnamefont {K.~J.}\ \bibnamefont {Schafer}},\ }\bibfield
  {title} {\enquote {\bibinfo {title} {Laser-imposed phase in resonant
  absorption of an isolated attosecond pulse},}\ }\href {\doibase
  10.1103/PhysRevA.88.033409} {\bibfield  {journal} {\bibinfo  {journal} {Phys.
  Rev. A}\ }\textbf {\bibinfo {volume} {88}},\ \bibinfo {pages} {033409}
  (\bibinfo {year} {2013})}\BibitemShut {NoStop}%
\bibitem [{\citenamefont {Ott}\ \emph {et~al.}(2013)\citenamefont {Ott},
  \citenamefont {Kaldun}, \citenamefont {Raith}, \citenamefont {Meyer},
  \citenamefont {Laux}, \citenamefont {Evers}, \citenamefont {Keitel},
  \citenamefont {Greene},\ and\ \citenamefont {Pfeifer}}]{Ott10052013}%
  \BibitemOpen
  \bibfield  {author} {\bibinfo {author} {\bibfnamefont {C.}~\bibnamefont
  {Ott}}, \bibinfo {author} {\bibfnamefont {A.}~\bibnamefont {Kaldun}},
  \bibinfo {author} {\bibfnamefont {P.}~\bibnamefont {Raith}}, \bibinfo
  {author} {\bibfnamefont {K.}~\bibnamefont {Meyer}}, \bibinfo {author}
  {\bibfnamefont {M.}~\bibnamefont {Laux}}, \bibinfo {author} {\bibfnamefont
  {J.}~\bibnamefont {Evers}}, \bibinfo {author} {\bibfnamefont {C.~H.}\
  \bibnamefont {Keitel}}, \bibinfo {author} {\bibfnamefont {C.~H.}\
  \bibnamefont {Greene}}, \ and\ \bibinfo {author} {\bibfnamefont
  {T.}~\bibnamefont {Pfeifer}},\ }\bibfield  {title} {\enquote {\bibinfo
  {title} {{Lorentz meets Fano in spectral line shapes: A universal phase and
  its laser control}},}\ }\href {\doibase 10.1126/science.1234407} {\bibfield
  {journal} {\bibinfo  {journal} {Science}\ }\textbf {\bibinfo {volume}
  {340}},\ \bibinfo {pages} {716--720} (\bibinfo {year} {2013})}\BibitemShut
  {NoStop}%
\bibitem [{\citenamefont {Kaldun}\ \emph {et~al.}(2014)\citenamefont {Kaldun},
  \citenamefont {Ott}, \citenamefont {Bl\"attermann}, \citenamefont {Laux},
  \citenamefont {Meyer}, \citenamefont {Ding}, \citenamefont {Fischer},\ and\
  \citenamefont {Pfeifer}}]{PhysRevLett.112.103001}%
  \BibitemOpen
  \bibfield  {author} {\bibinfo {author} {\bibfnamefont {A.}~\bibnamefont
  {Kaldun}}, \bibinfo {author} {\bibfnamefont {C.}~\bibnamefont {Ott}},
  \bibinfo {author} {\bibfnamefont {A.}~\bibnamefont {Bl\"attermann}}, \bibinfo
  {author} {\bibfnamefont {M.}~\bibnamefont {Laux}}, \bibinfo {author}
  {\bibfnamefont {K.}~\bibnamefont {Meyer}}, \bibinfo {author} {\bibfnamefont
  {T.}~\bibnamefont {Ding}}, \bibinfo {author} {\bibfnamefont {A.}~\bibnamefont
  {Fischer}}, \ and\ \bibinfo {author} {\bibfnamefont {T.}~\bibnamefont
  {Pfeifer}},\ }\bibfield  {title} {\enquote {\bibinfo {title} {{Extracting
  phase and amplitude modifications of laser-coupled Fano resonances}},}\
  }\href {\doibase 10.1103/PhysRevLett.112.103001} {\bibfield  {journal}
  {\bibinfo  {journal} {Phys. Rev. Lett.}\ }\textbf {\bibinfo {volume} {112}},\
  \bibinfo {pages} {103001} (\bibinfo {year} {2014})}\BibitemShut {NoStop}%
\bibitem [{\citenamefont {Meyer}\ \emph {et~al.}(2015)\citenamefont {Meyer},
  \citenamefont {Liu}, \citenamefont {M\"uller}, \citenamefont {Mewes},
  \citenamefont {Dreuw}, \citenamefont {Buckup}, \citenamefont {Motzkus},\ and\
  \citenamefont {Pfeifer}}]{Meyer22122015}%
  \BibitemOpen
  \bibfield  {author} {\bibinfo {author} {\bibfnamefont {K.}~\bibnamefont
  {Meyer}}, \bibinfo {author} {\bibfnamefont {Z.}~\bibnamefont {Liu}}, \bibinfo
  {author} {\bibfnamefont {N.}~\bibnamefont {M\"uller}}, \bibinfo {author}
  {\bibfnamefont {J.-M.}\ \bibnamefont {Mewes}}, \bibinfo {author}
  {\bibfnamefont {A.}~\bibnamefont {Dreuw}}, \bibinfo {author} {\bibfnamefont
  {T.}~\bibnamefont {Buckup}}, \bibinfo {author} {\bibfnamefont
  {M.}~\bibnamefont {Motzkus}}, \ and\ \bibinfo {author} {\bibfnamefont
  {T.}~\bibnamefont {Pfeifer}},\ }\bibfield  {title} {\enquote {\bibinfo
  {title} {Signatures and control of strong-field dynamics in a complex
  system},}\ }\href {\doibase 10.1073/pnas.1509201112} {\bibfield  {journal}
  {\bibinfo  {journal} {Proc. Natl. Acad. Sci. U.S.A.}\ }\textbf {\bibinfo
  {volume} {112}},\ \bibinfo {pages} {15613--15618} (\bibinfo {year}
  {2015})}\BibitemShut {NoStop}%
\bibitem [{\citenamefont {Udem}\ \emph {et~al.}(2002)\citenamefont {Udem},
  \citenamefont {Holzwarth},\ and\ \citenamefont {H\"ansch}}]{Nature.416.233}%
  \BibitemOpen
  \bibfield  {author} {\bibinfo {author} {\bibfnamefont {T.}~\bibnamefont
  {Udem}}, \bibinfo {author} {\bibfnamefont {R.}~\bibnamefont {Holzwarth}}, \
  and\ \bibinfo {author} {\bibfnamefont {T.~W.}\ \bibnamefont {H\"ansch}},\
  }\bibfield  {title} {\enquote {\bibinfo {title} {Optical frequency
  metrology},}\ }\href {http://dx.doi.org/10.1038/416233a} {\bibfield
  {journal} {\bibinfo  {journal} {Nature (London)}\ }\textbf {\bibinfo {volume}
  {416}},\ \bibinfo {pages} {233--237} (\bibinfo {year} {2002})}\BibitemShut
  {NoStop}%
\bibitem [{\citenamefont {Cundiff}(2002)}]{0022-3727-35-8-201}%
  \BibitemOpen
  \bibfield  {author} {\bibinfo {author} {\bibfnamefont {S.~T.}\ \bibnamefont
  {Cundiff}},\ }\bibfield  {title} {\enquote {\bibinfo {title} {Phase
  stabilization of ultrashort optical pulses},}\ }\href
  {http://stacks.iop.org/0022-3727/35/i=8/a=201} {\bibfield  {journal}
  {\bibinfo  {journal} {J. Phys. D}\ }\textbf {\bibinfo {volume} {35}},\
  \bibinfo {pages} {R43} (\bibinfo {year} {2002})}\BibitemShut {NoStop}%
\bibitem [{\citenamefont {Cundiff}\ and\ \citenamefont
  {Ye}(2003)}]{RevModPhys.75.325}%
  \BibitemOpen
  \bibfield  {author} {\bibinfo {author} {\bibfnamefont {S.~T.}\ \bibnamefont
  {Cundiff}}\ and\ \bibinfo {author} {\bibfnamefont {J.}~\bibnamefont {Ye}},\
  }\bibfield  {title} {\enquote {\bibinfo {title} {\textit{Colloquium}:
  Femtosecond optical frequency combs},}\ }\href {\doibase
  10.1103/RevModPhys.75.325} {\bibfield  {journal} {\bibinfo  {journal} {Rev.
  Mod. Phys.}\ }\textbf {\bibinfo {volume} {75}},\ \bibinfo {pages} {325--342}
  (\bibinfo {year} {2003})}\BibitemShut {NoStop}%
\bibitem [{\citenamefont {Udem}\ \emph {et~al.}(1999)\citenamefont {Udem},
  \citenamefont {Reichert}, \citenamefont {Holzwarth},\ and\ \citenamefont
  {H\"ansch}}]{PhysRevLett.82.3568}%
  \BibitemOpen
  \bibfield  {author} {\bibinfo {author} {\bibfnamefont {T.}~\bibnamefont
  {Udem}}, \bibinfo {author} {\bibfnamefont {J.}~\bibnamefont {Reichert}},
  \bibinfo {author} {\bibfnamefont {R.}~\bibnamefont {Holzwarth}}, \ and\
  \bibinfo {author} {\bibfnamefont {T.~W.}\ \bibnamefont {H\"ansch}},\
  }\bibfield  {title} {\enquote {\bibinfo {title} {{Absolute optical frequency
  measurement of the cesium ${\mathit{D}}_{1}$ line with a mode-locked
  laser}},}\ }\href {\doibase 10.1103/PhysRevLett.82.3568} {\bibfield
  {journal} {\bibinfo  {journal} {Phys. Rev. Lett.}\ }\textbf {\bibinfo
  {volume} {82}},\ \bibinfo {pages} {3568--3571} (\bibinfo {year}
  {1999})}\BibitemShut {NoStop}%
\bibitem [{\citenamefont {Diddams}\ \emph {et~al.}(2001)\citenamefont
  {Diddams}, \citenamefont {Udem}, \citenamefont {Bergquist}, \citenamefont
  {Curtis}, \citenamefont {Drullinger}, \citenamefont {Hollberg}, \citenamefont
  {Itano}, \citenamefont {Lee}, \citenamefont {Oates}, \citenamefont {Vogel},\
  and\ \citenamefont {Wineland}}]{Diddams03082001}%
  \BibitemOpen
  \bibfield  {author} {\bibinfo {author} {\bibfnamefont {S.~A.}\ \bibnamefont
  {Diddams}}, \bibinfo {author} {\bibfnamefont {T.}~\bibnamefont {Udem}},
  \bibinfo {author} {\bibfnamefont {J.~C.}\ \bibnamefont {Bergquist}}, \bibinfo
  {author} {\bibfnamefont {E.~A.}\ \bibnamefont {Curtis}}, \bibinfo {author}
  {\bibfnamefont {R.~E.}\ \bibnamefont {Drullinger}}, \bibinfo {author}
  {\bibfnamefont {L.}~\bibnamefont {Hollberg}}, \bibinfo {author}
  {\bibfnamefont {W.~M.}\ \bibnamefont {Itano}}, \bibinfo {author}
  {\bibfnamefont {W.~D.}\ \bibnamefont {Lee}}, \bibinfo {author} {\bibfnamefont
  {C.~W.}\ \bibnamefont {Oates}}, \bibinfo {author} {\bibfnamefont {K.~R.}\
  \bibnamefont {Vogel}}, \ and\ \bibinfo {author} {\bibfnamefont {D.~J.}\
  \bibnamefont {Wineland}},\ }\bibfield  {title} {\enquote {\bibinfo {title}
  {{An optical clock based on a single trapped $^{199}\mathrm{Hg}^+$ ion}},}\
  }\href {\doibase 10.1126/science.1061171} {\bibfield  {journal} {\bibinfo
  {journal} {Science}\ }\textbf {\bibinfo {volume} {293}},\ \bibinfo {pages}
  {825--828} (\bibinfo {year} {2001})}\BibitemShut {NoStop}%
\bibitem [{\citenamefont {Baltu\v{s}ka}\ \emph {et~al.}(2003)\citenamefont
  {Baltu\v{s}ka}, \citenamefont {Udem}, \citenamefont {Uiberacker},
  \citenamefont {Hentschel}, \citenamefont {Goulielmakis}, \citenamefont
  {Gohle}, \citenamefont {Holzwarth}, \citenamefont {Yakovlev}, \citenamefont
  {Scrinzi}, \citenamefont {H\"ansch},\ and\ \citenamefont
  {Krausz}}]{Nature.421.611}%
  \BibitemOpen
  \bibfield  {author} {\bibinfo {author} {\bibfnamefont {A.}~\bibnamefont
  {Baltu\v{s}ka}}, \bibinfo {author} {\bibfnamefont {T.}~\bibnamefont {Udem}},
  \bibinfo {author} {\bibfnamefont {M.}~\bibnamefont {Uiberacker}}, \bibinfo
  {author} {\bibfnamefont {M.}~\bibnamefont {Hentschel}}, \bibinfo {author}
  {\bibfnamefont {E.}~\bibnamefont {Goulielmakis}}, \bibinfo {author}
  {\bibfnamefont {C.}~\bibnamefont {Gohle}}, \bibinfo {author} {\bibfnamefont
  {R.}~\bibnamefont {Holzwarth}}, \bibinfo {author} {\bibfnamefont {V.~S.}\
  \bibnamefont {Yakovlev}}, \bibinfo {author} {\bibfnamefont {A.}~\bibnamefont
  {Scrinzi}}, \bibinfo {author} {\bibfnamefont {T.~W.}\ \bibnamefont
  {H\"ansch}}, \ and\ \bibinfo {author} {\bibfnamefont {F.}~\bibnamefont
  {Krausz}},\ }\bibfield  {title} {\enquote {\bibinfo {title} {Attosecond
  control of electronic processes by intense light fields},}\ }\href
  {http://dx.doi.org/10.1038/nature01414} {\bibfield  {journal} {\bibinfo
  {journal} {Nature (London)}\ }\textbf {\bibinfo {volume} {421}},\ \bibinfo
  {pages} {611--615} (\bibinfo {year} {2003})}\BibitemShut {NoStop}%
\bibitem [{\citenamefont {Stowe}\ \emph {et~al.}(2006)\citenamefont {Stowe},
  \citenamefont {Cruz}, \citenamefont {Marian},\ and\ \citenamefont
  {Ye}}]{PhysRevLett.96.153001}%
  \BibitemOpen
  \bibfield  {author} {\bibinfo {author} {\bibfnamefont {M.~C.}\ \bibnamefont
  {Stowe}}, \bibinfo {author} {\bibfnamefont {F.~C.}\ \bibnamefont {Cruz}},
  \bibinfo {author} {\bibfnamefont {A.}~\bibnamefont {Marian}}, \ and\ \bibinfo
  {author} {\bibfnamefont {J.}~\bibnamefont {Ye}},\ }\bibfield  {title}
  {\enquote {\bibinfo {title} {High resolution atomic coherent control via
  spectral phase manipulation of an optical frequency comb},}\ }\href {\doibase
  10.1103/PhysRevLett.96.153001} {\bibfield  {journal} {\bibinfo  {journal}
  {Phys. Rev. Lett.}\ }\textbf {\bibinfo {volume} {96}},\ \bibinfo {pages}
  {153001} (\bibinfo {year} {2006})}\BibitemShut {NoStop}%
\bibitem [{\citenamefont {Pe'er}\ \emph {et~al.}(2007)\citenamefont {Pe'er},
  \citenamefont {Shapiro}, \citenamefont {Stowe}, \citenamefont {Shapiro},\
  and\ \citenamefont {Ye}}]{PhysRevLett.98.113004}%
  \BibitemOpen
  \bibfield  {author} {\bibinfo {author} {\bibfnamefont {A.}~\bibnamefont
  {Pe'er}}, \bibinfo {author} {\bibfnamefont {E.~A.}\ \bibnamefont {Shapiro}},
  \bibinfo {author} {\bibfnamefont {M.~C.}\ \bibnamefont {Stowe}}, \bibinfo
  {author} {\bibfnamefont {M.}~\bibnamefont {Shapiro}}, \ and\ \bibinfo
  {author} {\bibfnamefont {J.}~\bibnamefont {Ye}},\ }\bibfield  {title}
  {\enquote {\bibinfo {title} {Precise control of molecular dynamics with a
  femtosecond frequency comb},}\ }\href {\doibase
  10.1103/PhysRevLett.98.113004} {\bibfield  {journal} {\bibinfo  {journal}
  {Phys. Rev. Lett.}\ }\textbf {\bibinfo {volume} {98}},\ \bibinfo {pages}
  {113004} (\bibinfo {year} {2007})}\BibitemShut {NoStop}%
\bibitem [{\citenamefont {Stowe}\ \emph {et~al.}(2008)\citenamefont {Stowe},
  \citenamefont {Pe'er},\ and\ \citenamefont {Ye}}]{PhysRevLett.100.203001}%
  \BibitemOpen
  \bibfield  {author} {\bibinfo {author} {\bibfnamefont {M.~C.}\ \bibnamefont
  {Stowe}}, \bibinfo {author} {\bibfnamefont {A.}~\bibnamefont {Pe'er}}, \ and\
  \bibinfo {author} {\bibfnamefont {J.}~\bibnamefont {Ye}},\ }\bibfield
  {title} {\enquote {\bibinfo {title} {Control of four-level quantum coherence
  via discrete spectral shaping of an optical frequency comb},}\ }\href
  {\doibase 10.1103/PhysRevLett.100.203001} {\bibfield  {journal} {\bibinfo
  {journal} {Phys. Rev. Lett.}\ }\textbf {\bibinfo {volume} {100}},\ \bibinfo
  {pages} {203001} (\bibinfo {year} {2008})}\BibitemShut {NoStop}%
\bibitem [{\citenamefont {Marian}\ \emph {et~al.}(2004)\citenamefont {Marian},
  \citenamefont {Stowe}, \citenamefont {Lawall}, \citenamefont {Felinto},\ and\
  \citenamefont {Ye}}]{Marian17122004}%
  \BibitemOpen
  \bibfield  {author} {\bibinfo {author} {\bibfnamefont {A.}~\bibnamefont
  {Marian}}, \bibinfo {author} {\bibfnamefont {M.~C.}\ \bibnamefont {Stowe}},
  \bibinfo {author} {\bibfnamefont {J.~R.}\ \bibnamefont {Lawall}}, \bibinfo
  {author} {\bibfnamefont {D.}~\bibnamefont {Felinto}}, \ and\ \bibinfo
  {author} {\bibfnamefont {J.}~\bibnamefont {Ye}},\ }\bibfield  {title}
  {\enquote {\bibinfo {title} {United time-frequency spectroscopy for dynamics
  and global structure},}\ }\href {\doibase 10.1126/science.1105660} {\bibfield
   {journal} {\bibinfo  {journal} {Science}\ }\textbf {\bibinfo {volume}
  {306}},\ \bibinfo {pages} {2063--2068} (\bibinfo {year} {2004})}\BibitemShut
  {NoStop}%
\bibitem [{\citenamefont {Cavaletto}\ \emph {et~al.}(2014)\citenamefont
  {Cavaletto}, \citenamefont {Harman}, \citenamefont {Ott}, \citenamefont
  {Buth}, \citenamefont {Pfeifer},\ and\ \citenamefont
  {Keitel}}]{nphoton.2014.113}%
  \BibitemOpen
  \bibfield  {author} {\bibinfo {author} {\bibfnamefont {S.~M.}\ \bibnamefont
  {Cavaletto}}, \bibinfo {author} {\bibfnamefont {Z.}~\bibnamefont {Harman}},
  \bibinfo {author} {\bibfnamefont {C.}~\bibnamefont {Ott}}, \bibinfo {author}
  {\bibfnamefont {C.}~\bibnamefont {Buth}}, \bibinfo {author} {\bibfnamefont
  {T.}~\bibnamefont {Pfeifer}}, \ and\ \bibinfo {author} {\bibfnamefont
  {C.~H.}\ \bibnamefont {Keitel}},\ }\bibfield  {title} {\enquote {\bibinfo
  {title} {Broadband high-resolution x-ray frequency combs},}\ }\href
  {http://www.nature.com/nphoton/journal/v8/n7/full/nphoton.2014.113.html}
  {\bibfield  {journal} {\bibinfo  {journal} {Nat. Photonics}\ }\textbf
  {\bibinfo {volume} {8}},\ \bibinfo {pages} {520} (\bibinfo {year}
  {2014})}\BibitemShut {NoStop}%
\bibitem [{\citenamefont {Liu}\ \emph {et~al.}(2014)\citenamefont {Liu},
  \citenamefont {Ott}, \citenamefont {Cavaletto}, \citenamefont {Harman},
  \citenamefont {Keitel},\ and\ \citenamefont
  {Pfeifer}}]{1367-2630-16-9-093005}%
  \BibitemOpen
  \bibfield  {author} {\bibinfo {author} {\bibfnamefont {Z.}~\bibnamefont
  {Liu}}, \bibinfo {author} {\bibfnamefont {C.}~\bibnamefont {Ott}}, \bibinfo
  {author} {\bibfnamefont {S.~M.}\ \bibnamefont {Cavaletto}}, \bibinfo {author}
  {\bibfnamefont {Z.}~\bibnamefont {Harman}}, \bibinfo {author} {\bibfnamefont
  {C.~H.}\ \bibnamefont {Keitel}}, \ and\ \bibinfo {author} {\bibfnamefont
  {T.}~\bibnamefont {Pfeifer}},\ }\bibfield  {title} {\enquote {\bibinfo
  {title} {Generation of high-frequency combs locked to atomic resonances by
  quantum phase modulation},}\ }\href
  {http://stacks.iop.org/1367-2630/16/i=9/a=093005} {\bibfield  {journal}
  {\bibinfo  {journal} {New J. Phys.}\ }\textbf {\bibinfo {volume} {16}},\
  \bibinfo {pages} {093005} (\bibinfo {year} {2014})}\BibitemShut {NoStop}%
\bibitem [{\citenamefont {Diels}\ and\ \citenamefont
  {Rudolph}(2006)}]{diels2006ultrashort}%
  \BibitemOpen
  \bibfield  {author} {\bibinfo {author} {\bibfnamefont {J.~C.}\ \bibnamefont
  {Diels}}\ and\ \bibinfo {author} {\bibfnamefont {W.}~\bibnamefont
  {Rudolph}},\ }\href@noop {} {\emph {\bibinfo {title} {Ultrashort laser pulse
  phenomena: fundamentals, techniques, and applications on a femtosecond time
  scale}}}\ (\bibinfo  {publisher} {Academic Press},\ \bibinfo {address}
  {Burlington, MA},\ \bibinfo {year} {2006})\BibitemShut {NoStop}%
\bibitem [{\citenamefont {Theodosiou}(1984)}]{PhysRevA.30.2881}%
  \BibitemOpen
  \bibfield  {author} {\bibinfo {author} {\bibfnamefont {C.~E.}\ \bibnamefont
  {Theodosiou}},\ }\bibfield  {title} {\enquote {\bibinfo {title} {{Lifetimes
  of alkali-metal\char22{}atom Rydberg states}},}\ }\href {\doibase
  10.1103/PhysRevA.30.2881} {\bibfield  {journal} {\bibinfo  {journal} {Phys.
  Rev. A}\ }\textbf {\bibinfo {volume} {30}},\ \bibinfo {pages} {2881--2909}
  (\bibinfo {year} {1984})}\BibitemShut {NoStop}%
\bibitem [{\citenamefont {Safronova}\ \emph {et~al.}(2004)\citenamefont
  {Safronova}, \citenamefont {Williams},\ and\ \citenamefont
  {Clark}}]{PhysRevA.69.022509}%
  \BibitemOpen
  \bibfield  {author} {\bibinfo {author} {\bibfnamefont {M.~S.}\ \bibnamefont
  {Safronova}}, \bibinfo {author} {\bibfnamefont {C.~J.}\ \bibnamefont
  {Williams}}, \ and\ \bibinfo {author} {\bibfnamefont {C.~W.}\ \bibnamefont
  {Clark}},\ }\bibfield  {title} {\enquote {\bibinfo {title} {Relativistic
  many-body calculations of electric-dipole matrix elements, lifetimes, and
  polarizabilities in rubidium},}\ }\href {\doibase 10.1103/PhysRevA.69.022509}
  {\bibfield  {journal} {\bibinfo  {journal} {Phys. Rev. A}\ }\textbf {\bibinfo
  {volume} {69}},\ \bibinfo {pages} {022509} (\bibinfo {year}
  {2004})}\BibitemShut {NoStop}%
\bibitem [{\citenamefont {Johnson}(2007)}]{johnson2007atomic}%
  \BibitemOpen
  \bibfield  {author} {\bibinfo {author} {\bibfnamefont {W.~R.}\ \bibnamefont
  {Johnson}},\ }\href@noop {} {\emph {\bibinfo {title} {Atomic Structure
  Theory: Lectures on Atomic Physics}}}\ (\bibinfo  {publisher} {Springer},\
  \bibinfo {address} {Berlin Heidelberg, New York},\ \bibinfo {year}
  {2007})\BibitemShut {NoStop}%
\bibitem [{\citenamefont {Scully}\ and\ \citenamefont
  {Zubairy}(1997)}]{Scully:QuantumOptics}%
  \BibitemOpen
  \bibfield  {author} {\bibinfo {author} {\bibfnamefont {M.~O.}\ \bibnamefont
  {Scully}}\ and\ \bibinfo {author} {\bibfnamefont {M.~S.}\ \bibnamefont
  {Zubairy}},\ }\href@noop {} {\emph {\bibinfo {title} {Quantum Optics}}}\
  (\bibinfo  {publisher} {Cambridge University Press},\ \bibinfo {address}
  {Cambridge},\ \bibinfo {year} {1997})\BibitemShut {NoStop}%
\bibitem [{\citenamefont {Foot}(2005)}]{Foot:AtomicPhysics}%
  \BibitemOpen
  \bibfield  {author} {\bibinfo {author} {\bibfnamefont {C.~J.}\ \bibnamefont
  {Foot}},\ }\href@noop {} {\emph {\bibinfo {title} {Atomic Physics}}}\
  (\bibinfo  {publisher} {Oxford University Press},\ \bibinfo {address}
  {Oxford},\ \bibinfo {year} {2005})\BibitemShut {NoStop}%
\bibitem [{\citenamefont {Kiffner}\ \emph {et~al.}(2010)\citenamefont
  {Kiffner}, \citenamefont {Macovei}, \citenamefont {Evers},\ and\
  \citenamefont {Keitel}}]{Kiffner_review}%
  \BibitemOpen
  \bibfield  {author} {\bibinfo {author} {\bibfnamefont {M.}~\bibnamefont
  {Kiffner}}, \bibinfo {author} {\bibfnamefont {M.}~\bibnamefont {Macovei}},
  \bibinfo {author} {\bibfnamefont {J.}~\bibnamefont {Evers}}, \ and\ \bibinfo
  {author} {\bibfnamefont {C.~H.}\ \bibnamefont {Keitel}},\ }\bibfield  {title}
  {\enquote {\bibinfo {title} {Vacuum-induced processes in multilevel atoms},}\
  }in\ \href@noop {} {\emph {\bibinfo {booktitle} {Prog. Opt.}}},\
  Vol.~\bibinfo {volume} {55},\ \bibinfo {editor} {edited by\ \bibinfo {editor}
  {\bibfnamefont {E.}~\bibnamefont {Wolf}}}\ (\bibinfo  {publisher}
  {Elsevier},\ \bibinfo {address} {Amsterdam},\ \bibinfo {year} {2010})\
  Chap.~\bibinfo {chapter} {3}, p.~\bibinfo {pages} {85}\BibitemShut {NoStop}%
\bibitem [{\citenamefont {Ilinova}\ and\ \citenamefont
  {Derevianko}(2012)}]{PhysRevA.86.013423}%
  \BibitemOpen
  \bibfield  {author} {\bibinfo {author} {\bibfnamefont {E.}~\bibnamefont
  {Ilinova}}\ and\ \bibinfo {author} {\bibfnamefont {A.}~\bibnamefont
  {Derevianko}},\ }\bibfield  {title} {\enquote {\bibinfo {title} {Dynamics of
  a three-level $\ensuremath{\Lambda}$-type system driven by trains of
  ultrashort laser pulses},}\ }\href {\doibase 10.1103/PhysRevA.86.013423}
  {\bibfield  {journal} {\bibinfo  {journal} {Phys. Rev. A}\ }\textbf {\bibinfo
  {volume} {86}},\ \bibinfo {pages} {013423} (\bibinfo {year}
  {2012})}\BibitemShut {NoStop}%
\bibitem [{\citenamefont {Horn}\ and\ \citenamefont
  {Johnson}(1991)}]{horn_johnson_1991}%
  \BibitemOpen
  \bibfield  {author} {\bibinfo {author} {\bibfnamefont {R.~A.}\ \bibnamefont
  {Horn}}\ and\ \bibinfo {author} {\bibfnamefont {C.~R.}\ \bibnamefont
  {Johnson}},\ }\enquote {\bibinfo {title} {Matrix equations and the kronecker
  product},}\ in\ \href {\doibase 10.1017/CBO9780511840371.005} {\emph
  {\bibinfo {booktitle} {Topics in Matrix Analysis}}}\ (\bibinfo  {publisher}
  {Cambridge University Press},\ \bibinfo {address} {Cambridge},\ \bibinfo
  {year} {1991})\ pp.\ \bibinfo {pages} {239--297}\BibitemShut {NoStop}%
\bibitem [{\citenamefont {Rey{-}de{-}Castro}\ \emph {et~al.}(2013)\citenamefont
  {Rey{-}de{-}Castro}, \citenamefont {Leghtas},\ and\ \citenamefont
  {Rabitz}}]{PhysRevLett.110.223601}%
  \BibitemOpen
  \bibfield  {author} {\bibinfo {author} {\bibfnamefont {R.}~\bibnamefont
  {Rey{-}de{-}Castro}}, \bibinfo {author} {\bibfnamefont {Z.}~\bibnamefont
  {Leghtas}}, \ and\ \bibinfo {author} {\bibfnamefont {H.}~\bibnamefont
  {Rabitz}},\ }\bibfield  {title} {\enquote {\bibinfo {title} {Manipulating
  quantum pathways on the fly},}\ }\href {\doibase
  10.1103/PhysRevLett.110.223601} {\bibfield  {journal} {\bibinfo  {journal}
  {Phys. Rev. Lett.}\ }\textbf {\bibinfo {volume} {110}},\ \bibinfo {pages}
  {223601} (\bibinfo {year} {2013})}\BibitemShut {NoStop}%
\bibitem [{\citenamefont {Liao}\ \emph {et~al.}(2015)\citenamefont {Liao},
  \citenamefont {Sandhu}, \citenamefont {Camp}, \citenamefont {Schafer},\ and\
  \citenamefont {Gaarde}}]{PhysRevLett.114.143002}%
  \BibitemOpen
  \bibfield  {author} {\bibinfo {author} {\bibfnamefont {C.-T.}\ \bibnamefont
  {Liao}}, \bibinfo {author} {\bibfnamefont {A.}~\bibnamefont {Sandhu}},
  \bibinfo {author} {\bibfnamefont {S.}~\bibnamefont {Camp}}, \bibinfo {author}
  {\bibfnamefont {K.~J.}\ \bibnamefont {Schafer}}, \ and\ \bibinfo {author}
  {\bibfnamefont {M.~B.}\ \bibnamefont {Gaarde}},\ }\bibfield  {title}
  {\enquote {\bibinfo {title} {Beyond the single-atom response in absorption
  line shapes: Probing a dense, laser-dressed helium gas with attosecond pulse
  trains},}\ }\href {\doibase 10.1103/PhysRevLett.114.143002} {\bibfield
  {journal} {\bibinfo  {journal} {Phys. Rev. Lett.}\ }\textbf {\bibinfo
  {volume} {114}},\ \bibinfo {pages} {143002} (\bibinfo {year}
  {2015})}\BibitemShut {NoStop}%
\bibitem [{\citenamefont {Gradshteyn}\ and\ \citenamefont
  {Ryzhik}(2007)}]{Gradshteyn}%
  \BibitemOpen
  \bibfield  {author} {\bibinfo {author} {\bibfnamefont {I.~S.}\ \bibnamefont
  {Gradshteyn}}\ and\ \bibinfo {author} {\bibfnamefont {I.~M.}\ \bibnamefont
  {Ryzhik}},\ }\href@noop {} {\emph {\bibinfo {title} {Table of Integrals,
  Series, and Products}}},\ edited by\ \bibinfo {editor} {\bibfnamefont
  {A.}~\bibnamefont {Jeffrey}}\ and\ \bibinfo {editor} {\bibfnamefont
  {D.}~\bibnamefont {Zwillinger}}\ (\bibinfo  {publisher} {Academic Press},\
  \bibinfo {address} {Burlington, MA},\ \bibinfo {year} {2007})\ p.~\bibinfo
  {pages} {48}\BibitemShut {NoStop}%
\end{thebibliography}


\end{document}